\documentclass[prd,aps,preprintnumbers,floatfix,showpacs,preprint,tightenlines,superscriptaddress]{revtex4}
\usepackage{epsfig}
\usepackage{graphics}
\usepackage{latexsym}
\usepackage{amsmath}
\usepackage{amssymb}
\usepackage{rotating}

\begin{document}

\preprint{ADP-05-04/T614}
\preprint{JLAB-THY-05-305}
\preprint{DESY 05-039}
%\preprint{
%\vbox{
%\hbox{ADP-04-XX/TXXX}
%\hbox{JLAB-THY-04-XX}
%\hbox{DESY 04-XXX}
%}}

\title{Search for the pentaquark resonance signature in lattice QCD}

\author{B.~G.~Lasscock}
\author{J.~Hedditch}
\affiliation{    Special Research Centre for the
                 Subatomic Structure of Matter,
                 and Department of Physics,
                 University of Adelaide, Adelaide SA 5005,
                 Australia}
\author{D.~B.~Leinweber}
\affiliation{    Special Research Centre for the
                 Subatomic Structure of Matter,
                 and Department of Physics,
                 University of Adelaide, Adelaide SA 5005,
                 Australia}
\author{W.~Melnitchouk}
\affiliation{Jefferson Lab, 12000 Jefferson Ave.,
             Newport News, VA 23606 USA}
\author{A.~W.~Thomas}
\affiliation{    Special Research Centre for the
                 Subatomic Structure of Matter,
                 and Department of Physics,
                 University of Adelaide, Adelaide SA 5005,
                 Australia}
\affiliation{Jefferson Lab, 12000 Jefferson Ave.,
             Newport News, VA 23606 USA}
\author{A.~G.~Williams}
\affiliation{    Special Research Centre for the
                 Subatomic Structure of Matter,
                 and Department of Physics,
                 University of Adelaide, Adelaide SA 5005,
                 Australia}
\author{R.~D.~Young}
\affiliation{    Special Research Centre for the
                 Subatomic Structure of Matter,
                 and Department of Physics,
                 University of Adelaide, Adelaide SA 5005,
                 Australia}
\affiliation{Jefferson Lab, 12000 Jefferson Ave.,
             Newport News, VA 23606 USA}
\author{J.~M.~Zanotti}
\affiliation{    Special Research Centre for the
                 Subatomic Structure of Matter,
                 and Department of Physics,
                 University of Adelaide, Adelaide SA 5005,
                 Australia}
\affiliation{	 John von Neumann-Institut f\"ur Computing NIC/DESY,
		 15738 Zeuthen, Germany}

\begin{abstract}
Claims concerning the possible discovery of the $\Theta^+$ pentaquark,
with minimal quark content $uudd\bar{s}$, have motivated our comprehensive
study into possible pentaquark states using lattice QCD.  We review
various pentaquark interpolating fields in the literature and create
a new candidate ideal for lattice QCD simulations. 
Using these interpolating fields we attempt to
isolate a signal for a five-quark resonance. Calculations are
performed using improved actions on a large $20^{3} \times 40$ lattice
in the quenched approximation.  The
standard lattice resonance signal of increasing attraction between 
baryon constituents
for increasing quark mass is not observed for
% >>>
% spin-1/2 pentaquark states. 
spin-${1\over 2}$ pentaquark states.
% <<<
We conclude that evidence supporting the existence of a
spin-${1\over 2}$ pentaquark resonance does not exist in quenched QCD.
\end{abstract}

\vspace{3mm}

\maketitle

%\newpage

%%%%%%%%%%%%%%%%%%%%%%%%%%%%%%%%%%%%%%%%%%%%%%%%%%%%%%%%%%%%%%%%%%%%%%%%%%
\section{Introduction}
%%%%%%%%%%%%%%%%%%%%%%%%%%%%%%%%%%%%%%%%%%%%%%%%%%%%%%%%%%%%%%%%%%%%%%%%%%

The recently reported observations of a baryon state with strangeness 
% >>>
% $S=+1$ some $\sim 100$~MeV above the $NK$ threshold has sparked 
$S=+1$ some 100~MeV above the $NK$ threshold has sparked 
% <<<
considerable interest in excited hadron spectroscopy.
Because this state has minimal quark content $uudd\bar s$, its discovery 
would be the first direct evidence for baryons with an exotic quark 
structure --- namely, baryons whose quantum numbers cannot be described
in terms of a 3-quark configuration alone.

% .........................................................................
\subsection{Phenomenology}
\label{ssec:I-phen}

The experimental findings were reported in real 
\cite{Nakano:2003qx,Stepanyan:2003qr,Kubarovsky:2003fi,Barth:2003es} and quasi-real photoproduction 
experiments \cite{Airapetian:2003ri}, and further positive sightings were reported
in $K$-nucleus collisions \cite{Barmin:2003vv}, $pp$ \cite{Abdel-Bary:2004ts} and $pA$ 
\cite{Aleev:2004sa,Aslanyan:2004gs} reactions, and in neutrino \cite{Asratyan:2003cb,NOMAD} and deep 
inelastic electron scattering \cite{Chekanov:2004kn}, for a total of around a
dozen positive results.
Currently only the charge and strangeness of this state, which has been 
labeled $\Theta^+$, are known; its spin, parity and isospin are as yet 
undetermined, although there are hints \cite{Lorenzon:2004rw} that it may be
isospin zero.
The mass of the $\Theta^+$ is found to be around
$M_{\Theta^+} = 1540$~MeV.
However, its most striking feature is its exceptionally narrow width.
In most cases the width has been smaller than the experimental
resolution, while analysis of $NK$ scattering data suggests that the
width cannot be greater than $\sim 1$~MeV \cite{Nussinov:2003ex,Arndt:2003xz,Haidenbauer:2003rw,Cahn:2003wq,Gibbs:2004ji}.
Such a narrow state, 100~MeV above threshold, presents a challenge to
most theoretical models
% >>>
\cite{Jennings:2003wz,Close:2004tp,Burns:2004wy,Maltman:2004sn}.
% <<<

Subsequently, a number of null results have been reported from $e^+ e^-$ 
\cite{Schael:2004nm,DELPHI,Armstrong:2004gp,Bai:2004gk,Abe:2004wf,Aubert:2004bm} and $p\bar p$ \cite{Litvintsev:2004yw} 
colliders, as well as from $pp$ \cite{E690}, $\gamma p$ \cite{Stenson:2004yz},
hadron-$p$ \cite{SELEX,Napolitano:2004mn}, hadron-nucleus \cite{Longo:2004gd}, $pA$ 
\cite{Knopfle:2004tu,Antipov:2004jz}, $\mu A$ \cite{COMPASS}, and nucleus-nucleus 
\cite{Pinkenburg:2004ux} fixed target experiments.
The production mechanism for the $\Theta^+$ in these reactions would 
be via fragmentation, and although the fragmentation functions are not 
known, these results suggest that if the $\Theta^+$ exists, its 
production mechanism, along with its quantum numbers, is exotic.
For more detailed accounts of the current experimental status of 
pentaquark searches see Refs.~\cite{Hicks:2004ge,Hicks:2005pm,Dzierba:2004db}.

While the experimental verification of the $\Theta^+$ and the 
determination of its quantum numbers await definitive confirmation,
it is timely to examine the theoretical predictions for the masses of
$S=+1$ pentaquark states.
Numerous model studies have been carried out recently aimed at revealing
the dynamical nature of the $\Theta^+$, ranging from Skyrmion models
\cite{Diakonov:1997mm,Praszalowicz:2003ik}, QCD sum rules \cite{Sugiyama:2003zk,Zhu:2003ba}, hadronic models 
\cite{Sibirtsev:2004bg,Oh:2004wp} and quark models
% >>>
\cite{Jaffe:2003sg,Stancu:2003if,Karliner:2003sy,Carlson:2003pn,Maltman:2004sn},
% <<<
to name just a few.

% .........................................................................
\subsection{Lattice pentaquark studies}
\label{ssec:I-lat}

While models can often be helpful in obtaining a qualitative 
understanding of data, we would like to see what QCD predicts for the 
masses of the pentaquark states.
Currently lattice QCD is the only quantitative method of obtaining 
hadronic properties directly from QCD, and several, mainly exploratory, 
studies of pentaquark masses have been performed 
\cite{Csikor:2003ng,Sasaki:2003gi,Mathur:2004jr,Ishii:2004qe,Ishii:2004ib,Alexandrou:2004ws,
Chiu:2004gg,Takahashi:2004sc,Csikor:2004us,Sasaki:2004vz}.

A crucial issue in lattice QCD analyses of excited hadrons is exactly 
what constitutes a signal for a resonance.
As evidence of a resonance, most lattice studies to date have sought 
to find the empirical mass splitting between the $\Theta^+$ and the
$NK$ threshold at the unphysically large quark masses used in lattice 
simulations.
This leads to the assumption that in the negative parity channel the 
% >>>
$\Theta^+$ will be about 100~MeV above the S-wave $NK$ 
% $\Theta^+$ will be about 100~MeV more massive than the S-wave $NK$ 
threshold.
% However as we will argue below, at sufficiently large quark masses
However, as we will argue below, at sufficiently large quark masses
% <<<
the true signal for a $\Theta^+$ resonance on the lattice, as with
all other excited states studied on the lattice \cite{Leinweber:2004it,Melnitchouk:2002eg,Zanotti:2003fx},
should be the presence of binding, in which case the resonance mass
% >>>
% would be {\em smaller} than the $NK$ threshold.
would be below the $NK$ threshold.
% <<<

% >>>
% In the positive parity channel where the $N$ and $K$ must be in a 
In the positive parity channel, where the $N$ and $K$ must be in a 
% <<<
relative P-wave, in a finite lattice volume the energy of the $NK$
state will typically be above the mass of the experimental $\Theta^+$
candidate.
Observation of a pentaquark mass below the P-wave $NK$ threshold would 
then be a clear signal for a $\Theta^+$ resonance.
% >>>
% With the exception of Chiu \& Hsieh \cite{Chiu:2004gg}, however, the positive 
% parity state found on the lattice is rejected as a candidate for the 
% $\Theta^+$ on the basis that its mass is too large.
In all of the lattice studies, with the exception of Chiu \& Hsieh
\cite{Chiu:2004gg}, the mass of the positive parity state has been found
to be too large to be interpreted as a candidate for the $\Theta^+$.
% <<<

One should note that in obtaining a relatively low mass positive parity
state, Chiu \& Hsieh \cite{Chiu:2004gg} perform a linear chiral extrapolation
of the pentaquark mass in $m_\pi^2$ using only the lightest few quark
masses, for which the errors are relatively large.
Although linear extrapolations of hadron properties are common in the 
literature, these invariably neglect the non-analyticities in $m_\pi^2$ 
arising from the long-range structure of hadrons associated with the 
% >>>
% pion cloud.
pion cloud \cite{Young:2002cj,Leinweber:2003dg}.
% <<<
Csikor {\it et al.} \cite{Csikor:2003ng} and Sasaki \cite{Sasaki:2003gi} use a slightly 
modified extrapolation, in which the squared pentaquark mass is fitted 
and extrapolated as a function of $m_\pi^2$.
% >>>
% As a cautionary note, since the chiral behavior of pentaquark masses 
As a cautionary note, since the chiral behaviour of pentaquark masses 
% <<<
is at present unknown, extrapolation of the lattice results to physical 
quark masses can lead to large systematic uncertainties, which are 
generally underestimated in the literature.

The ordering of the $NK$ and $\Theta^+$ states in the negative parity 
channel presents some challenges for lattice analyses.
Sasaki \cite{Sasaki:2004vz} and Csikor {\it et al.} \cite{Csikor:2004us} argue that 
if the $\Theta^+$ is more massive than the $NK$ threshold, then one 
% >>>
% needs to extract from the correlators more than the lightest state that 
% the operators have overlap with.
needs to extract from the correlators more than the lightest state with
which the operators have overlap.
% <<<
It has been suggested \cite{Sasaki:2004vz} that if one can find an operator 
that has negligible coupling to the $NK$ state, then one can fit the 
correlation function at intermediate Euclidean times to extract the mass 
of the heavier state.
The idea of simply choosing an operator that does not couple to the $NK$ 
threshold is problematic, however, because there is no way to determine 
{\it a priori} the extent to which an operator couples to a particular 
state.
Our approach instead will be to use a number of different interpolating 
fields, which will enhance the ability to couple to different states.
This approach has also been adopted by Fleming \cite{Fleming:2005jz}, and by
the MIT group \cite{MIT}.

Extracting multiple states in lattice QCD is usually achieved by 
performing a correlation matrix analysis, which we adopt in this work,
or via Bayesian techniques.
The analysis of Sasaki \cite{Sasaki:2003gi} uses a single interpolating field
and employs a standard analysis with an exponential fit to the 
correlation function.
Csikor {\it et al.} \cite{Csikor:2003ng}, Takahashi {\it et al.} \cite{Takahashi:2004sc} and 
% >>>
% Chiu \& Hsieh \cite{Chiu:2004gg}, have, on the other hand, performed
Chiu \& Hsieh \cite{Chiu:2004gg} have, on the other hand, performed
% <<<
correlation matrix analyses using several different interpolating
fields.
In the negative parity sector, these authors extract from their 
correlation matrices both a ground state and an excited state.
In all of these studies the positive parity state is found to lie 
significantly higher than the negative parity ground state.

Since the S-wave $NK$ scattering state lies very near the lowest
energy state observed on the lattice, the issue of extracting a
genuine $\Theta^+$ resonance from lattice simulations presents an
important challenge, and a number of ideas have been proposed to
distinguish between a true resonance state and the scattering of
the free $N$ and $K$ states in a finite volume
\cite{Mathur:2004jr,Ishii:2004qe,Ishii:2004ib,Alexandrou:2004ws}.
% >>>
% Using the Bayesian fitting techniques, the Kentucky Group \cite{Mathur:2004jr} 
% has examined the volume dependence of the residue of the ground state, 
Using the Bayesian fitting techniques, Mathur {\em et al.} \cite{Mathur:2004jr} 
have examined the volume dependence of the residue of the ground state, 
% <<<
noting that each state --- the pentaquark, $N$, and $K$ --- is volume 
% >>>
% normalized such that the residue of the $NK$ state is proportional to 
normalised such that the residue of the $NK$ state is proportional to 
% <<<
the inverse spatial lattice volume.
The analysis suggests that the lowest-lying state is the $NK$
scattering state, but leaves open the question of the existence of
a higher-lying pentaquark resonance state.

% >>>
% However, conflicting results are reported by Alexandou {\em et al.} \cite{Alexandrou:2004ws}
% where ratios of the residues indicate single particle states. At the 
However, conflicting results are reported by Alexandou {\em et al.}
\cite{Alexandrou:2004ws}, who find that ratios of the weights in the
correlation function are more consistent with single particle states
than scattering states. At the 
% <<<
same time, the broad distribution of $u$ and $\bar{s}$ quarks presented there suggests
% >>>
% to us the formation of a $NK$ scattering state.
to us the formation of an $NK$ scattering state.
% <<<
%On the other hand, Alexandou {\em et al.} \cite{ALEXANDROU} find
%that the ratios of the weights of the pentaquark correlators on three
%different lattice volumes are more consistent with a single particle
%state, although they do not identify the lower $KN$ scattering state
%in their analysis.

Using a single interpolating field, Ishii {\it et al.} \cite{Ishii:2004qe,Ishii:2004ib}
have introduced different boundary conditions in the quark propagators
in an attempt to separate a genuine pentaquark resonance state from
the $NK$ scattering state.
Here the quark propagators mix such that the effective mass of the
$NK$ scattering state changes, while the mass of a genuine resonance
is unchanged.
% >>>
% Again, the lowest-lying state displays the properties of a $NK$ 
Again, the lowest-lying state displays the properties of an $NK$ 
% <<<
scattering state, but leaves open the issue of whether a higher-lying
pentaquark resonance exists.

In identifying the nature of excited states, one should also explore 
the possibility that the excited state could be a two-particle state.
% >>>
% Since we expect our interpolating fields may couple to all possible 
Since we expect that our interpolating fields may couple to all possible 
% <<<
two-particle states to some degree, we compare the results of our 
correlation matrix analysis to all the possible two-particle states.
% >>>
% In the negative parity channel this includes the S-wave $NK$, $NK^*$, $\Delta K^{*}$ ( isospin-1 only ), 
% and $N'K$, where $N'$ is the lowest positive parity excitation of the 
% nucleon.
% In the positive parity channel we consider the S-wave $N^*K$ and
% $N^* K^*$ two-particle states, in addition to the P-wave $NK$ and $NK^*$
% states, where $N^*$ is the lowest-lying negative parity excitation of
% the nucleon.
In the negative parity sector this includes the S-wave $NK$, $NK^*$, and
$\Delta K^*$ (isospin-1 only) channels, as well as the $N'K$, where $N'$
is the lowest positive parity excitation of the nucleon.
In the positive parity channel we consider the S-wave $N^* K$
two-particle state, where $N^*$ is the lowest-lying negative
parity excitation of the nucleon, in addition to the P-wave $NK$ and
$NK^*$ states.
% <<<

% .........................................................................
\subsection{Lattice resonance signatures}
\label{ssec:I-res}

Our approach to assessing the existence of a genuine pentaquark 
resonance is complementary to the aforementioned approaches.
In the following we search for evidence of attraction between the 
constituents of the pentaquark state, which is vital to the formation of
a resonance.
Doing so requires careful measurement of the effective mass splitting 
between the pentaquark state and the sum of the free $N$ and $K$
masses measured on the {\it same lattice}.
As discussed in detail below, attraction between the constituents of 
every baryon resonance ever calculated on the lattice 
\cite{Leinweber:2004it,Melnitchouk:2002eg,Zanotti:2003fx} has been sufficient to render the
resonance mass {\it lower} than the sum of the free decay channel
masses when calculated at sufficiently large quark mass.
If the behaviour of the pentaquark is similar to that of every
other resonance on the lattice, then searching for a signal of a
pentaquark resonance above the $NK$ threshold at the large quark masses typically considered in lattice QCD
will mean that one is simply looking in the wrong place.

% Added from the cons section
One might have some concern as to whether the standard lattice
resonance signature should appear for exotic pentaquark states where
quark-antiquark annihilation cannot reduce the quark content to a
% >>>
% ``three-quark state''.  Clearly the approach to the infinite quark mass
``three-quark state''.
Clearly the approach to the infinite quark mass
limit will be different.  
% <<<
%
%Whereas resonances with standard quantum
%numbers will approach a resonance to two-particle ratio of 3/5, the
%exotic pentaquark to two-particle ratio will approach 1.  
However, the
heavy quark limit is far from the quark masses explored in this
% >>>
% investigation where evidence of nontrival fock-space components in the
% hadronic wavefunctions is abundant. 
investigation, where evidence of nontrival Fock-space components (such as
those including $q\bar q$ loops) in the hadronic wave functions is 
abundant. 
% <<<
For example, the quenched and unquenched $\Delta$ masses differ by more than
$100$ MeV at the quark masses considered here with the mass lying {\it lower} in
the presence of dynamical fermions.
% >>>
% We consider a quark mass as light as $0.05$ GeV, which is much less than the
We consider quark masses as light as $0.05$ GeV, which is much less than the
% <<<
hadronic scale, $1.5$ GeV, associated with pentaquark quantum numbers.
% >>>
% In short it is inappropriate to refer to the traditional resonances explored in lattice QCD
% as ``three quark states''.
In short, the traditional resonances explored in lattice QCD cannot be
considered simply as ``three-quark states'', so that there is little
reason to expect the lattice resonance signature to be qualitatively
different for ``ordinary'' and pentaquark baryons.
% <<<
%One can still ask
%the question as to whether the heavy-quark limit is approached from
%above or below.  
%An approach from above would, indicate the absence of attraction, vital to
%the existence of a resonance.
%

In the process of searching for attraction it is essential to explore a 
large number of interpolating fields having the quantum numbers of the 
putative pentaquark state.
In Sec.~II we consider a comparatively large collection of pentaquark 
% >>>
% interpolating fields and create new interpolators designed to maximize 
interpolating fields and create new interpolators designed to maximise 
% <<<
the opportunity to observe attraction in the pentaquark state.
We study two types of interpolating fields: those based on a nucleon 
plus kaon configuration, and those constructed from two diquarks
coupled to an $\bar s$ quark.

The technical details of the lattice simulations are discussed in 
Sec.~III, where we outline the construction of correlation functions
from interpolating fields, and the correlation matrix analysis,
as well as the actions used in this study.
It is essential to use a form of improved action, as large scaling 
violations in the standard Wilson action could lead to a false
signature of attraction.
Our simulations are therefore performed with the nonperturbatively
${\cal O}(a)$-improved FLIC fermion action 
\cite{Zanotti:2001yb,Zanotti:2004dr,Boinepalli:2004fz}, which displays
nearly perfect scaling, providing continuum limit results at finite 
lattice spacing \cite{Zanotti:2004dr}.
% >>>
%% The simulations are carried out on large $20^3 \times 40$ lattices 
%The simulations are carried out on large, $20^3 \times 40$, lattices 
The simulations are carried out on a large, $20^3 \times 40$, lattice, 
% <<<
with the ${\cal O}(a^2)$--tadpole-improved Luscher-Weisz plaquette
plus rectangle gauge action \cite{Luscher:1984xn}.
The lattice spacing is 0.128~fm, as determined via the Sommer scale,
% >>>
% $r_0 = 0.49$~fm, in an analysis incorporating the lattice coulomb 
$r_0 = 0.49$~fm, in an analysis incorporating the lattice Coulomb 
% <<<
potential \cite{Edwards:1997xf}.

In Sec.~IV we present our results for the even and odd parity pentaquark 
states, in both the isoscalar and isovector channels.
% >>>
% As we will see, there is no evidence of attraction between the meson and baryon in the 
% pentaquark state; to the contrary, we find evidence of repulsion.
As we will see, there is no evidence of attraction in the 
pentaquark channel; on the contrary, we find evidence of repulsion.
% <<<
As the quark masses increase and the quark distributions become more 
% >>>
% localized, the mass splitting between the lowest-lying pentaquark state 
localised, the mass splitting between the lowest-lying pentaquark state 
% <<<
and the sum of the free $N$ and $K$ masses is generally observed to {\it increase}.
Moreover, the standard lattice resonance signature of the resonance mass 
lying {\it lower} than the sum of the free decay channel masses at 
sufficiently large quark mass is absent.
As we conclude in Sec.~V, evidence supporting the existence of a
spin-${1\over 2}$ pentaquark resonance does not exist in quenched QCD.

%%%%%%%%%%%%%%%%%%%%%%%%%%%%%%%%%%%%%%%%%%%%%%%%%%%%%%%%%%%%%%%%%%%%%%%%%
\section{Interpolating Fields}
\label{sec:if}

In this section we review the interpolating fields which have been used
in recent pentaquark studies, in both the QCD sum rule approach and in
lattice QCD calculations.
% >>>
% We then propose new interpolators designed to maximze the opportunity to
% observe attraction between the meson and baryon at large quark masses.
We then propose new interpolators designed to maximise the opportunity to
observe attraction between the pentaquark constituents at large quark masses.
% <<<
Two general types of interpolating fields are considered: those based on an
``$NK$'' configuration (either $n K^+$ or $p K^0$), and those based on a 
``diquark-diquark-$\bar s$'' configuration.
We examine both of these types, and discuss the relations between them.

% .......................................................................
\subsection{$NK$-type interpolating fields}
\label{ssec:NK}

The simplest ``$NK$''-type interpolating field is referred herein as the
``colour-singlet'' form,
\begin{equation}
\label{eq:NK:sing}
\chi_{NK} = {1 \over \sqrt{2}} \epsilon^{abc}
	(u^{T a} C \gamma_5 d^b)
	\left\{ u^c (\bar s^e i \gamma_5 d^e)\
		\mp\ (u \leftrightarrow d)
	\right\}\ ,
\end{equation}
where the $\mp$ corresponds to the isospin $I=0$ and $1$ channels, 
respectively.
One can easily verify that the field $\chi_{NK}$ transforms negatively 
under the parity transformation $q \to \gamma_0 \, q$, and therefore the 
negative parity state will propagate in the upper-left Dirac quadrant of 
the correlation function, contrary to the more standard 
``positive-parity'' interpolators \cite{Leinweber:2004it}.
Note that the colour-index sum here corresponds to a ``molecular''
(or ``fall-apart'') state with both the ``$N$'' and ``$K$'' components
of Eq.~(\ref{eq:NK:sing}) colour singlets. %
For negative parity the field $\chi_{NK}$ couples the
( {\em large} $\times$ {\em large} ) $\times$ {\em large}$\ \times$ ( {\em large} $\times$ {\em large} ) 
components of Dirac spinors, and should therefore produce a strong signal.
For the positive parity projection (see Sec.~\ref{ssec:2pt} below), it 
involves one lower (or small) component, coupling \newline
( {\em large} $\times$ {\em large} ) $\times$ {\em small} $\times$  
	    ( {\em large} $\times$ {\em large} ),
which is known to lead to a weaker signal in this channel 
\cite{Leinweber:2004it}.

Some authors \cite{Csikor:2003ng,Takahashi:2004sc} have argued that $\chi_{NK}$ is a poor 
choice of interpolator for accessing the pentaquark resonance, and that 
an interpolator that suppresses the colour-singlet $NK$ channel may 
provide better overlap with a pentaquark resonance, should it exist.
% >>>
% Csikor {\it et al.} \cite{Csikor:2003ng}, the Kentucky group \cite{Mathur:2004jr}, 
Csikor {\it et al.} \cite{Csikor:2003ng}, Mathur {\it et al.} \cite{Mathur:2004jr}, 
% <<<
Takahashi {\it et al.} \cite{Takahashi:2004sc} and Chiu {\it et al.} \cite{Chiu:2004gg}
(in lattice calculations), and Zhu \cite{Zhu:2003ba} (in a QCD sum rule 
calculation), have considered a slightly modified form in which the 
colour indices between the $N$ and $K$ components of the interpolating
field are mixed,
\begin{equation}
\label{eq:NK:fused}
\chi_{\widetilde{NK}} = {1 \over \sqrt{2}} \epsilon^{abc}
        (u^{T a} C \gamma_5 d^b)
	\left\{ u^e (\bar s^e i \gamma_5 d^c)\
		\mp\ (u \leftrightarrow d)
	\right\}\ ,
\end{equation}
for $I=0$ and 1, respectively.
We refer to this alternative colour assignment as a ``colour-fused'' 
interpolator, whereby the coloured three-quark and $q\bar q$ pair are 
fused to form a colour-singlet hadron.
Of course, for 1/3 of the combinations the colours $e$ and $c$ will 
coincide, so that the ``colour-singlet''--``colour-singlet'' state will 
also arise from the field $\chi_{\widetilde{NK}}$.  
Upon constructing the correlation functions associated with each of
% >>>
% these interpolators, one encounters a sum of 324 colour combinations.
%%>>> these interpolators, one encounters a sum of $(2\times 3\times 3)^2=324$
%%<<< colour combinations.
these interpolators, one encounters a sum of $\left( 3! \times 3 \right)^{2} = 324$
colour combinations with a non-trivial contribution to the correlation function.
% <<<
% >>>
% However, only 1/9 of these terms will have the same colour combination 
% for the colour-singlet correlator and the colour-fused correlator.
However, only 1/9 of these terms will be common to the colour-singlet
and colour-fused correlators.
% <<<
It will be interesting therefore to see if increased binding between the
pentaquark constituents can be observed.

In Zhu's QCD sum rule (QCDSR) calculation \cite{Zhu:2003ba} interpolating 
fields based on the Ioffe current were also considered, such as
\begin{eqnarray}
%
%\chi^{I=1}_{\widetilde{NK}}
%&=& {1 \over \sqrt{2}} \epsilon^{abc}
%	(u^{T a} C \gamma_\mu d^b) \gamma_\mu \gamma_5 u^e
%	(\bar s^e i \gamma_5 d^c)\
% -\ (u \leftrightarrow d)\ ,
%\label{eq:I=1}					\\
%
\gamma_5\, \chi^{\rm QCDSR}_{\widetilde{NK}}
&=& {1 \over \sqrt{2}} \epsilon^{abc}
	(u^{T a} C \gamma_\mu u^b) \gamma_5 \, \gamma_\mu d^e
	(\bar s^e i \gamma_5 d^c) + (u \leftrightarrow d)\ .
\end{eqnarray}
%  
% He notes that the three quarks and the remaining $\bar q q$ pair are
% both in a colour adjoint representation, although this choice is not
% unique \cite{COLOUR}.
%
It is well known that the Ioffe current,
\begin{equation}
\epsilon^{abc} (u^{T a} C \gamma_\mu\, u^b) \gamma_5 \, \gamma_\mu d^c
\, ,
\end{equation}
can be written as a linear combination of the standard lattice 
interpolator,
\begin{equation}
\epsilon^{abc} (u^{T a} C \gamma_5\, d^b) u^c \, ,
\label{eq:good}
\end{equation}
and an alternate interpolator whose overlap with the ground state
is suppressed by a factor of $\sim$~100 \cite{Leinweber:1994nm}
\begin{equation}
\epsilon^{abc} (u^{T a} C  d^b) \gamma_5\, u^c \, .
\label{eq:bad}
\end{equation}
In the QCD sum rule approach, the interpolator of Eq.~(\ref{eq:bad}) 
can be used to subtract excited state contributions, while the nucleon 
is accessed via the interpolator of Eq.~(\ref{eq:good}) 
\cite{Leinweber:1994nm,Leinweber:1995fn}.
% >>>
% However, in the lattice approach, the former interpolator of Eq. (\ref{eq:bad}) plays little 
However, in the lattice approach, the interpolator of Eq.~(\ref{eq:bad}) plays little 
% <<<
to no role in accessing the lowest-lying state properties for either 
positive or negative parities \cite{Melnitchouk:2002eg}.

% .......................................................................
\subsection{Diquark-type interpolating fields}
\label{ssec:diq}

The other type of interpolating field which has been discussed is
one in which the non-strange quarks are coupled into two sets of
diquarks, together with the antistrange quark.
Jaffe and Wilczek \cite{Jaffe:2003sg} suggested that the lowest energy diquark 
state would have two scalar diquarks in a relative P-wave coupled to the 
$\bar s$.
The lowest mass pentaquark would then be one containing two scalar 
diquarks.
The configuration of two isospin $I=0$ diquarks gives a purely $I=0$ 
interpolating field,
\begin{eqnarray}
\label{eq:JW}
\chi_{JW} &=& \epsilon^{abc}\epsilon^{aef}\epsilon^{bgh}
	(u^{T e} C \gamma_5 d^f)(u^{T g} C \gamma_5 d^h)C \bar s^{T c}\ .
\end{eqnarray}
If the interpolating field is local, the two diquarks in $\chi_{JW}$
are identical, but because their colour indices are antisymmetrised, 
this field vanishes identically.
The field $\chi_{JW}$ would be non-zero if the diquark fields were
non-local.
On the other hand, the use of non-local fields significantly increases
the computational cost of lattice calculations.
While this remains an important avenue to explore in future studies, 
in this work we focus on the construction of pentaquark operators from 
local fields.

% >>>
% A variant of the scalar field $\chi_{JW}$ can be utilized by observing 
A variant of the scalar field $\chi_{JW}$ can be utilised by observing 
% <<<
that the antisymmetric tensors in Eq.~(\ref{eq:JW}) can be expanded in 
terms of Kronecker-$\delta$ symbols,
\begin{eqnarray}
\epsilon^{abc} \epsilon^{dec}
&=& \delta^{ad} \delta^{be} - \delta^{ae} \delta^{bd}\ ,
\end{eqnarray}
which enables the interpolating field $\chi_{JW}$ to be rewritten as
\begin{eqnarray}
\chi_{JW} &=& \epsilon^{abc}(u^{T a} C \gamma_5 d^b)
		\left\{ (u^{T c} C\gamma_{5} d^e)
			- (u^{T e} C\gamma_{5} d^c)
% >>>
%		\right\} C \bar{ s^{T e} }\ .
		\right\} C \bar{s}^{T e}\ .
% <<<
\end{eqnarray}
One can then define two interpolating fields,
\begin{eqnarray}
\label{eq:SS:sing}
\chi_{SS}
&=& {1 \over \sqrt{2}} \epsilon^{abc}
	(u^{T a} C \gamma_5 d^b)
	(u^{T c} C \gamma_5 d^e)
	C \bar s^{T e}\ ,			\\
\label{eq:SS:fused}
\chi_{\widetilde{SS}}
&=& {1 \over \sqrt{2}} \epsilon^{abc}
	(u^{T a} C \gamma_5 d^b)
	(u^{T e} C \gamma_5 d^c)
	C \bar s^{T e}\ ,
\end{eqnarray}
which are equal but do not individually vanish.
Clearly these fields transform negatively under parity, and, as with
$\chi_{NK}$ and $\chi_{\widetilde{NK}}$, couple
( {\em large} $\times$ {\em large} ) $\times$ ( {\em large} $\times$ {\em large} ) $\times$ {\em large}
components for negative parity states, making them ideal for lattice
simulations \cite{Leinweber:2004it}.

To determine the isospin of $\chi_{SS}$, one can express the second 
diquark as a sum of colour symmetric and antisymmetric components,
\begin{eqnarray}
\label{eq:SS-rearr}
\chi_{SS} &=& {1 \over \sqrt{2}}\ \epsilon^{abc}
  (u^{T a} C \gamma_5 d^b)
  \left\{ \frac{1}{2} (u^{T c} C \gamma_5 d^e - u^{T e} C \gamma_5 d^c )
  \right.				\cr
& & \hspace*{3.2cm} 
  \left.
       +\ \frac{1}{2} (u^{T c} C \gamma_5 d^e + u^{T e} C \gamma_5 d^c )
  \right\} C \bar s^{T e}\  .
\end{eqnarray}
The first term in the braces in Eq.~(\ref{eq:SS-rearr}), which is 
isoscalar, is equivalent to the field $\chi_{JW}$, and vanishes for
the reasons discussed above.
The second term, which does not vanish, is isovector, so that the field 
$\chi_{SS}$ is pure isospin $I=1$.

An interesting question is how much, if any, overlap exists between
the diquark-type field $\chi_{SS}$ and the $NK$-type fields in 
Sec.~\ref{ssec:NK}.
This can be addressed by performing a Fierz transformation on the fields.
For the field $\chi_{SS}$, one finds:
\begin{eqnarray}
\chi_{SS}
&=& {1 \over 4} \epsilon^{abc}
	(u^{T a} C \gamma_5 d^b)
\big\{ - (\bar s^e u^e) \gamma_5 d^c\
	+\ (\bar s^e \gamma_\mu u^e) \gamma^\mu \gamma_5 d^c\
	+\ {1 \over 2} (\bar s^e \sigma_{\mu\nu} u^e)
	  \sigma^{\mu\nu} \gamma_5 d^c			\nonumber\\
& & \hspace*{4cm}
	+\ (\bar s^e \gamma_\mu \gamma_5 u^e) \gamma^\mu d^c\
	-\ (\bar s^e \gamma_5 u^e) d^c
\big\}\ .
\label{eq:fierz}
\end{eqnarray}
The last two terms in Eq.~(\ref{eq:fierz}) resemble $NK$-type 
interpolating fields, similar to those introduced in
Sec.~\ref{ssec:NK}, while the second term corresponds to an $NK^*$-type
configuration.

Note that for the $NK$-like terms in Eq.~(\ref{eq:fierz}) the colour 
structure corresponds to the colour-singlet operator $\chi_{NK}$.
It has been suggested that the colour-singlet $NK$ interpolating field 
% >>>
% would have significant ovelap with the $NK$ scattering state and hence 
would have significant overlap with the $NK$ scattering state and hence 
% <<<
not couple strongly to a pentaquark resonance.
On the other hand, Fierz transforming the field $\chi_{\widetilde{SS}}$, 
in analogy with Eq.~(\ref{eq:fierz}), would give rise to an $NK$-like 
term corresponding to the colour-fused $\chi_{\widetilde{NK}}$ operator.
Since the fields $\chi_{SS}$ and $\chi_{\widetilde{SS}}$ are equivalent,
however, this demonstrates that selection of the operator 
$\chi_{\widetilde{SS}}$ (with the colour-fused $NK$ configuration)
over the operator $\chi_{SS}$ (with the colour-singlet $NK$ 
configuration) would be spurious.

Several authors \cite{Sasaki:2003gi,Ishii:2004qe,Ishii:2004ib,Chiu:2004gg} have also used a variant of the 
field $\chi_{JW}$ in lattice simulations, in which a scalar diquark is 
coupled to a pseudoscalar diquark \cite{Sugiyama:2003zk},
\begin{eqnarray}
\label{sas}
\chi_{PS} &=& \epsilon^{abc}\epsilon^{aef}\epsilon^{bgh}
		(u^{T e} C d^f)
		(u^{T g} C \gamma_5 d^h)
		C \bar s^{T c}\ .
\end{eqnarray}
In this case the two diquarks are not identical and so the field does 
not vanish.
Since both diquarks in $\chi_{PS}$ are isoscalar, this field has
isospin zero and transforms positively under parity.
For positive parity it couples
( {\em large} $\times$ {\em small} ) $\times$ ( {\em large} $\times$ {\em large} ) $\times$ {\em large}
components of Dirac spinors and is therefore suitable for lattice simulations.
For the negative parity projection, it couples 
( {\em large} $\times$ {\em small} ) $\times$ ( {\em large}  $\times$ {\em large} ) $\times$ {\em small},
so that the signal will be doubly suppressed relative to the other negative parity state interpolators.
Since it has been used in the literature, for completeness we also 
include $\chi_{PS}$ in our analysis.
To be consistent with the parity assignments of the other interpolating
fields discussed above, one can use a modified form,
\begin{eqnarray}
\chi_{PS} &\to& \gamma_5\ \chi_{PS}\ ,
\end{eqnarray}
which then transforms negatively under parity.
The effect of this is merely to switch the positive and negative
parity components propagating in the $\{ (1,1), (2,2) \}$ and
$\{ (3,3), (4,4) \}$ elements of the correlation function
(see Sec.~\ref{ssec:2pt} below).

\section{Lattice Techniques}
In this section we discuss the techniques used to extract bound state 
masses in lattice calculations.
We first outline the computation of two-point correlation functions,
both at the baryon level, and at the quark level in terms of the 
interpolating fields introduced in Sec.~\ref{sec:if}.
To study more than one interpolating field, we perform a correlation 
matrix analysis to isolate the individual states, and describe the basic 
steps in this analysis.
Finally, we present some details of the lattice simulations, including 
the gauge and fermion actions used.
Throughout this work we shall use the Pauli representation of the Dirac 
$\gamma$-matrices defined in Appendix B of Sakurai \cite{SAKURAI}.
In particular, the $\gamma$-matrices are Hermitian with
$\sigma_{\mu\nu} = [\gamma_{\mu},\ \gamma_{\nu}]/(2i)$.	

\subsection{ Two-point correlation functions }
\label{ssec:2pt}
% ---------------------------------------------------------------------
\subsubsection{Baryon level}
\label{sssec:Blevel}

Let us define a function, $\mathcal{G}$, of the interpolating field 
$\chi$ at Euclidean time $t$ and momentum ${\vec p}$ as
\begin{eqnarray}
\mathcal{G}(t,{\vec p})
&=& \sum_{\vec x}\ \exp({-i {\vec p} \cdot {\vec x}}) 
\left\langle 0 \left|
   T\ \chi(x)\ \bar\chi(0)\
\right| 0 \right\rangle\ ,
\end{eqnarray}
%
% >>>
% where we have suppressed the Dirac indices on the interpolating fields.
% Inserting a compete set of intermediate momentum, energy and spin states
where we have suppressed the Dirac indices.
Inserting a complete set of intermediate momentum, energy and spin states
% <<<
$| B,p',s \rangle$,
\begin{eqnarray}
\mathcal{G}(t,{\vec p})
= { \sum_{s,p',B} }\sum_{\vec x} \exp({-i {\vec p} \cdot {\vec x}}) \
  \langle 0 |\ \chi(x)\ | B,p',s \rangle
  \langle B,p',s |\ \bar\chi(0)\ | 0 \rangle\ ,
\end{eqnarray}
and using translational invariance, one can write the function 
$\mathcal{G}$ as
\begin{eqnarray}
%
% \mathcal{G}(t,{\vec p}) &=&\
% { \sum_{s,p',B} }\sum_{\vec x} \exp({-i {\vec p} \cdot {\vec x}})\
% \times \cr
% & & \langle 0 | \
% { \exp^{\hat{H}t}\exp^{-i\hat{\vec{P}}} }
% \chi(0)\
% { \exp^{ i\hat{\vec{P}}}\exp^{-\hat{H}t} } 
% { | B,p',s \rangle \langle B,p',s |}\
% \bar\chi(0)
% | 0 \rangle \cr
% & & \cr
%
\mathcal{G}(t,{\vec p})
&=& \sum_{s,B}\exp(-E_{B}t)\
    \langle 0 | \ \chi(0)\ | B,p,s \rangle
    \langle B,p,s |\ \bar\chi(0)\ | 0 \rangle\ ,
\end{eqnarray}
where $E_B = \sqrt{M_B^2 + \vec p^2}$ is the energy of the state 
$| B \rangle$ and $M_B$ is its mass.

Despite having a specific intrinsic parity, the interpolating field 
$\chi$ in fact has overlap with both positive and negative parity 
states.
%
% For a given baryon interpolating field, the function $\mathcal{G}$ 
% contains information on both positive and negative parity states.
%
The overlap of $\chi$ with the intermediate state $| B^\mp \rangle$,
where the superscript denotes parity $-1$ or $+1$, at the source
can be parameterised by a coupling strength $\lambda^{B^\mp}$,
\begin{eqnarray}
\langle 0 | \chi(0) | B^-, p, s \rangle
&=& \lambda^{B^-} \sqrt{ {M_{B^-} \over E_{B^-}} }\
    u_{B^-}(p,s)\ ,
\label{eq:lam-}					\\
\langle 0 | \chi(0) | B^+, p, s \rangle
&=& \lambda^{B^+} \sqrt{ {M_{B^+} \over E_{B^+}} }\
    \gamma_5\ u_{B^+}(p,s)\ ,
\label{eq:lam+}
\end{eqnarray}
where $E_{B^\mp}$ and $M_{B^\mp}$ correspond to the energies and
% >>>
% masses of the positive and negative parity states, respectively.
masses of the negative and positive parity states, respectively.
% <<<
Note that in Eq.~(\ref{eq:lam+}) the matrix $\gamma_5$ premultiplies
the spinor $u_{B^+}$, since the interpolating fields that we use in
this analysis all transform negatively under parity.
This is in contrast to the more standard case where the fields have
positive parity \cite{Leinweber:2004it}, in which case the definitions of 
$\lambda^{B^\mp}$ in Eqs.~(\ref{eq:lam-}) and (\ref{eq:lam+}) are 
interchanged.

At large Euclidean times the function $\mathcal{G}$ can be written
\begin{eqnarray}
\mathcal{G}(t,{\vec p}) \to
   \lambda^{B^-} \bar{\lambda}^{B^-}
   { \left( \gamma \cdot p + M_{B^-} \right) \over 2 M_{B^-} }
   \exp(-E_{B^-} t)
+\ \lambda^{B^+} \bar{\lambda}^{B^+}
   { \left( \gamma \cdot p - M_{B^+} \right) \over 2 M_{B^+} }
   \exp(-E_{B^+} t)\ ,
\end{eqnarray}
where $p$ is the on-shell four-momentum vector,
and $\bar{\lambda}^{B^\mp}$ is the coupling strength of the field
% >>>
% $\bar{\chi}$ at the the source to the state $|B^\mp \rangle$.
$\bar{\chi}$ at the source to the state $|B^\mp \rangle$.
% <<<
In our lattice simulations the source is smeared, so that 
% >>>
% $\lambda^{B^\pm}$ and $\bar{\lambda}^{B^\pm}$ are not necessarily equal.
$\lambda^{B^\mp}$ and $\bar{\lambda}^{B^\mp}$ are not necessarily equal.
% <<<
%
With fixed boundary conditions in the time direction, one can project
% >>>
% out states with definite parity using the matrix \cite{Melnitchouk:2002eg,Lee:1998cx},
out states with definite parity using the matrix \cite{Melnitchouk:2002eg,Lee:1998cx}
% <<<
%
\begin{eqnarray}
\label{eq:pProjOp}
\Gamma^{\mp}
= {1 \over 2}
  \left( 1 \mp {M_{B^\pm} \over E_{B^\pm}} \gamma_4 \right)\ .
\end{eqnarray}
The parity-projected two-point correlation function can then be
expressed as the spinor trace of the function $\mathcal{G}(t,{\vec p})$,
\begin{eqnarray}
G^{\mp}(t,{\vec p})
&=& {\rm tr} [\Gamma^{\mp} \mathcal{G}(t,{\vec p})]\
\label{eq:2pt:had}					\\
&=& \sum_B \lambda^{B^\mp} \bar{\lambda}^{B^\mp}
    \exp{(-E_{B^\mp}t)}\ .
\end{eqnarray}

Because of the exponential suppression (at large Euclidean times)
of states with energies greater than the ground state energy, in the 
large-$t$ limit the correlation function for $\vec p=0$ (which we
use in practice in this analysis) will be dominated by the ground
state with mass $m_0^\mp$,
\begin{eqnarray}
G^{\mp}(t,\vec 0) &\stackrel{t\to\infty}{=}&
\lambda^\mp_0 \bar{\lambda}^\mp_0 \exp{(-m^{\mp}_{0}t)}\ ,
\end{eqnarray} 
where $\bar{\lambda}^\mp_0$ and $\lambda^\mp_0$ are the ground
% >>>
% state couplings for the positive and negative parity states,
% at the source and sink respectively.
state couplings for the negative and positive parity states
at the source and sink, respectively.
% <<<
The effective mass of the baryon $B^\mp$ is then defined in terms
of ratios of the correlation function at successive time slices,
\begin{eqnarray}
M^\mp_{\rm eff}(t)
&=& \ln{ \left( G^{\mp}(t,\vec 0) \over G^{\mp}(t+1,\vec 0) \right) }\ .
\end{eqnarray}  
In the large-$t$ limit one therefore picks out the state with the
lowest mass,
\begin{eqnarray}
M^\mp_{\rm eff}(t)
&\stackrel{t\to\infty}{=}& m_0^\mp\ .
\end{eqnarray}  
In order to study masses of excited states, one can in principle attempt 
to fit the correlation function at finite $t$ with a sum of exponentials 
corresponding to the ground state plus one or more excited states.
In practice this is difficult, however, due to the large statistics 
required for a reliable extraction of the masses.
An alternative approach is to use several interpolating fields, and 
extract masses of an orthogonal set of operators using a correlation 
matrix analysis, as we discuss in Sec.~\ref{ssec:corrmatrix} below.

% ---------------------------------------------------------------------
\subsubsection{Quark level}
\label{sssec:Qlevel}

The two-point correlation functions discussed above are all derived 
at the hadronic level.
They can be expressed in a form suitable for actual lattice 
simulations by substituting the interpolating fields in Sec.~\ref{sec:if} 
and performing the appropriate Wick contractions of the time-ordered 
products of fields.
We use the notation
$\langle 0 | T u^{a}_{\alpha}(x)\bar{u}^{b}_{\beta}(0) | 0 \rangle
% >>>
% = {\rm tr}[U^{ab}_{\alpha\beta}(x,0)]$ for the $u$ quark, and similarly 
= U^{ab}_{\alpha\beta}(x,0)$ for the $u$ quark, and similarly 
% <<<
for the $d$ and $s$ quarks, where $\alpha$ and $\beta$ represent Dirac 
spinor indices.

% >>>
% For the ``molecular'' $KN$ interpolating field $\chi_{NK}$ in 
For the ``molecular'' $NK$ interpolating field $\chi_{NK}$ in 
% <<<
Eq.~(\ref{eq:NK:sing}) the diagonal ($pK^0/pK^0$ and $nK^+/nK^+$) 
correlation function is given by
\begin{eqnarray}
\mathcal{G}^{NK/NK}
&=& \sum_{a'b'c'd'e'} \epsilon^{a'b'c'}\delta^{d'e'} \
\sum_{abcde} \epsilon^{abc}\delta^{de} \cr
& &
\hspace*{-2cm}
\times
\left\{
- {\rm tr}\left[ \gamma_5 S^{dd' *}(x,0) \gamma_5
		 \left( \gamma_5 D^{ee'}(x,0) \gamma_5 \right)^T
	  \right]
\right.						\cr
& &
\hspace*{-1cm}
\times
\left\{
  U^{cc'}(x,0)\
  {\rm tr}\left[ U^{aa'}(x,0) \left( C\gamma_5 D^{bb'}(x,0) \gamma_5 C \right)^T
	  \right]
\right.						\cr
& &
\hspace*{-0.5cm}
\left.
- U^{ca'}(x,0) \left( C \gamma_5 D^{bb'}(x,0) \gamma_5 C \right)^T\ U^{ac'}(x,0)
\right\} 					\cr
& &
\hspace*{-1.5cm}
+\ U^{cc'}\
{\rm tr}\left[ U^{aa'}(x,0) \left( \gamma_5 D^{eb'}(x,0) C \gamma_5 \right)^T\
	       \gamma_5 S^{dd' *}(x,0) \gamma_5
	       \left( C \gamma_5 D^{be'}(x,0) \gamma_5 \right)^T
	\right]					\cr
& &
\hspace*{-1.5cm}
\left.
-\
  U^{ca'}(x,0) \left( \gamma_5 D^{eb'}(x,0) \gamma_5 C \right)^T\
  \gamma_5 S^{dd' *}(x,0) \gamma_5
  \left( C \gamma_5 D^{be'}(x,0) \gamma_5 \right)^T\
  U^{ac'}(x,0)
\right\}\ .					\cr
& &
\label{eq:cf_NK:sing}
\end{eqnarray}

The propagators in Eq.~(\ref{eq:cf_NK:sing}) are defined from source $0$ to point
$x$, and we have used the relation
$(\gamma_5 S^{ab}(x,0) \gamma_5)^*_{\alpha\beta}
 = S^{ba}(0,x)_{\beta\alpha}$
%>>>>
%for the $s$ quark propagators.
 for the anti-strange quark propagator.
%<<<<
%
% The strange quark propagator is fixed to $\kappa = 0.12885$ in all our 
% calculations.
%
Note that the first two terms in Eq.~(\ref{eq:cf_NK:sing}) correspond to 
a product of the $N$ and $K$ correlation functions, whereas the last two 
terms have a mixed flavour and colour structure.
%
% >>>
% The correlation function for the operator of Eq.(\ref{eq:NK:fused}) with mixed
The correlation function for the operator of Eq.~(\ref{eq:NK:fused}) with mixed
% <<<
colour labels can be obtained from $\mathcal{G}^{NK/NK}$ by interchanging
the $c \leftrightarrow e$ and $c' \leftrightarrow e'$ colour indices.

For the $p K^0/n K^+$ interference correlation function, one has
\begin{eqnarray}
\mathcal{G}^{pK^0/nK^+}
&=& \sum_{a'b'c'd'e'} \epsilon^{a'b'c'}\delta^{d'e'} \
\sum_{abcde} \epsilon^{abc}\delta^{de} \cr
& &
\hspace*{-2cm}
\times
\left\{
- {\rm tr} \left[ U^{ab'}(x,0) \left( C\gamma_5 D^{ba'}(x,0) C\gamma_5 \right)^T
	   \right]\
\right.					\cr
& &
\hspace*{-1cm}
\times\
  U^{ce'}(x,0) \gamma_5
  \left( \gamma_5 S^{dd' *}(x,0) \gamma_5 \right)^T\
  \gamma_5 D^{ec'}(x,0)			\cr
& &
\hspace*{-1.5cm}
+\
  U^{cb'}(x,0)
  \left( C\gamma_5 D^{ba'}(x,0) C\gamma_5 \right)^T
  U^{ae'}(x,0) \gamma_5
  \left( \gamma_5 S^{dd' *}(x,0) \gamma_5 \right)^T
  \gamma_5 D^{ec'}(x,0)			\cr
& &
\hspace*{-1.5cm}
+\
  U^{ce'}(x,0) \gamma_5
  \left( \gamma_5 S^{dd' *}(x,0) \gamma_5 \right)^T\
  \gamma_5 D^{ea'}(x,0) C\gamma_5
  \left( U^{ab'}(x,0) \right)^T
  C\gamma_5 D^{bc'}(x,0)		\cr
& &
\hspace*{-1.5cm}
\left.
-\
  U^{cb'}(x,0)
  \left( \gamma_5 D^{ea'}(x,0) C\gamma_5 \right)^T
  \left( \gamma_5 S^{dd' *}(x,0) \gamma_5 \right)
  \left( U^{ae'}(x,0) \gamma_5 \right)^T
  C\gamma_5 D^{bc'}(x,0)
\right\}\ ,				\cr
& &
\end{eqnarray}
for the colour-singlet interpolating field $\chi_{NK}$, with a similar 
replacement $c \leftrightarrow e$, $c' \leftrightarrow e'$ for the 
colour-fused field $\chi_{\widetilde{NK}}$.

Similarly, the correlation function for the $\chi_{SS}$ ``diquark'' 
interpolating field in Eq.~(\ref{eq:SS:sing}) is given by
\begin{eqnarray}
\label{eq:Gss}
\mathcal{G}^{SS/SS}
&=& - \sum_{a'b'c'd'e'} \epsilon^{a'b'c'}\delta^{d'e'} \
      \sum_{abcde} \epsilon^{abc}\delta^{de}
% >>>
%      C \gamma_5 S^{ee' *}(x,0) \gamma_5 \tilde{C}	\cr
      C \gamma_5 S^{ee' *}(x,0) \gamma_5 C		\cr
% <<<
&\times&
\left\{ \
   {\rm tr} \left[ (U^{cc'}(x,0))^T C \gamma_5 D^{dd'}(x,0)\ C \gamma_5
	    \right]
   {\rm tr} \left[ (U^{aa'}(x,0))^T C \gamma_5 D^{bb'}(x,0)\ C \gamma_5
	    \right]
\right.							\cr
& &
-\ {\rm tr} \left[ (U^{ac'}(x,0))^T C \gamma_5 D^{bb'}(x,0)\ C \gamma_5
		   (U^{ca'}(x,0))^T C \gamma_5 D^{dd'}(x,0)\ C \gamma_5
	    \right]					\cr
& &
+\ {\rm tr} \left[ (U^{ac'}(x,0))^T C \gamma_5 D^{bd'}(x,0)\ C \gamma_5
	    \right]
   {\rm tr} \left[ (U^{ca'}(x,0))^T C \gamma_5 D^{db'}(x,0)\ C \gamma_5
	    \right]					\cr
& &
\left.
-\ {\rm tr} \left[ (U^{aa'}(x,0))^T C \gamma_5 D^{bd'}(x,0)\ C \gamma_5
	 	   (U^{cc'}(x,0))^T C \gamma_5 D^{db'}(x,0)\ C \gamma_5
	    \right]  
\right\} .						\cr
& &
\end{eqnarray}
%
% For the alternative diquark interpolating field $\chi_{\widetilde{SS}}$ in
% Eq.~(\ref{eq:diq_fierz}), the correlation function can be obtained from
% $\mathcal{G}^{JW/JW}$ in Eq.~(\ref{eq:Gss}) by once again interchanging
% $c \leftrightarrow e$ and $c' \leftrightarrow e'$.

Following the parity projection in Eq.~(\ref{eq:pProjOp}), the 
% >>>
% correlation functions $G^{\mp}_{ij}$ can be made real by including both 
correlation functions $G^{\mp}$ can be made real by including both 
% <<<
the $U$ and $U^*$ gauge field configurations in the ensemble averaging 
% >>>
% used to construct the lattice correlation functions, providing an 
% improved unbiased estimator which is strictly real.
% This can be is easily implemented at the correlation function level by 
used to construct the lattice correlation functions.
This provides an improved unbiased estimator which is strictly real.
This is easily implemented at the correlation function level by 
% <<<
observing that
$$
M^{-1}(\{U_\mu^*\}) = \left[ C\gamma_5 \, M^{-1}(\{U_\mu\})\, (C\gamma_5)^{-1}
		      \right]^*
$$ 
holds for quark propagators.
% >>>
% For a more detailed discussion of this issue see Ref.~\cite{Melnitchouk:2002eg,Leinweber:1990dv}.
For a more detailed discussion of this issue see Refs.~\cite{Melnitchouk:2002eg,Leinweber:1990dv}.
% <<<

\subsection{ Correlation matrix analysis }
\label{ssec:corrmatrix}
In the previous section we described how the mass of the ground state is 
extracted from the two-point correlation function by fitting a constant 
to the effective mass.
% >>>
% Excited state masses can be extracted either by fitting the effective 
% mass with several exponentials (which is, in general, quite difficult to
Excited state masses can be extracted either by fitting the correlation
function with several exponentials (which is, in general, quite difficult to
% <<<
do reliably), or by using more than one interpolating field.
In the latter approach, which was implemented in the $N^*$ spectrum 
analysis in Ref.~\cite{Melnitchouk:2002eg} and which we adopt in this work, a set of
linearly independent operators will, in general, overlap with 
more than one state.
We use a correlation matrix analysis to convert a set of $N$ linearly 
independent operators into a set of $N$ orthogonal operators.

% ...[MAYBE REPHRASE THE FOLLOWING \& INCLUDE IN PREAMBLE TO 
%	SEC.\ref{ssec:corrmatrix}??]...
% However, if there are more states contributing to the correlation functions than
% there are operators to describe them, the new orthogonal operators may still overlap 
% with an admixture of states.
% If this is the case then the logarithm of the eigenvalue will not correspond to the
% mass of a state, because the time dependence of the correlation function is still more 
% complicated than that.
% The correlation matrix analysis is still useful because the new operators have 
% contamination from other excited states suppressed.
% We have the prospect of  using euclidean time evolution to extract the mass of the 
% dominant state.

In principle, to access the entire spectrum of states would require an 
infinite tower of operators.
In practice we use a $2 \times 2$ correlation matrix
% >>>
(in particular, for the $NK$-type interpolating fields),
% <<<
which enables us to access two states in each channel.
If the analysis is performed at large enough Euclidean times, the 
contributions from the $N>2$ excited states will be exponentially 
suppressed, as found in the earlier $N^*$ analysis \cite{Melnitchouk:2002eg}.
%
%Another approach is to perform the correlation matrix analysis at a 
%range of times and accept that the new operators will access an admixture 
%of excited states.
%The correlation functions can then be evolved to large Euclidean times 
%and accepted or rejected on the basis of whether or not the new operators 
%give better access to the lightest states in the tower.

%
%   Explain the Idea of Creating new Operators
%
% \subsubsection{ Eigenvalue equations }

Generalising the two-point correlation function in Eq.~(\ref{eq:2pt:had})
% >>>
% to the case of two different interpolating fields $\chi_i$ and $\chi_j$ 
to the case of two different interpolating fields $\chi_i$ and $\bar\chi_j$ 
% <<<
at the sink and source, respectively, the momentum-space two-point 
correlation function {\em matrix} $G_{ij}$ (at $\vec p=0$) can be 
written as
\begin{eqnarray}
G_{ij}(t) &=& \sum_{\alpha=0}^{N-1} 
\lambda^\alpha_i \bar{\lambda}^\alpha_j \exp(-m_\alpha t)\ ,
\end{eqnarray}
where $\alpha$ denotes each of the $N$ states in the tower of excited 
states, and we have suppressed the parity labels.
% >>>
% If the operators $\chi_i$, $\chi_j$ were orthogonal, the matrix $G_{ij}$ 
% would be diagonal, with the only $t$ dependence coming from the 
If the operators $\chi_i$, $\chi_j$ are orthogonal, the matrix $G_{ij}$ 
will be diagonal, with the only $t$ dependence coming from the 
% <<<
exponential factor, in which case one would have a recurrence relation,
\begin{eqnarray}
G_{ij}(t) &=& \exp(-m_\alpha\ \Delta t)\ G_{ij}(t+\Delta t) \, \delta_{ij}\ .
\end{eqnarray}
In general the operators will not be orthogonal, and a new set 
of operators must be created from a linear combination of the old 
operators using the eigenvalue equation.
% >>>
%In the event that the number of states excited by the interpolators
In the event that the number of states
% <<<
matches the number of interpolators, an orthogonal set of interpolators
can be constructed
% >>>
% by diagonalizing the correlation matrix 
% subject to the condition:
by diagonalising the correlation matrix 
subject to the condition
% <<<
% 
\begin{eqnarray}
\label{eq:eig1}
G_{ij}(t+\Delta t) \, u^\alpha_j = \lambda^\alpha \, G_{ik}(t) \, u^\alpha_k\ ,
\end{eqnarray}
or, 
\begin{eqnarray}
\label{eq:eig1a}
\left( G^{-1}(t) \, G(t+\Delta t) \right)_{ij} \, u^\alpha_j
% >>>
% = \lambda^\alpha \, u^\alpha_i\ .
= \lambda^\alpha \, u^\alpha_i\ ,
% <<<
\end{eqnarray}
%
% >>>
% where $u_j^\alpha$ are real eigenvectors, and the corresponding eigenvalue $\lambda^{\alpha}= \exp(-m_\alpha\ \Delta t)$.
where $u_j^\alpha$ are real eigenvectors, and the corresponding
eigenvalue is $\lambda^{\alpha}= \exp(-m_\alpha\ \Delta t)$.
% <<<

% >>>
% A real symmetric matrix is diagonalised by its eigenvectors, however, 
A real symmetric matrix is diagonalised by its eigenvectors.
However, 
% <<<
since our smearing prescriptions are different at the source and the 
sink, the correlation matrix is real but non-symmetric.
% >>>
% Consequently, one has to solve the additional left-eigenvalue equation:
Consequently, one has to solve the additional left-eigenvalue equation
% <<<
%
\begin{eqnarray}
\label{eq:eig2}
v^\alpha_i \, G_{ij}(t+\Delta t) &=& \lambda^\alpha \, v^\alpha_k \, G_{kj}(t)\ ,
\end{eqnarray}  
for eigenvectors $v_i^\alpha$, or equivalently
\begin{eqnarray}
\label{eq:eig2a}
v^\alpha_i \, \left( G(t+\Delta t) \, G^{-1}(t) \right)_{ij}
&=& \lambda^\alpha \, v^\alpha_j\ .
\end{eqnarray}
The eigenvectors $u^\alpha$ and $v^\alpha$ diagonalise the correlation 
matrix at times $t$ and $t+\Delta t$,
\begin{eqnarray}
v^\alpha_i \, G_{ij}(t+\Delta t) \, u^\beta_j
&=& \lambda^\alpha \, v^\alpha_i \, G_{ij}(t) \, u^\beta_j	\cr
&=& \lambda^\beta \,  v^\alpha_i \, G_{ij}(t) \, u^\beta_j\ ,
\end{eqnarray}  
and if $\lambda^{\alpha} \neq \lambda^{\beta}$ for $\alpha \neq \beta$ then,
\begin{eqnarray}
v^\alpha_i \, G_{ij}(t+\Delta t) \, u^\beta_j &\propto& \delta^{\alpha\beta}\ .
\end{eqnarray}  
The projected correlation matrix $v^\alpha_i G_{ij}(t) u^\alpha_j$ thus 
%corresponds to a new set of orthogonal operators.
describes the single state $\alpha$.

In the present analysis, for each state considered our aim will be
% >>>
% to optimize the results at every quark mass.
to optimise the results at every quark mass.
% <<<
We use the covariance matrix to find where the $\chi^2$/dof for a least
squares fit to the effective masses is $< 1.5$ for all quark masses.
Stepping back one time slice, we then apply the correlation matrix
analysis.
If the correlation matrix analysis is successful, {\em i.e.}, the
correlation matrix is invertible and the eigenvalues are real and
positive, we proceed to the next step.
If the correlation matrix analysis fails, we take another step back
in time, and continue stepping back until the analysis is successful
for a given quark mass.

The mass of the state derived from the projected correlation matrix
is then compared with the mass obtained using the standard analysis
techniques.
Any mixing of the ground state with excited states will result in
masses from the unprojected operators which lie between the true
ground and excited state masses.
Therefore, in the case of the ground state mass, if the new mass is
% >>>
% smaller then we use the result derived from the correlation matrix;
{\it smaller} then we use the result derived from the correlation matrix;
% <<<
otherwise, we keep the standard analysis result.
For an excited state, on the other hand, the result from the
correlation matrix analysis is used if the new mass is {\em larger}
than that from the standard analysis.

\subsection{ Lattice simulations }
\label{ssec:sims}
Having outlined the techniques used to extract excited baryon masses and 
% >>>
% the choice of interpolatig fields, we next describe the gauge and 
the choice of interpolating fields, we next describe the gauge and 
% <<<
fermion actions used in this analysis.
A more detailed account of the actions has been given by Zanotti
{\em et al.} \cite{Zanotti:2001yb}.

% ........................................................................
\subsubsection{Gauge action}

For the gauge fields, we use the Luscher-Weisz mean-field improved 
plaquette plus rectangle action \cite{Luscher:1984xn}.
The gauge action is given by
\begin{eqnarray}
S_{\rm G} \!&=&\! \frac{5\beta}{3}
      \sum_{\rm{sq}}\frac{1}{3}{\cal R}e\, {\rm{tr}}[1-U_{\rm{sq}}(x)]
\ -\ \frac{\beta}{12u_{0}^2}
      \sum_{\rm{rect}}\frac{1}{3}{\cal R}e\, {\rm{tr}}[1-U_{\rm{rect}}(x)]\, ,
\label{gaugeaction}
\end{eqnarray}
where the operators $U_{\rm{sq}}(x)$ and $U_{\rm{rect}}(x)$ are defined as
\begin{subequations}
\begin{eqnarray}
U_{\rm{sq}}(x)
&=& U_{\mu}(x) \, U_{\nu}(x+\hat{\mu}) \,
    U^{\dagger}_{\mu}(x+\hat{\nu}) \, U^\dagger_{\nu}(x)\ ,        \\
U_{\rm{rect}}(x)
&=& U_{\mu}(x) \, U_{\nu}(x+\hat{\mu}) \,
    U_{\nu}(x+\hat{\nu}+\hat{\mu}) \,
    U^{\dagger}_{\mu}(x+2\hat{\nu}) \,
    U^{\dagger}_{\nu}(x+\hat{\nu}) \,
    U^\dagger_{\nu}(x)                   			\nonumber\\
&+& U_{\mu}(x) \, U_{\mu}(x+\hat{\mu}) \,
    U_{\nu}(x+2\hat{\mu}) \,
    U^{\dagger}_{\mu}(x+\hat{\mu}+\hat{\nu}) \,
    U^{\dagger}_{\mu}(x+\hat{\nu}) \, U^\dagger_{\nu}(x)\ .
\label{actioneqn}
\end{eqnarray}
\end{subequations}%
The link product $U_{\rm{rect}}(x)$ denotes the rectangular $1\times2$
and $2\times1$ plaquettes, and for the tadpole improvement factor we use
the plaquette measure,
\begin{equation}
u_0
= \left\langle \frac{1}{3}{\cal R}e \, {\rm{tr}}\langle U_{\rm{sq}}\rangle
  \right\rangle^{1/4}\ .
\label{uzero}
\end{equation}
The gauge configurations are generated using the Cabibbo-Marinari
pseudoheat-bath algorithm with three diagonal SU(2) subgroups looped
over twice.
The simulations are performed using a parallel algorithm with appropriate
link partitioning, as described in Ref.~\cite{Bonnet:2000db}.

The calculations are performed on a $20^3\times 40$ lattice at $\beta=4.53$.
The scale is set via the Sommer scale $r_{0}$, obtained from the static quark 
potential \cite{Edwards:1997xf},
$$
V({\rm \bf r}) = V_0 +\sigma r -e \left[\frac{1}{{\rm \bf r}}\right]
 + l\left(\left[\frac{1}{{\rm \bf r}}\right] - \frac{1}{r}\right)\ ,
$$
where $V_0$, $\sigma$, $e$ and $l$ are fit parameters, and
$\left[\frac{1}{{\rm \bf r}} \right]$ denotes the tree-level lattice
% >>>
% Coulomb term
Coulomb term,
% <<<
%
\begin{eqnarray}
\left[\frac{1}{{\bf r}}\right]
&=& 4\pi\int\, \frac{d^3{\rm \bf k}}{(2\pi)^3}
  \cos({\rm \bf k}\cdot{\rm \bf r}) \, D_{44}(0,{\rm \bf k})\ ,
\end{eqnarray}
with $D_{44}(k)$ the time-time component of the gluon propagator.
Note that $D_{44}(k_4,{\rm \bf k})$ is gauge independent in the Breit
frame ($k_4 = 0$) since $k_4^2 / k^2=0$.
% >>>
% In the continuum limit, the Coulombic term reduces to
In the continuum limit, the Coulomb term reduces to
% <<<
%
\begin{eqnarray}
\left[\frac{1}{{\rm \bf r}}\right]
\rightarrow \frac{1}{r}\ .
\end{eqnarray}
%
%   scale determination here.
%
% >>>
% Knowledge of $V_{0},\sigma$ and $e$ allows one to calculate $r_{0}$ which is defined by,
Knowledge of $V_0,\sigma$ and $e$ allows one to calculate $r_0$, which is defined by
% <<<
\begin{eqnarray}
% r_{0}^{2}\frac{\partial V(r)}{\partial r} \vline_{_{_{_{_{_{_{{\textrm{\footnotesize $\ r=r_{0}$}  }}}}}}}} = 1.65 . 
% This is a nasty hack to put the subscript at the bottom of the vline, vspace seems to be inopperable in math mode.
r_0^2 \left. \frac{\partial V(r)}{\partial r} \right|_{r=r_0} = 1.65\ . 
% TRY THIS INSTEAD!
% <<<
\end{eqnarray}
% >>>
% Using the phenomenological value of $r_{0}=0.49\ {\rm fm}$, the spacing of our lattice is $0.128(2)\ {\rm fm}$.
Using the phenomenological value of $r_{0}=0.49\ {\rm fm}$, the spacing
of our lattice is $a = 0.128(2)\ {\rm fm}$.
% <<<

% ........................................................................
\subsubsection{Fat-link irrelevant fermion action}

For the quark fields, we use the Fat-Link Irrelevant Clover (FLIC) 
fermion action \cite{Zanotti:2001yb}, which provides a new form of 
nonperturbative ${\cal O}(a)$ improvement \cite{Leinweber:2002bw}.
This action has previously been used to study hadronic masses 
\cite{Zanotti:2001yb}, as well as the excited baryon spectrum 
\cite{Melnitchouk:2002eg}.
Here fat links are generated by smearing links with their nearest 
transverse neighbours in a gauge covariant manner (APE smearing).
This has the effect of reducing the problem of exceptional 
% >>>
% configurations common to Wilson-style actions, and minimising the
configurations common to Wilson-style actions \cite{Boinepalli:2004fz},
and minimising the
% <<<
effect of renormalisation on the action improvement terms.
Since only the irrelevant, higher-dimensional terms in the action are 
smeared, while the relevant, dimension-four operators are left untouched, 
% >>>
% the short-distance behavior of the quark and gluon interactions is 
the short-distance behaviour of the quark and gluon interactions is 
% <<<
retained.
The use of fat links \cite{DeGrand:1999gp} in the irrelevant operators also 
removes the need to fine tune the clover coefficient in removing all 
${\cal O}(a)$ artifacts.

The smearing procedure involves replacing a link, $U_{\mu}(x)$, with a 
sum of the link and $\alpha$ times its staples \cite{Falcioni:1984ei,Albanese:1987ds},
\begin{eqnarray}
&&\!\!U_{\mu}(x) \rightarrow U_{\mu}'(x) =
(1-\alpha) \, U_{\mu}(x) 				\cr
&&
\hspace*{1cm}
+\ \frac{\alpha}{6}\sum_{\nu=1 \atop \nu\neq\mu}^{4}
  \Big[ U_{\nu}(x) \,
        U_{\mu}(x+\nu a) \,
        U_{\nu}^{\dag}(x+\mu a) \,
     +\ U_{\nu}^{\dag}(x-\nu a) \,
        U_{\mu}(x-\nu a) \,
        U_{\nu}(x-\nu a +\mu a)
  \Big] \,,  \nonumber
\end{eqnarray}
followed by projection back to SU(3).
The unitary matrix $U_{\mu}^{\rm FL}$ which maximises
$$
{\cal R}e \, {\rm{tr}}\left[ U_{\mu}^{\rm FL}\, U_{\mu}^{' \dagger} \right]
$$
is selected by iterating over the three diagonal SU(2) subgroups of SU(3).
The entire procedure of smearing followed immediately by projection is 
repeated $n$ times.
% >>>
% The fat links used in this work are created with $\alpha = 0.7$ and $n=8$,
The fat links used in this work are created with $\alpha = 0.7$ and $n=6$,
% <<<
as discussed in Ref.~\cite{Zanotti:2001yb}.
The mean-field improved FLIC action is given by \cite{Zanotti:2001yb}
\begin{equation}
S_{\rm SW}^{\rm FL}
= S_{\rm W}^{\rm FL} - \frac{iC_{\rm SW}\ \kappa r}{2(u_{0}^{\rm FL})^4}\
             \bar{\psi}(x) \, \sigma_{\mu\nu} \, F_{\mu\nu}\psi(x)\ ,
\label{eq:S_SW}
\end{equation}
where $F_{\mu\nu}$ is constructed using fat links, and the plaquette 
measure $u_{0}^{\rm FL}$ is calculated via Eq.~(\ref{uzero}) using the 
fat links.
The factor $C_{\rm SW}$ is the (Sheikholeslami-Wohlert) clover 
coefficient \cite{Sheikholeslami:1985ij}, defined to be 1 at tree-level.
The quark hopping parameter is $\kappa = 1/(2m + 8r)$, and we use the 
conventional choice of the Wilson parameter, $r=1$.
In Eq.~(\ref{eq:S_SW}) the mean-field improved Fat-Link Irrelevant 
Wilson action is given by
\begin{eqnarray}
S_{\rm W}^{\rm FL}
=  &\sum_{x}& \bar{\psi}(x)\psi(x) 
+ \kappa \sum_{x,\mu} \bar{\psi}(x)
    \bigg[ \gamma_{\mu}
      \bigg( \frac{U_{\mu}(x)}{u_0} \psi(x+\hat{\mu})
- \frac{U^{\dagger}_{\mu}(x-\hat{\mu})}{u_0} \psi(x-\hat{\mu})
      \bigg)						\nonumber\\
& & \hspace*{3.2cm}
-\ r\ \bigg(
 \frac{U_{\mu}^{\rm FL}(x)}{u_0^{\rm  FL}} \psi(x+\hat{\mu})
+ \frac{U^{{\rm FL}\dagger}_{\mu}(x-\hat{\mu})}{u_0^{\rm FL}}
          \psi(x-\hat{\mu})
      \bigg)
    \bigg]\ .
\end{eqnarray}

As shown by Zanotti {\em et al.} \cite{Zanotti:2001yb}, the mean-field 
improvement parameter for the fat links is very close to 1, so that
the mean-field improved coefficient for $C_{\rm SW}$ is adequate 
\cite{Zanotti:2001yb}.
% >>>
% A futher advantage is that one can now use highly improved definitions
A further advantage is that one can now use highly improved definitions
% <<<
of $F_{\mu\nu}$ (involving terms up to $u_0^{12}$), which give 
impressive near-integer results for the topological charge 
\cite{Bilson-Thompson:2001ca,Bilson-Thompson:2002jk}.
In particular, we employ an ${\cal O}(a^4)$ improved definition of 
$F_{\mu\nu}$, as used by Bilson-Thompson {\it et al.} \cite{Bilson-Thompson:2001ca,Bilson-Thompson:2002jk}.

A fixed boundary condition in the time direction is implemented by 
setting $U_t(\vec x, N_t) = 0\ \forall\ \vec x$ in the hopping terms of 
the fermion action, and periodic boundary conditions are imposed in the 
spatial directions.
% >>>
% Gauge-invariant gaussian smearing \cite{Gusken:1989qx} in the spatial 
Gauge-invariant Gaussian smearing \cite{Gusken:1989qx} in the spatial 
% <<<
dimensions is applied at the source to increase the overlap of the 
interpolating operators with the ground states.
The source-smearing technique \cite{Gusken:1989qx} starts with a point 
source, $\psi_0({\vec x}_0, t_0)$, at space-time location
$({\vec x}_0, t_0) = (1,1,1,8)$, and proceeds via the iterative scheme,
\begin{equation}
\psi_i(x,t) = \sum_{x'} F(x,x') \, \psi_{i-1}(x',t) \, ,
\end{equation}
where
\begin{eqnarray}
F(x,x')
&=& \frac{1}{(1+\alpha)}
    \Big( \delta_{x, x'} 
	+ \frac{\alpha}{6} \sum_{\mu=1}^3
	  \big [  U_\mu(x) \, \delta_{x',x+\widehat\mu}\
	       +\ U_\mu^\dagger(x-\widehat\mu) \, \delta_{x', x-\widehat\mu}
	  \big ]
\Big) \, .
\end{eqnarray}
Repeating the procedure $N$ times gives the following fermion field:
\begin{equation}
\psi_N(x,t) = \sum_{x'} F^N(x,x') \, \psi_0(x',t) \, .
\end{equation}
The parameters $N$ and $\alpha$ govern the size and shape of the
% >>>
% smearing function and in our simulations we use $N=30$ and $\alpha=6$.
smearing function and in our simulations we use $N=35$ and $\alpha=6$.
% <<<

% >>>
% Six quark masses are used in the calculations \cite{Zanotti:2001yb}, and the strange 
Six quark masses are used in the calculations, and the strange 
% <<<
quark mass is taken to be the third largest ($\kappa = 1.2885$) quark mass in each case.
% >>>
% This $\kappa$ provides a pseudoscalar mass of $697$~MeV which compares well with the experimental value
% of $2M_{{\rm K}}^{2} - M_{\pi}^{2} = 693$~MeV motivated by leading order chiral perturbation theory. 
At this $\kappa$ the $s\bar s$ pseudoscalar mass is 697~MeV, which
compares well with the experimental value of
$2M_{\rm K}^2 - M_\pi^2 = 693$~MeV motivated by leading order
chiral perturbation theory. 
% <<<
The smallest pion mass considered is
$m_\pi = 464$~MeV.
We have also considered two smaller masses, but find that the signal for
these becomes quite noisy, and do not include them in the analysis.
The analysis for the $NK$-type interpolators is based on a sample of 200 configurations and
the analysis for the $PS$ and $SS$-type interpolators is based on an ensemble of 340 configurations.
The error
analysis is performed by a second-order, single-elimination jackknife,
% >>>
% with the $\chi^2$ per degree of freedom ($N_{\rm DF}$) obtained via
with the $\chi^2$ per degree of freedom obtained via
% <<<
covariance matrix fits.

\section{Results and Discussion}
In this section we present the results of our lattice simulations of 
pentaquark masses, in both the $J^P = {1\over 2}^-$ and ${1\over 2}^+$ 
channels, and for isospin $I=0$ and 1.
In addition to extracting the masses, we also study the mass differences
between the candidate pentaquark states and the free two-particle states.
This analysis is actually crucial in determining the nature of the 
states observed on the lattice, and the identification of true resonances.

% ........................................................................
\subsection{Signatures of a resonance}
\label{ssec:sigres}

At sufficiently small quark masses, the standard lattice technology will 
find, at large Euclidean times, the $NK$ decay channel as the ground 
state of all the five-quark interpolating fields discussed in 
Sec.~\ref{sec:if}.
In previous analyses of pentaquark masses, the observation of a signal 
at a mass slightly above the $NK$ threshold has been interpreted 
\cite{Csikor:2003ng,Sasaki:2003gi} as a candidate for a pentaquark resonance.
A robust test of whether a signal is a resonance or a scattering state
should involve an analysis of the binding of the state
as a function of the quark mass.
In a simple model with an attractive potential between the meson and baryon, the
resonance would be expected to sit lower in the potential with 
increasing quark mass.
%
%For example, the spherically symmetric square-well potential has the 
%number of bound states governed by the simple formula .... where $V_{0}$ is the strength of the attraction.
%
% >>>
% For sufficiently heavy quarks masses, a bound state will appear
For sufficiently large quark masses, a bound state will appear
% <<<
({\it lighter} than its decay products), and therefore become accessible using standard
lattice technology.
% >>>
% This behavior has in fact been observed in every lattice study of the $N^*$ 
This behaviour has in fact been observed in every lattice study of the $N^*$ 
% <<<
spectrum in every channel.
This feature is central to the study of the electromagnetic properties of
decuplet baryons \cite{Leinweber:1992hy} and their transitions \cite{Leinweber:1992pv,Alexandrou:2003ea,Alexandrou:2004xn} in lattice QCD.

\begin{figure}[tp]
\includegraphics[height=12.0cm,angle=90]{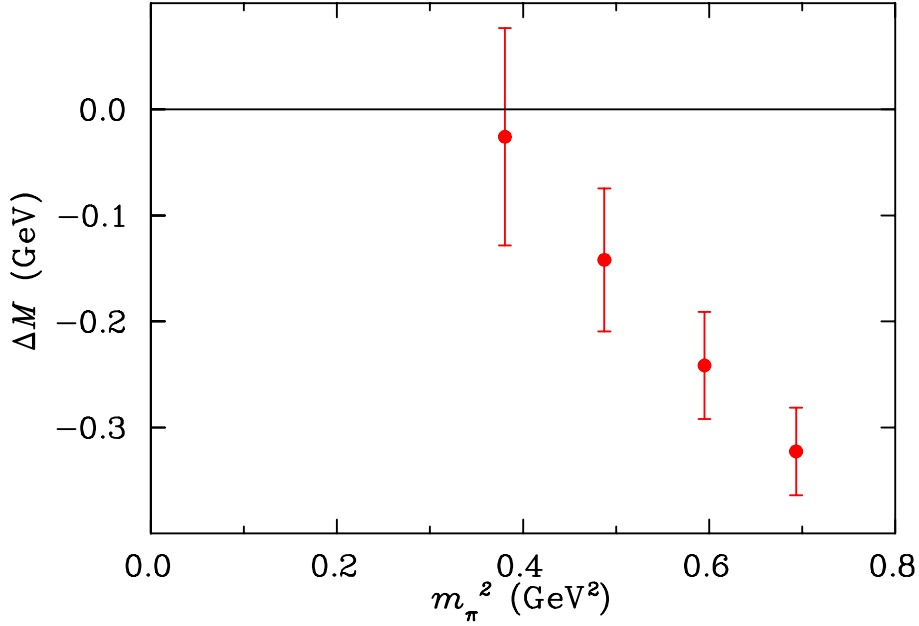} 
\caption{\label{fig:N1535}
	Mass difference between the lowest-lying negative parity 
	excited nucleon bound state,
	the $I(J^P)={1 \over 2}({1 \over 2}^-)\ N^*(1535)$,
	and the S-wave $N+\pi$ two-particle scattering state.}
\end{figure}

\begin{figure}[tp]
\includegraphics[height=12.0cm,angle=90]{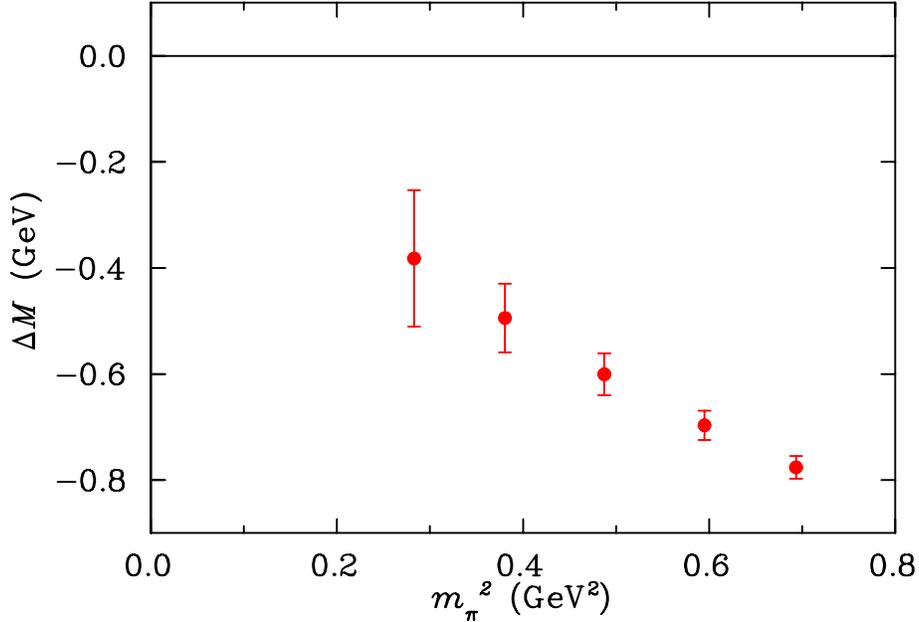}
\caption{\label{fig:Delta}
	Mass difference between the
	$I(J^P)={3 \over 2}({3 \over 2}^+)\ \Delta(1232)$
	and the $P$-wave $N+\pi$ two-particle scattering state. }
\end{figure}

As an example, consider the $J^P={1\over 2}^-$ parity partner of the 
nucleon, namely the $N^*(1535)$ baryon, in lattice QCD.
In the continuum, the $N^*(1535)$ is a resonance which decays to a 
nucleon and a pion.
On the lattice, however, the $N^*({1\over 2}^-)$ is stable at the
% >>>
(unphysically) large quark masses where its mass is smaller than 
% <<<
the sum of the nucleon and pion masses \cite{Melnitchouk:2002eg}.
To illustrate this we show in Fig.~\ref{fig:N1535} the mass splitting 
% >>>
% between the $N^*({1\over 2}^-)$ and the S-wave $N+\pi$ two-particle
between the $N^*({1\over 2}^-)$ and the non-interacting S-wave $N+\pi$
two-particle
% <<<
state, calculated on the same lattice.
For all values of $m_\pi$ illustrated the mass of the $N^*({1\over 2}^-)$ is below 
that of the $N+\pi$, and the mass difference clearly increases in magnitude with 
increasing $m_\pi$.
% >>>
% The binding becomes stronger at larger quark masses.
In other words, the binding becomes stronger at larger quark masses.
% <<<

% >>>
% A similar behavior is also observed in the case of the $\Delta(1232)$
A similar behaviour is also observed in the case of the $\Delta(1232)$
% <<<
isobar.
% >>>
% The mass difference between the $J^P={3\over 2}^+$ resonance and the
% P-wave $N+\pi$ two-particle mass, shown in Fig.~\ref{fig:Delta}, is
The mass difference between the $J^P={3\over 2}^+$ resonance
and the lowest available P-wave $N+\pi$ two-particle energy, shown in
Fig.~\ref{fig:Delta}, is
% <<<
negative, and, as in Fig.~\ref{fig:N1535}, increases with $m_\pi^2$.
In fact, this pattern is repeated in every other baryon channel probed in lattice QCD, such as
% >>>
% the $J^P={1\over 2}^+$ and ${3\over 2}^-$ channels \cite{Melnitchouk:2002eg} for example.
% Thus the standard lattice resonance signature for the $\Theta^{+}$ resonance is the existence of  
the $J^P={1\over 2}^+$ and ${3\over 2}^-$ channels, for example
\cite{Leinweber:2004it,Melnitchouk:2002eg,Zanotti:2003fx}.
Thus the standard lattice resonance signature for the
$\Theta^+$ resonance is the existence of  
% <<<
a state with a mass which becomes {\it smaller} than that of the $N + K$
two-particle state as the quark mass increases, with the mass difference
increasing at larger quark masses.

%%%%%%%%%%%%%%%%%%%%%%%%%%%%%%%%%%%%%%%%%%%%%%%%%%%%%%%%%%%%%%%%%%%%%%%
% >>>
% \subsection{Isoscalar negative parity states}
\subsection{Negative parity isoscalar states}
% <<<
\label{ssec:I0-}

We begin our discussion of the results with the isospin-0, negative
parity states.
The lowest energy of a system with a nucleon and a kaon would have the
$N$ and $K$ in a relative S-wave, in which case the overall parity is 
negative.
Isoscalar states can be constructed with the $NK$-type interpolating 
fields, $\chi_{NK}$ and $\chi_{\widetilde{NK}}$, as well as with the
$PS$-type field $\chi_{PS}$.

\begin{figure}[tp]
\includegraphics[height=12.0cm,angle=90]{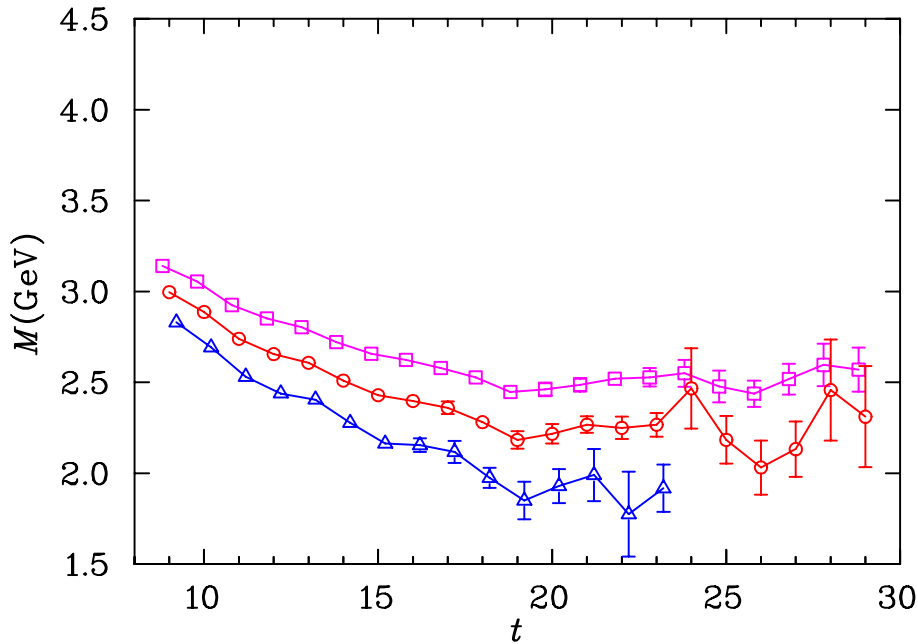} 
\caption{\label{fig:I0_NK.neg.sing}
	Effective mass of the $I(J^P)=0(\frac{1}{2}^-)$
	colour singlet $NK$-type pentaquark interpolator, $\chi_{NK}$.
	The data correspond to $\kappa=1.2780$ (squares),
	1.2885 (circles), and 1.2990 (triangles).}
\end{figure}

\begin{figure}[tp]
\includegraphics[height=12.0cm,angle=90]{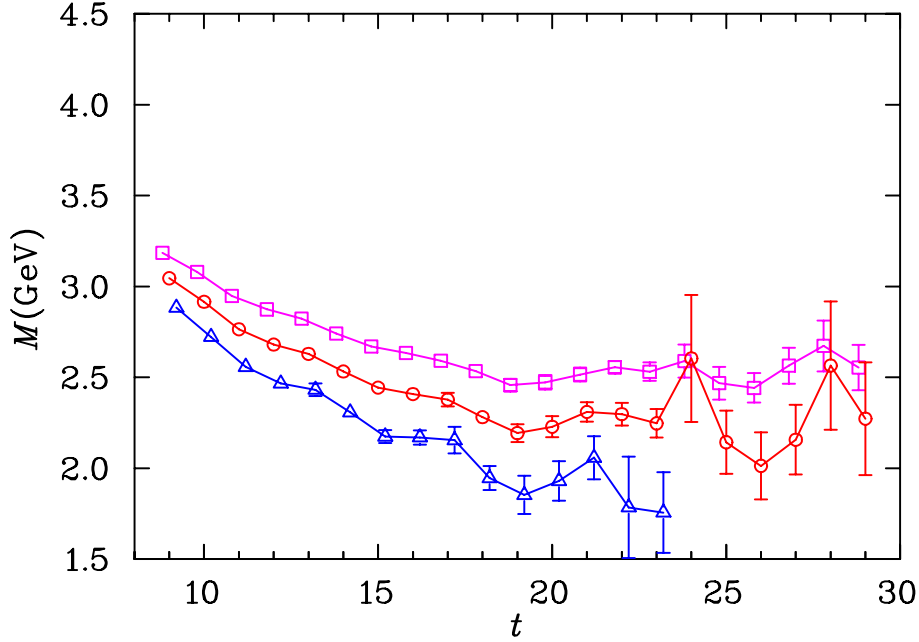} 
\caption{\label{fig:I0_NK.neg.fused}
% >>>
%	Effective mass of the $I(J^P)=0(\frac{1}{2}^-)$
%	colour fused $NK$-type pentaquark interpolator, $\chi_{\widetilde{NK}}$.
%	The data correspond to $\kappa=1.2780$ (squares),
%	1.2885 (circles), and 1.2990 (triangles).}
	As in Fig.~\ref{fig:I0_NK.neg.sing}, but for the
	$I(J^P)=0(\frac{1}{2}^-)$ colour fused $NK$-type pentaquark
	interpolator, $\chi_{\widetilde{NK}}$.}
% <<<
\end{figure}

\begin{figure}[tp]
\includegraphics[height=12.0cm,angle=90]{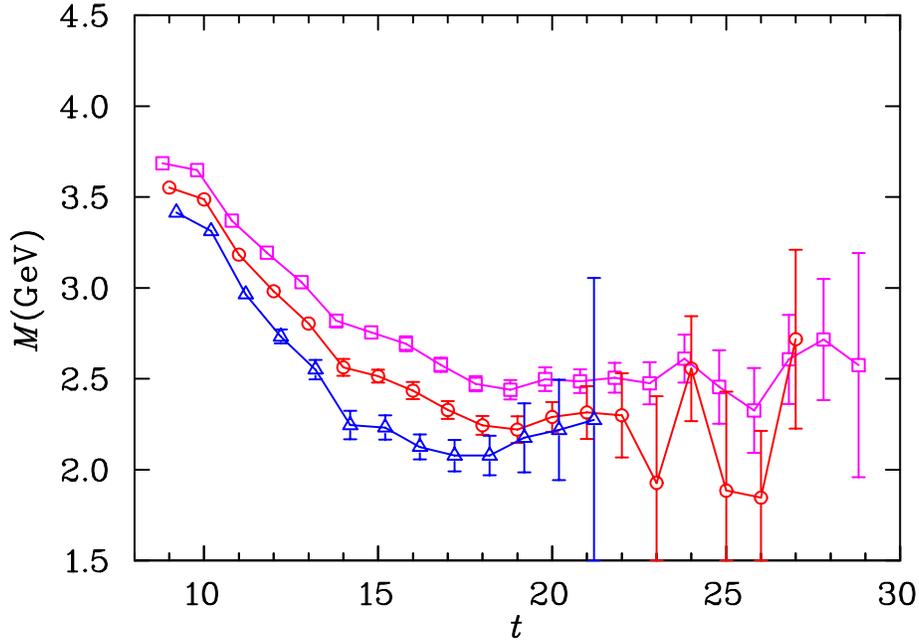} 
\caption{\label{fig:I0_PS.neg}
	As in Fig.~\ref{fig:I0_NK.neg.sing}, but for the
	$I(J^P)=0(\frac{1}{2}^-)$ $PS$-type
	pentaquark interpolator, $\chi_{PS}$.}
\end{figure}

The effective mass for the colour singlet $\chi_{NK}$ field is shown in 
Fig.~\ref{fig:I0_NK.neg.sing} for several representative $\kappa$ values. 
The results for the colour fused operator $\chi_{\widetilde{NK}}$ of Fig.~\ref{fig:I0_NK.neg.fused} are 
very similar to those for $\chi_{NK}$.
% from page 30.
% 
% Note that in order to avoid clutter in the figures we do not show the
% points at the larger $t$ values which have larger error bars, and have 
% little influence on the fit.
(Note that to avoid clutter in the figures we do not show the
points at the larger $t$ values which have larger error bars, and have 
little influence on the fits.)
% <<<
To determine whether these operators have significant overlap with
more than one state, a correlation matrix analysis is performed.
% >>>
% However, it was not possible to improve the ground state mass as
% described in Sec.~(\ref{ssec:corrmatrix}).
However, we find that it is not possible to improve the ground state
mass as described in Sec.~\ref{ssec:corrmatrix}.
% <<<
% >>>
% Consequently the standard ({\it i.e.}, non-correlation matrix)
Consequently the results using the standard ({\it i.e.}, non-correlation matrix)
% <<<
analysis techniques are reported in this channel.
For comparison, in Fig.~\ref{fig:I0_PS.neg} we also show the effective 
mass of the negative parity $\chi_{PS}$ diquark-type field.
As expected, because of the presence of two smaller components of
Dirac spinors in $\chi_{PS}$ compared with $\chi_{NK}$ and $\chi_{\widetilde{NK}}$ , the signal
here is much noisier than for either of the $NK$-type fields.
% >>>
This is despite the fact that almost twice as many configurations
are used for the $\chi_{PS}$ than for the $NK$-type fields.
% <<<

The pentaquark masses are extracted by fitting the effective masses
% >>>
% in Figs.~\ref{fig:I0_NK.neg.sing} and \ref{fig:I0_PS.neg} over appropriate
in Figs.~\ref{fig:I0_NK.neg.sing}--\ref{fig:I0_PS.neg} over appropriate
% <<<
$t$ intervals, chosen according to the criterion that the $\chi^2$ per 
degree of freedom % calculated with the covariance matrix analysis
% >>>
% is less than 1.5, and (since our data are normally distributed) 
is less than 1.5, and 
% <<<
preferably close to 1.0.
For the $\chi_{NK}$ and $\chi_{\widetilde{NK}}$ fields, fitting the effective mass in the window
% >>>
% $t=19-23$ is found to optimise the $\chi^2$/dof.\par
$t=19-23$ is found to optimise the $\chi^2$/dof.
% <<<
For the $\chi_{PS}$ field, the signal is lost at slightly earlier times,
and consequently we fit in the time interval $t=18-20$.
% >>>
% The results for the masses corresponding to the $\chi_{NK}$ and 
The results for the masses corresponding to the $\chi_{NK}$,
$\chi_{\widetilde{NK}}$ and 
% <<<
$\chi_{PS}$ fields are tabulated in Table~\ref{tab:I0.neg}.

\begin{table}[tp]     
\caption{\label{tab:I0.neg}
	The pion mass and the masses of the $I(J^P)=0(\frac{1}{2}^-)$ states extracted with the  
	colour singlet $NK$,
	colour fused $\widetilde{NK}$
	and $PS$-type pentaquark interpolating fields 
	for various values of $\kappa$. }
\begin{ruledtabular} 
\begin{tabular}{c|cccc}
$\kappa$ & $aM_{\pi}$  & $aM_{NK}$ & $aM_{\widetilde{NK}}$ & $aM_{PS}$ \cr
\hline
1.2780 & 0.540(2)  & 1.612(17) & 1.625(16) & 1.601(21) \cr
1.2830 & 0.500(2)  & 1.539(17) & 1.553(16) & 1.532(23) \cr
1.2885 & 0.453(2)  & 1.449(20) & 1.461(20) & 1.458(27) \cr
1.2940 & 0.400(3)  & 1.349(28) & 1.361(29) & 1.396(37) \cr
1.2990 & 0.345(3)  & 1.236(50) & 1.245(48) & 1.372(72) \cr
1.3025 & 0.300(3)  & 1.145(67) & 1.138(80) & 1.442(171) \cr
\end{tabular}
\end{ruledtabular}
\end{table}

\begin{table}[tp]     
\caption{\label{tab:2P.neg}
        Masses of the
	S-wave $N+K$, $N+K^*$, $\Delta+K^{*}$ and $N'+K$ two-particle states.
}
\begin{ruledtabular} 
\begin{tabular}{c|cccc}
$\kappa$ & $aM^{\rm S-wave}_{N+K}$          & $aM^{\rm S-wave}_{N+K^{*}}$
         & $aM^{\rm S-wave}_{\Delta+K^{*}}$ & $aM^{\rm S-wave}_{N'+K}$ 
\cr
\hline
1.2780 & 1.553(10) & 1.762(14) & 1.893(13) & 2.180(27) \cr
1.2830 & 1.485(11) & 1.704(16) & 1.848(14) & 2.136(32) \cr
1.2885 & 1.404(13) & 1.635(19) & 1.799(16) & 2.092(39) \cr
1.2940 & 1.314(18) & 1.561(26) & 1.751(19) & 2.061(54) \cr
1.2990 & 1.216(32) & 1.485(41) & 1.709(24) & 2.056(87) \cr
1.3025 & 1.130(52) & 1.421(68) & 1.682(29) & 2.093(144) \cr
\end{tabular}
\end{ruledtabular}
\end{table}

\begin{figure}[tp]
\includegraphics[height=12.0cm,angle=90]{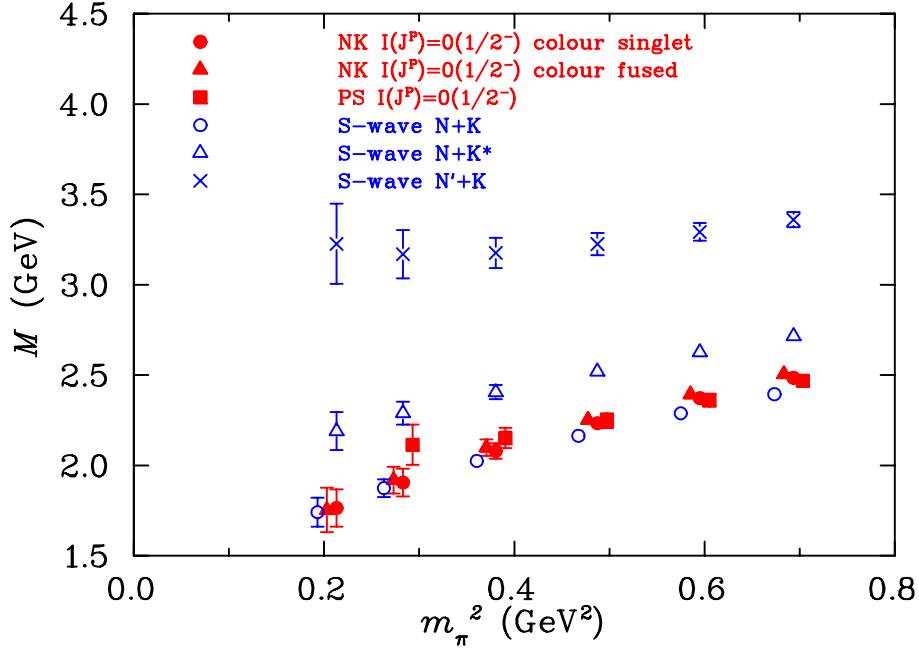}
\caption{\label{fig:I0.neg}
	Masses of the $I(J^P)=0(\frac{1}{2}^-)$ states extracted with the 
	colour singlet $NK$, colour fused $\widetilde{NK}$
	and $PS$-type pentaquark interpolating fields
	as a function of $m_\pi^2$.
	For comparison, the masses of the S-wave $N+K$, $N+K^*$
	and $N'+K$ two-particle states are also shown.
	Some of the points have been offset for clarity.}
\end{figure}

The pentaquark masses are compared with masses of several two-particle
% >>>
% states reported in Table~(\ref{tab:2P.neg}).
states, which are reported in Table~\ref{tab:2P.neg}.
% <<<
The lowest-energy two-particle states in the $I(J^P)=0(\frac{1}{2}^-)$
channel are the S-wave $N+K$, the S-wave $N+K^*$, the P-wave $N^*+K$, 
where $N^*$ is the lowest negative parity excitation of the nucleon, 
and the S-wave $N'+K$, where $N'$ is the first positive-parity excited 
state of the nucleon.
% Because of the spherically symmetric nature of our source smearing,
% we expect that our operators will not couple strongly to P-wave states, 
% and should therefore not have significant overlap with the P-wave 
% $N^*+K$ state.
The contributions to the correlation function from the P-wave states are likely to be
% >>>
% suppressed because there is a contribution to the P-wave signal from two
suppressed, however, because there is a contribution to the P-wave signal
from two
% <<<
{\it small} components of the spinors.
We therefore focus on the S-wave states in 
Table~\ref{tab:2P.neg}.

% >>>
% Furthermore, the positive parity $N'$ state, which is calculated 
The positive parity $N'$ state, which is calculated 
% <<<
from the interpolating field
$\epsilon^{abc} (u^{a T} C d^b) \gamma_5 u^c$, is known to have poor 
overlap with the nucleon ground state, as well as with the low-lying
${1\over 2}^+$ excitations, such as the Roper $N^*(1440)$.
In fact, it gives a mass greater than $\sim 2$~GeV, significantly 
above the low-lying ${1\over 2}^+$ excitations \cite{Melnitchouk:2002eg}.
We expect, therefore, that our pentaquark fields will not have
strong overlap with the $N'+K$ state either.

The results for the extracted masses from Table~\ref{tab:I0.neg}
are displayed in Fig.~\ref{fig:I0.neg} as a function of $m_\pi^2$.
% >>>
% The masses of the pentaquark states extracted from the $\chi_{NK}$ and $\chi_{PS}$
% fields agree within the errors, although the errors on the $\chi_{PS}$
The masses of the pentaquark states extracted from the $\chi_{NK}$,
$\chi_{\widetilde{NK}}$ and $\chi_{PS}$ fields agree within the errors
(the $\chi_{NK}$ and $\chi_{\widetilde{NK}}$ masses in particular are
very close), although the errors on the $\chi_{PS}$
% <<<
become large at the smaller quark masses.
The pentaquark masses are either consistent with or lie above the mass
of the lowest two-particle state, namely the S-wave $N+K$.
% >>>
% Table~\ref{tab:I0.neg} also shows that the 
% mass extracted with the colour fused $\chi_{\widetilde{NK}}$ interpolating field is 
% degenerate with the colour singlet $\chi_{NK}$ interpolating field.
% <<<
%
% As expected, the S-wave $N+K^*$ and especially the $N'+K$ two-particle
% states lie significantly above the other states.

\begin{figure}[tp]
\includegraphics[height=12.0cm,angle=90]{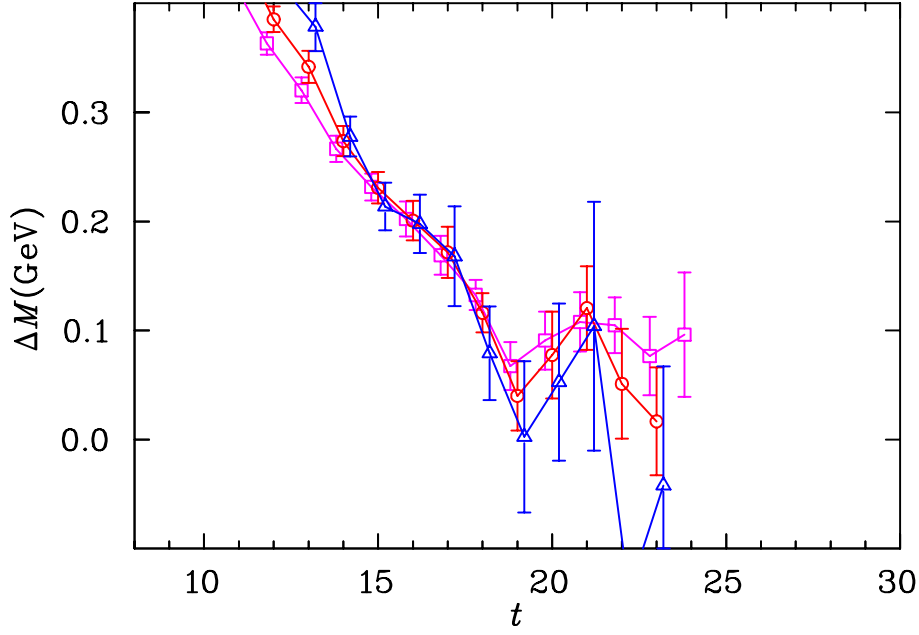} 
\caption{\label{fig:difI0_NK.neg.meff}
% >>>
%	Effective mass splitting between the $I(J^P)=0(\frac{1}{2}^-)$ state extracted with the 
%	colour singlet $NK$-type pentaquark interpolator, $\chi_{NK}$ and the S-wave $N+K$ two particle state.
	Effective mass difference between the $I(J^P)=0(\frac{1}{2}^-)$
	state extracted with the colour singlet $NK$-type pentaquark
	interpolator, $\chi_{NK}$, and the S-wave $N+K$ two-particle
	state.
% <<<
	The data correspond to $\kappa=1.2780$ (squares),
	1.2885 (circles), and 1.2990 (triangles).}
\end{figure}

\begin{figure}[tp]
\includegraphics[height=12.0cm,angle=90]{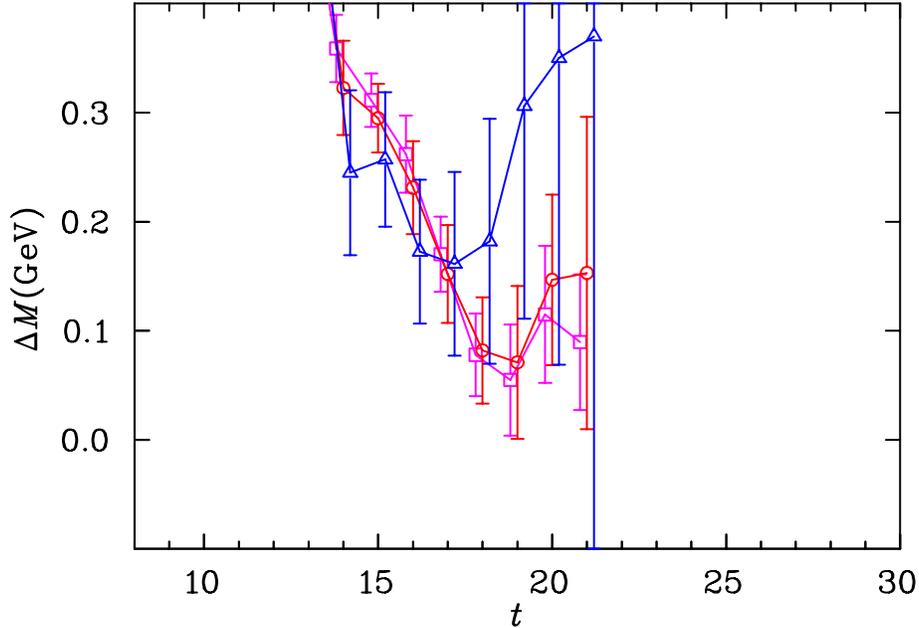} 
\caption{\label{fig:difI0_PS.neg.meff}
	As in Fig.~\ref{fig:difI0_NK.neg.meff}, but for the
	$I(J^P)=0(\frac{1}{2}^-)$ $PS$-type
	interpolating field, $\chi_{PS}$.}
\end{figure}

% >>>
\begin{table}[tp]     
\caption{\label{tab:difI0.neg}
	Mass differences between the $I(J^P)=0(\frac{1}{2}^-)$ states
	extracted with the $NK$ and $PS$-type  pentaquark
	interpolating fields 
	and the S-wave $N+K$ two-particle state.}
\begin{ruledtabular} 
\begin{tabular}{c|cc}
$\kappa$ & $aM_{NK} - aM_{N+K}^{\rm S-wave}$ & $aM_{PS} - aM_{N+K}^{\rm S-wave}$ \cr 
\hline
1.2780 & 0.056(9) & 0.051(19) \cr
1.2830 & 0.052(11) & 0.052(21) \cr
1.2885 & 0.048(13) & 0.060(25) \cr
1.2940 & 0.043(17) & 0.086(37) \cr
1.2990 & 0.025(30) & 0.148(75) \cr
1.3025 & -0.011(67) & 0.286(177) \cr
\end{tabular}
\end{ruledtabular}
\end{table}

% <<<

\begin{figure}[tp]
\includegraphics[height=12.0cm,angle=90]{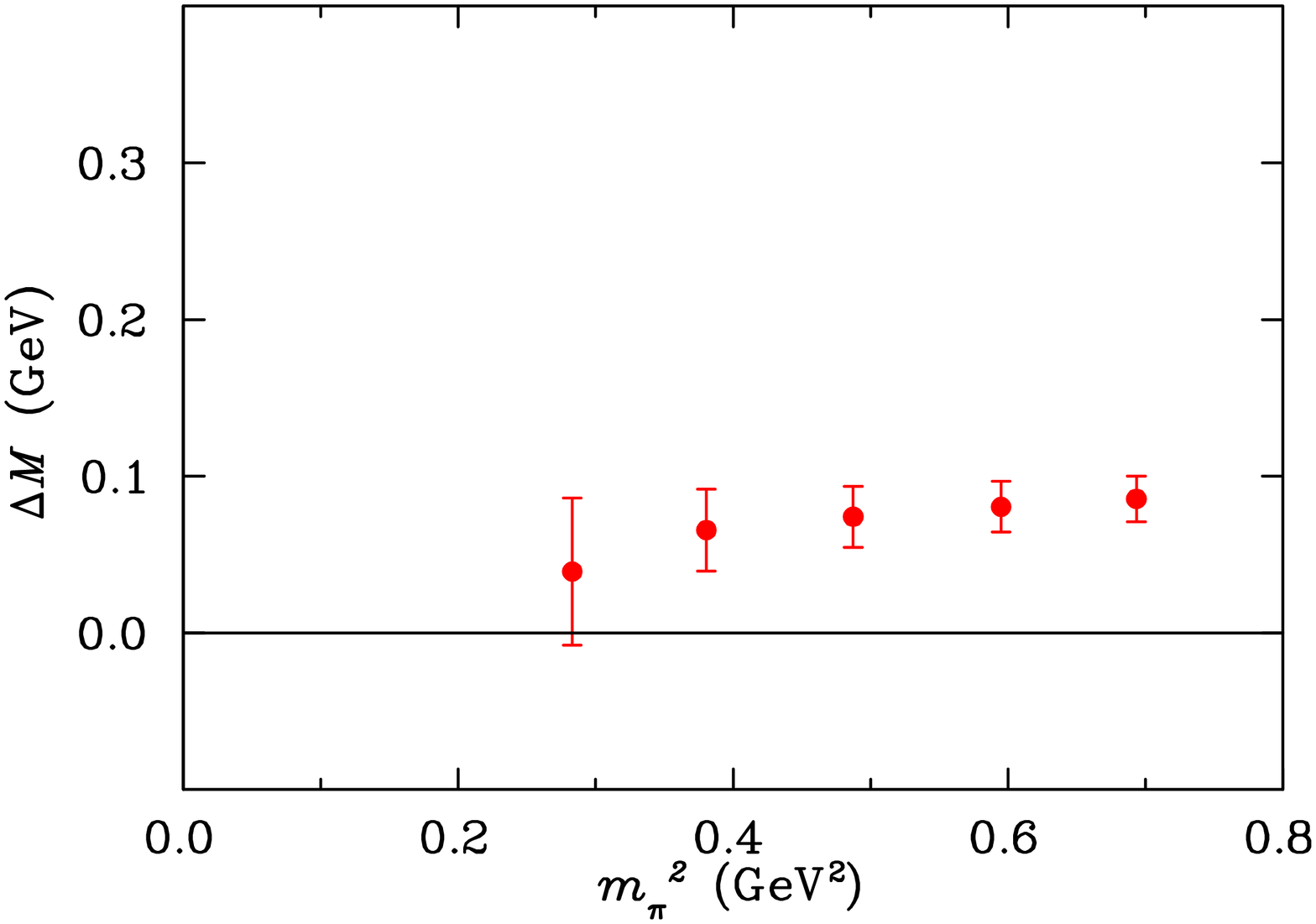}
\caption{\label{fig:difI0_NK.neg}
	Mass difference between the $I(J^P)=0(\frac{1}{2}^-)$ state extracted with the 
	$NK$-type pentaquark interpolating field and the S-wave $N+K$ two-particle state.}
\end{figure}

\begin{figure}[tp]
\includegraphics[height=12.0cm,angle=90]{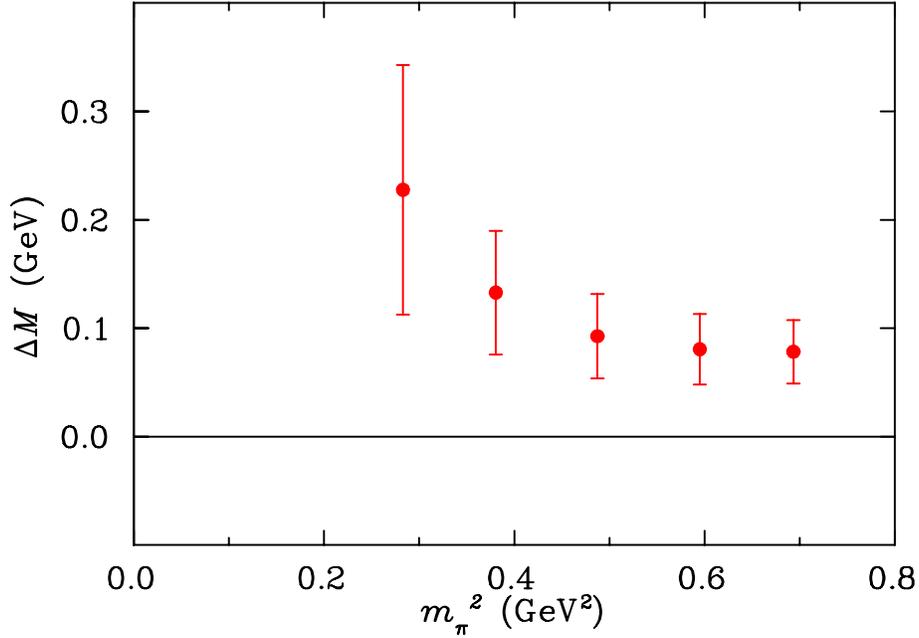}
\caption{\label{fig:difI0_PS.neg}
	Mass difference between the $I(J^P)=0(\frac{1}{2}^-)$ state extracted with the 
	$PS$-type pentaquark interpolating field and the S-wave
	$N+K$ two-particle state.}
\end{figure}

The mass differences $\Delta M$ between the low-lying pentaquark states and
the two-particle scattering states can be better resolved by fitting
the effective mass for the mass difference directly.
This allows for cancellation of
systematic errors
since the pentaquark and two-particle states are generated
% >>>
% from the same gauge field configuations, and hence their systematic
from the same gauge field configurations, and hence their systematic
% <<<
errors are strongly correlated.
% >>>
% Figs.~\ref{fig:difI0_NK.neg.meff} and \ref{fig:difI0_PS.neg.meff} illustrate the effective mass plots for the mass splittings.
Figures~\ref{fig:difI0_NK.neg.meff} and \ref{fig:difI0_PS.neg.meff}
illustrate the effective mass plots for the mass differences.
% <<<
%
% >>>
% We emphasize that the scale of these figures is six times more precise than that of the previous figures of this subsection.
Note that the scale of these figures is enlarged by a factor of six
compared with Figs.~\ref{fig:I0_NK.neg.sing}--\ref{fig:I0_PS.neg}.
% <<<
% >>>
% The mass splittings between the state extracted from the colour singlet $NK$ interpolator and the S-wave $N+K$
% two-particle state are fitted at time slices $t=19-24$, while that
% between the $PS$ extracted state at $t=18-20$.
% The results of the mass difference analysis are presented in
% Table~\ref{tab:difI0.neg}, and in Figs.~\ref{fig:difI0_NK.neg} and
% \ref{fig:difI0_PS.neg} for the $\chi_{NK}$ and $\chi_{PS}$ fields,
% respectively.
% The mass splitting analysis clearly reveals that the masses derived
% from the $NK$ pentaquark operator are consistently higher than the
% lowest-mass two-particle state.
% The mass difference $\Delta M$ is $\sim 100$~MeV at the larger quark
% masses, and weakly dependent on $m_\pi^2$, with a possible trend
% towards a larger $\Delta M$ with increasing $m_\pi^2$, suggesting 
% %
% in the more accurate results,
% %
% repulsion not attraction.
% Note the size of the error bars for the mass difference is reduced
% compared with the error bars on the masses in Fig.~\ref{fig:I0.neg}.
% For the pseudoscalar-scalar diquark-type field, the mass difference
% between the S-wave $N+K$ is consistent with zero within errors, apart
% from the lowest quark mass, where it maybe positive.
%
The mass difference between the state extracted from the colour singlet
$NK$ interpolator and the S-wave $N+K$ two-particle state is fitted at
time slices $t=19-21$, while that between the $PS$ extracted state and
the $N+K$ state is fitted at $t=18-20$.

The results of the mass difference analysis are presented in
Table~\ref{tab:difI0.neg}, and in Figs.~\ref{fig:difI0_NK.neg} and
\ref{fig:difI0_PS.neg} for the $\chi_{NK}$ and $\chi_{PS}$ fields,
respectively.
We see clearly that the masses derived from the $NK$ pentaquark operator
are consistently higher than the lowest-mass two-particle state.
The mass difference $\Delta M$ is $\sim 100$~MeV at the larger quark
masses, and weakly dependent on $m_\pi^2$, with a possible trend
towards a larger $\Delta M$ with increasing $m_\pi^2$.
Note the size of the error bars for the mass difference is reduced
compared with the error bars on the masses in Fig.~\ref{fig:I0.neg}.
% <<<

Since the difference between the reported experimental $\Theta^+$ mass
and the physical S-wave $N+K$ continuum is $\sim 100$~MeV, naively one
% >>>
% may be tempted to interpret the result in Fig.~\ref{fig:difI0_NK.neg}
may be tempted to interpret the results in Figs.~\ref{fig:difI0_NK.neg}
and \ref{fig:difI0_PS.neg.meff}
% <<<
as a signature of the $\Theta^+$ on the lattice.
% >>>
% However, the behavior of the pentaquark--$(N+K)$ mass difference in
% Fig.~\ref{fig:difI0_NK.neg} is in marked contrast to that of all other
However, the behaviour of the pentaquark--$(N+K)$ mass difference
is in marked contrast to that of all other
% <<<
excited $N^*$ states studied on the lattice \cite{Melnitchouk:2002eg,Zanotti:2003fx,Leinweber:2004it},
as discussed in Sec.~\ref{ssec:sigres} above, for which $\Delta M$ is
{\em negative}.
%
% and {\em decreasing} with increasing $m_\pi^2$.
%Our results on the other hand provide evidence for a repulsive
%potential, rather than attraction.
The lack of any binding leads us to conclude that the observed signal
% >>>
% is unlikely to be a bound state resonance, and may instead correspond
is unlikely to be a resonance, and may instead correspond
% <<<
to an $N+K$ two-particle state.
The volume dependent analysis in Ref.~\cite{Mathur:2004jr} indeed concluded
that their signal, which is consistent with our results, corresponds
to an $NK$ scattering state.

%%%%%%%%%%%%%%%%%%%%%%%%%%%%%%%%%%%%%%%%%%%%%%%%%%%%%%%%%%%%%%%%%%%%%%%%
% >>>
% \subsection{Isovector negative parity states}
\subsection{Negative parity isovector states}
% <<<
\label{ssec:I1-}

For the isospin-1, negative parity sector, we consider three operators
which can create $I(J^P) = 1({1\over 2}^-)$ states: the isovector
combinations of the colour singlet $\chi_{NK}$ and colour fused
$\chi_{\widetilde{NK}}$, and the $SS$-type operator,
$\chi_{SS}$.
As for the isoscalar case, we perform a $2\times 2$ correlation matrix
analysis for the $NK$-type fields, and here we do find improved 
access to the lowest lying state.

% >>>
% Using the paradigm for optimizing the results described in
Using the paradigm for optimising the results described in
% <<<
Sec.~\ref{ssec:corrmatrix}, we perform the correlation matrix analysis
for the largest 4 quark masses starting at $t=20$ with $\Delta t = 3$.
% >>>
% Here the ground state mass was found to be lower with the correlation
Here the ground state mass is found to be lower with the correlation
% <<<
matrix than with the standard analysis, indicating that the
% >>>
% contamination of the ground state from excited states was reduced.
contamination of the ground state from excited states is reduced.
% <<<
For the second lightest quark mass we fit at $t=18$, and for the
lightest quark mass at $t=17$, with $\Delta t =3$ in both cases.
% >>>
% For these two lightest quark masses, the ground state mass was not
% lowered, so here the standard analysis techniques were used.
% For the excited state, the correlation matrix determined masses which
% were all higher than with the naive analysis for all quark masses, thus improving the analysis.
For these two lightest quark masses, the ground state mass is not
lowered, so here the standard analysis techniques are used.
For the excited state, the masses from the correlation matrix are all
higher than with the naive analysis for all quark masses, thus
improving the analysis.
% <<<

\begin{figure}[tp]
\includegraphics[height=12.0cm,angle=90]{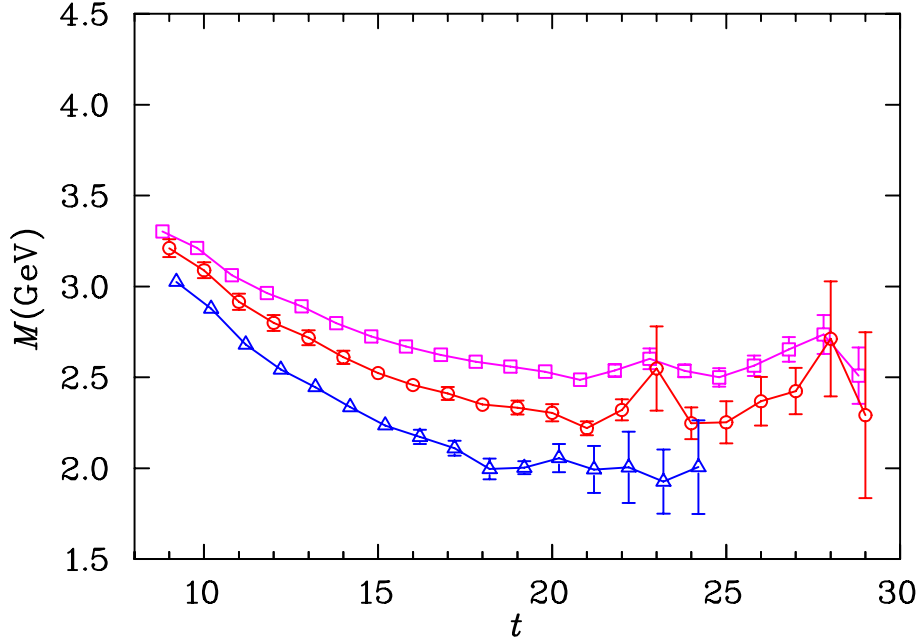} 
\caption{\label{fig:I1_NK1.neg}
% >>>
%	Effective mass of the $I(J^P)=1(\frac{1}{2}^-)$ state corresponding to the $NK$
	Effective mass of the $I(J^P)=1(\frac{1}{2}^-)$ state
	corresponding to the $NK$-type
% <<<
	pentaquark ``state 1'' for several values of $\kappa$,
	$\kappa=1.2780$ (squares), 1.2885 (circles), and 1.2990
	(triangles).}
\end{figure}

\begin{figure}[tp]
\includegraphics[height=12.0cm,angle=90]{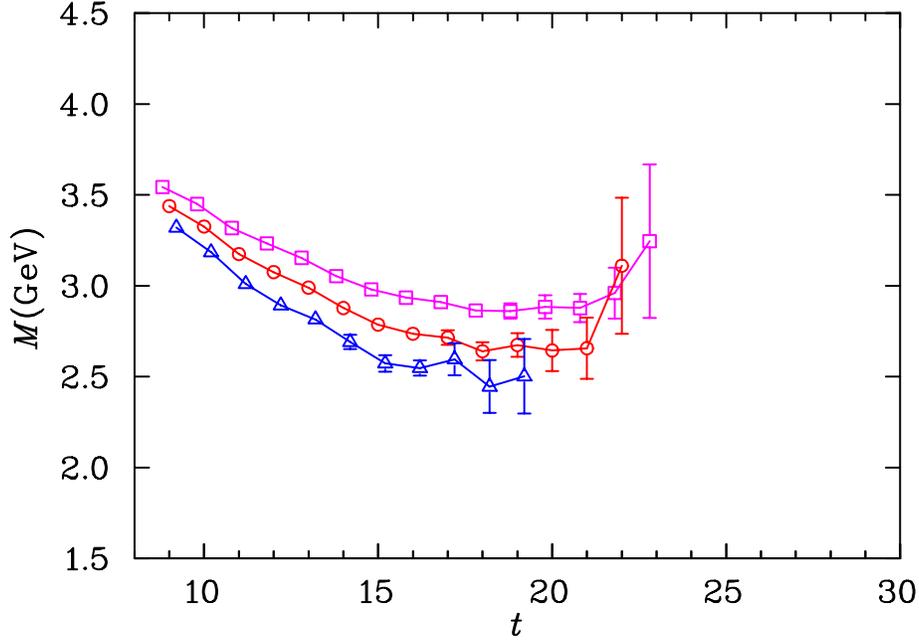} 
\caption{\label{fig:I1_NK2.neg}
% >>>
%	As in Fig.~\ref{fig:I1_NK1.neg} but for the
%	$I(J^P)=1(\frac{1}{2}^-)$ $NK$ pentaquark ``state 2''.}
	As in Fig.~\ref{fig:I1_NK1.neg}, but for the
	$I(J^P)=1(\frac{1}{2}^-)$ $NK$-type pentaquark ``state 2''.}
% <<<
\end{figure}

\begin{figure}[tp]
\includegraphics[height=12.0cm,angle=90]{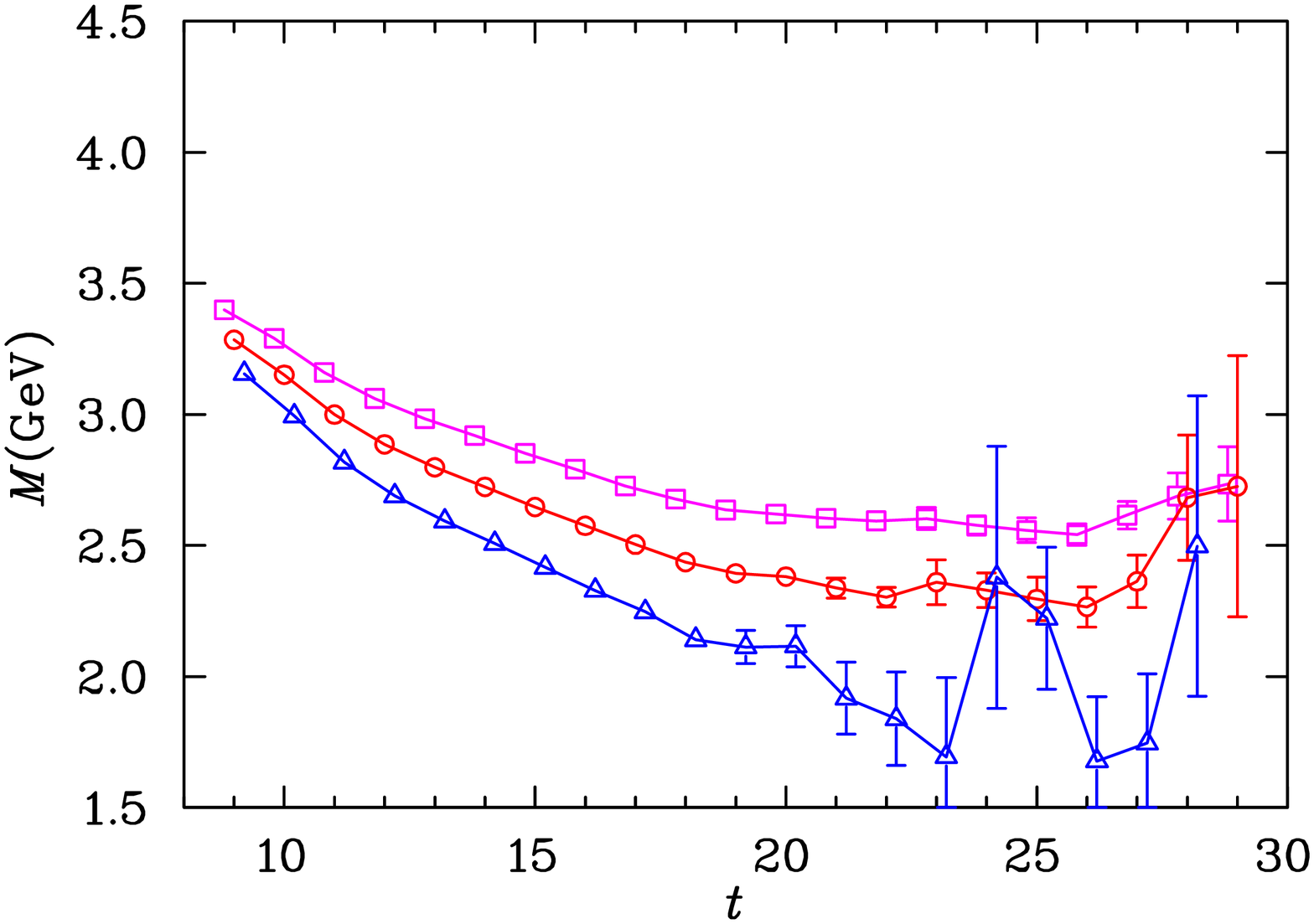} 
\caption{\label{fig:I1_SS.neg}
	As in Fig.~\ref{fig:I1_NK1.neg}, but for the
	$I(J^P)=1(\frac{1}{2}^-)$ $SS$-type interpolator, $\chi_{SS}$. }
\end{figure}

% >>>
% The effective masses for the two projected $NK$-type correlation matrix states, based on the  
The effective masses for the two projected $NK$-type correlation matrix states,
% <<<
which we refer to as
% >>>
% ``state 1'' and ``state 2'', are shown in Figs.~\ref{fig:I1_NK1.neg} and 
``state 1'' (for the ground state) and ``state 2'' (for the excited state),
are shown in Figs.~\ref{fig:I1_NK1.neg} and 
% <<<
\ref{fig:I1_NK2.neg}, respectively.
For comparison, we also show the effective mass plot for the
$SS$-type field $\chi_{SS}$ in Fig.~\ref{fig:I1_SS.neg}.
%The ground state masses extracted with the $NK$ and $SS$-type interpolators were fitted at time slices
%$t=22-26$.
% >>>
% The ground state mass extracted with the $NK$-type  interpolator was fitted at time slices
% $t=22-26$, while mass  extracted with the $SS$-type  interpolator was fitted at time slices
% $t=19-28$.
The ground state mass extracted with the $NK$-type  interpolator is fitted
at time slices $t=22-26$, while the mass extracted with the $SS$-type
interpolator is fitted at time slices $t=19-28$.
% <<<

\begin{table}[tp]     
\caption{\label{tab:I1.neg}
	Masses of the $I(J^P)=1(\frac{1}{2}^-)$ states extracted with the  
	$NK$ and $SS$-type 
	pentaquark interpolating fields for various values of $\kappa$. }
\begin{ruledtabular} 
\begin{tabular}{c|ccc}
$\kappa$ & $aM_{NK(1)}$ & $aM_{NK(2)}$ & $aM_{SS}$ \cr
\hline
1.2780 & 1.649(15)  & 1.859(22)  & 1.692(8) \cr
1.2830 & 1.578(18)  & 1.797(25)  & 1.619(9) \cr
1.2885 & 1.497(27)  & 1.720(31)  & 1.530(11) \cr
1.2940 & 1.408(48)  & 1.629(47)  & 1.434(16) \cr
1.2990 & 1.313(66)  & 1.577(77)  & 1.334(26) \cr
1.3025 & 1.251(144) & 1.554(175) & 1.245(51) \cr
\end{tabular}
\end{ruledtabular}
\end{table}

\begin{figure}[tp]
\includegraphics[height=12.0cm,angle=90]{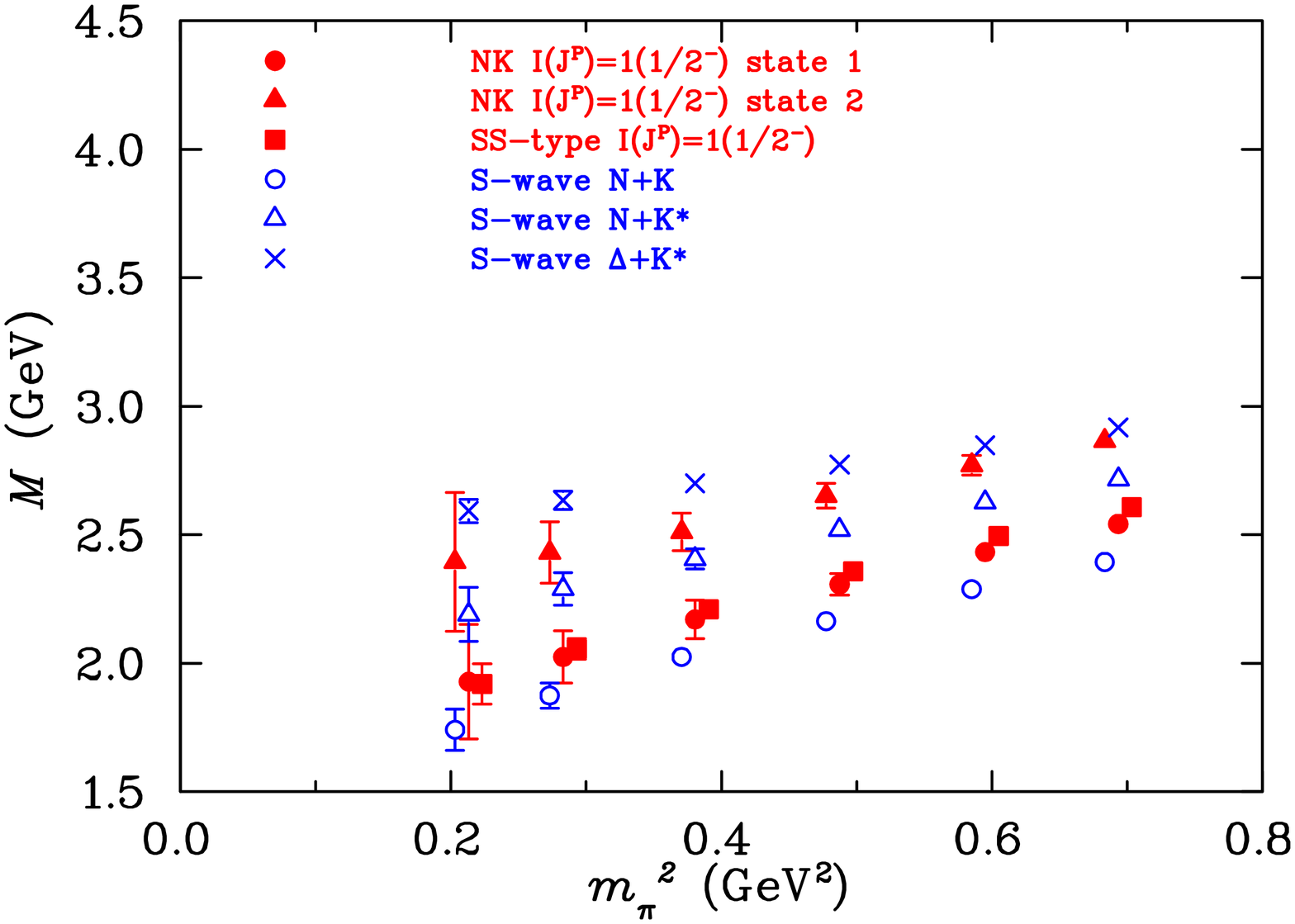} 
\caption{\label{fig:I1.neg}
	Masses of the $I(J^P)=1(\frac{1}{2}^-)$ states extracted with the 
	$NK$ and $SS$-type pentaquark
	interpolating fields as a function of $m_\pi^2$, compared with the
	masses of the S-wave $N+K$, $N+K^*$ and $\Delta+K^{*}$ two-particle
	states.  Some of the points have been offset for clarity.}
\end{figure}

The resulting extracted masses are tabulated in Table~\ref{tab:I1.neg}
and shown in Fig.~\ref{fig:I1.neg}.
% >>>
% A clear mass splitting of $\sim 400$~MeV is seen between the ground state
% and excited state for the $NK$-type operators, and the ground state mass
A clear mass splitting of $\sim 400$~MeV is seen between the ground state
and the excited state for the $NK$-type operators.  The ground state mass
% <<<
is consistent with that obtained from the $\chi_{SS}$ operator for the four smallest quark masses, 
but is slightly smaller for the two largest quark masses.
As for the isoscalar channel, the ground state masses are either
consistent with or slightly above the masses of the lowest two-particle
state, the S-wave $N+K$.
The excited state lies slightly above the S-wave two-particle $N+K^*$
% >>>
% threshold, and appears to be an admixture of $N+K^{*}$ and $\Delta + K^{*}$ scattering states.
threshold, which suggests that it may be an admixture of $N+K^*$ and
$\Delta + K^*$ scattering states.
% <<<

\begin{table}[tp]     
\caption{\label{tab:difI1.neg}
	Mass differences between the $I(J^P)=1(\frac{1}{2}^-)$ states
	extracted with the $NK$ and $SS$-type  pentaquark
	interpolating fields 
% >>>
%	and the S-wave $N+K$, $N+K^{*}$ and  $N+K$ two-particle states respectively. }
	and the S-wave $N+K$, $N+K^{*}$ and  $N+K$ two-particle states,
	respectively. }
% <<<
\begin{ruledtabular} 
\begin{tabular}{c|ccc}
% >>>
% $\kappa$ & $aM_{1} - aM_{N+K}^{\rm S-wave}$ & $aM_{2} - aM_{N+K^{*}}^{\rm S-wave}$ & $aM_{SS} - aM_{N+K}^{\rm S-wave}$ \cr 
$\kappa$ & $aM_{NK(1)} - aM_{N+K}^{\rm S-wave}$ & $aM_{NK(2)} - aM_{N+K^{*}}^{\rm S-wave}$ & $aM_{SS} - aM_{N+K}^{\rm S-wave}$ \cr 
% <<<
\hline
1.2780 & 0.067(9) & 0.109(16) & 0.100(11) \cr
1.2830 & 0.063(13) & 0.106(18) & 0.090(15) \cr
1.2885 & 0.059(21) & 0.099(24) & 0.071(20) \cr
1.2940 & 0.056(35) & 0.082(39) & 0.048(26) \cr
1.2990 & -0.006(53) & 0.102(76) & -0.030(78) \cr
1.3025 & -0.161(223) & 0.146(176) & -0.021(75) \cr
\end{tabular}
\end{ruledtabular}
\end{table}

\newpage

\begin{figure}[tp]
\includegraphics[height=12.0cm,angle=90]{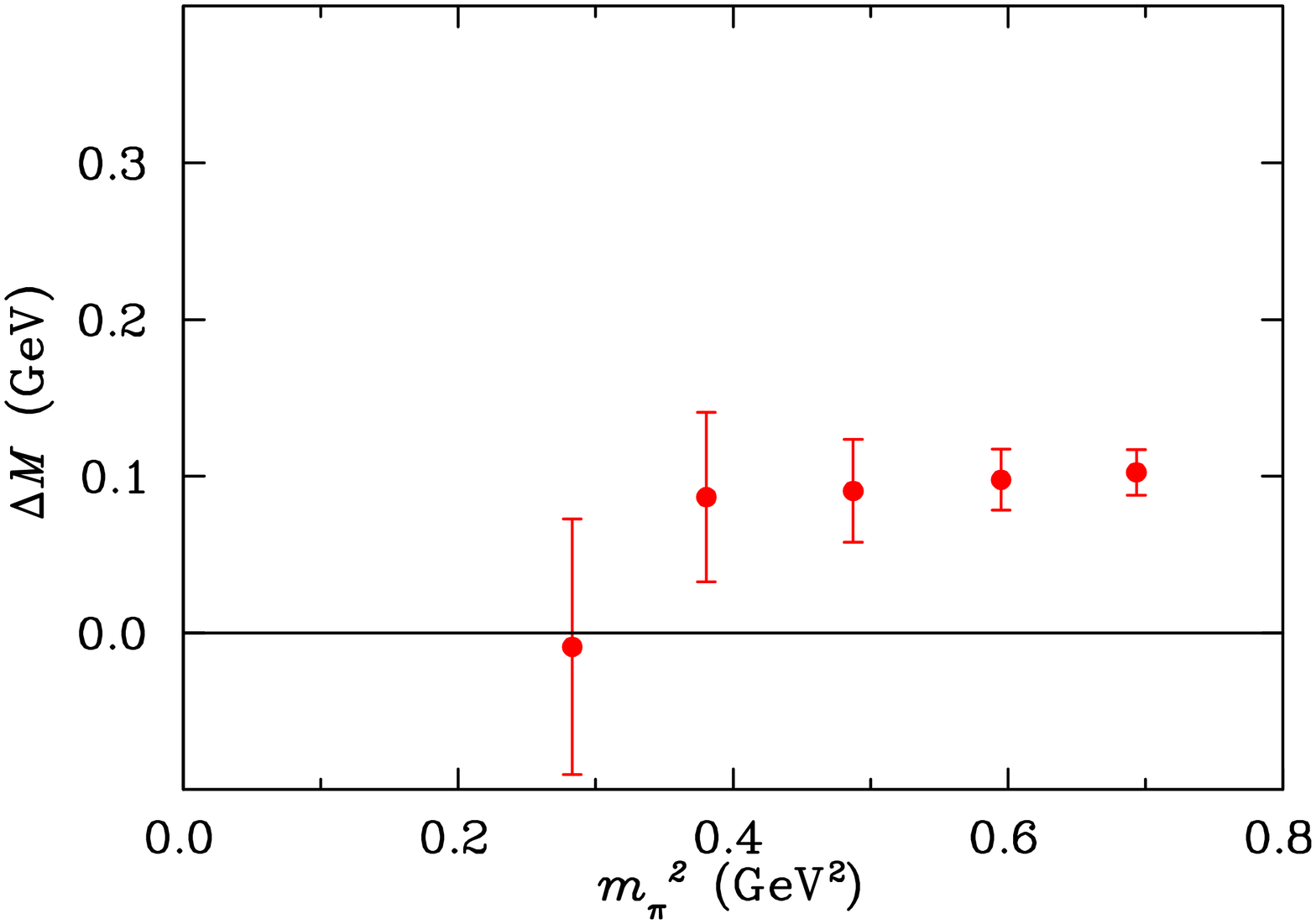} 
\caption{\label{fig:difI1_NK1.neg}
	Mass difference between the $I(J^P)=1(\frac{1}{2}^-)$ state corresponding to the 
	$NK$-type pentaquark ``state 1'' and the S-wave $N+K$
	two-particle state.}
\end{figure}

\begin{figure}[tp]
\includegraphics[height=12.0cm,angle=90]{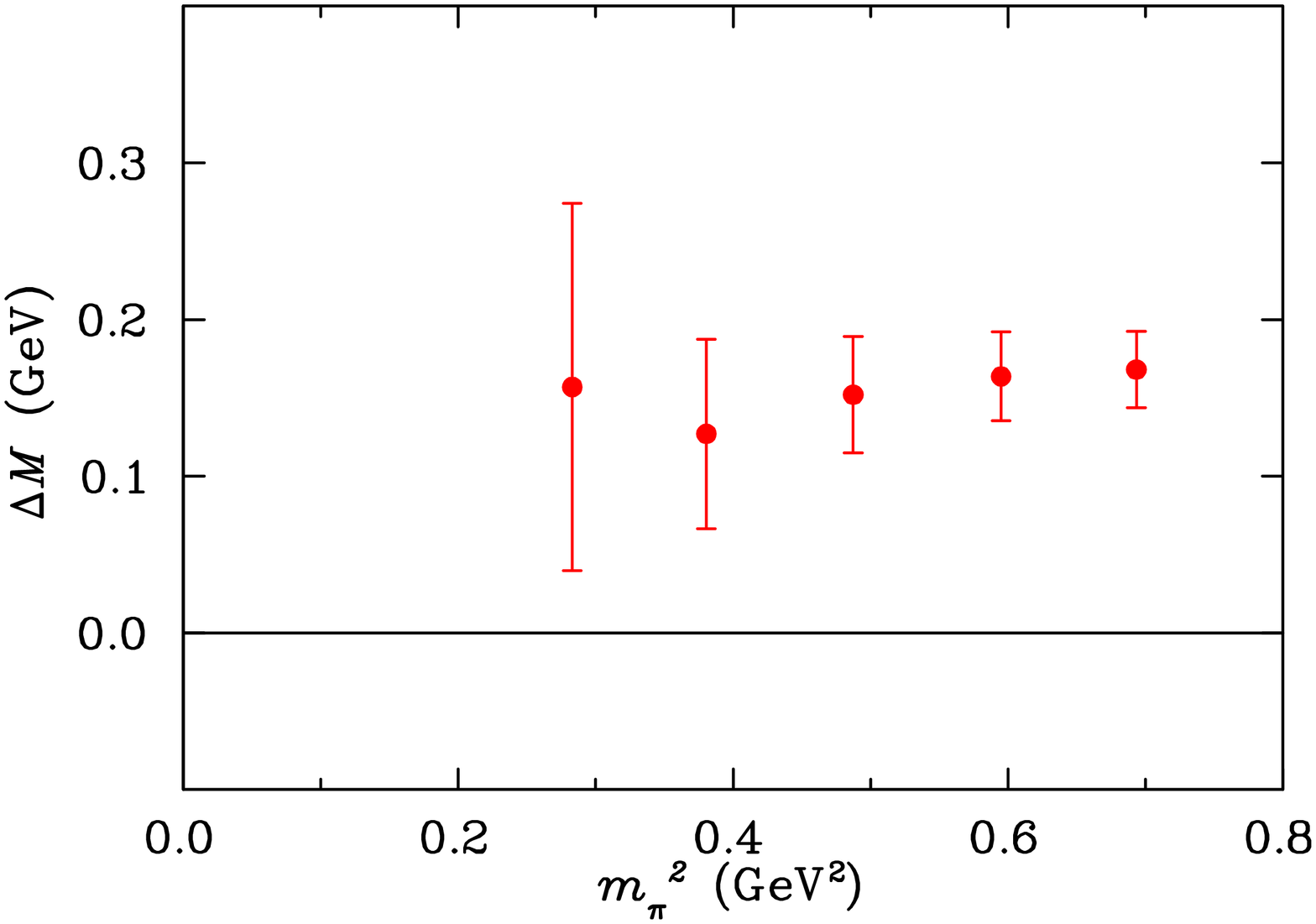} 
\caption{\label{fig:difI1_NK2.neg}
	Mass difference between the $I(J^P)=1(\frac{1}{2}^-)$ state corresponding to the 
	$NK$-type pentaquark ``state 2'' and the S-wave $N+K^{*}$
	two-particle state.}
\end{figure}

\begin{figure}[tp]
\includegraphics[height=12.0cm,angle=90]{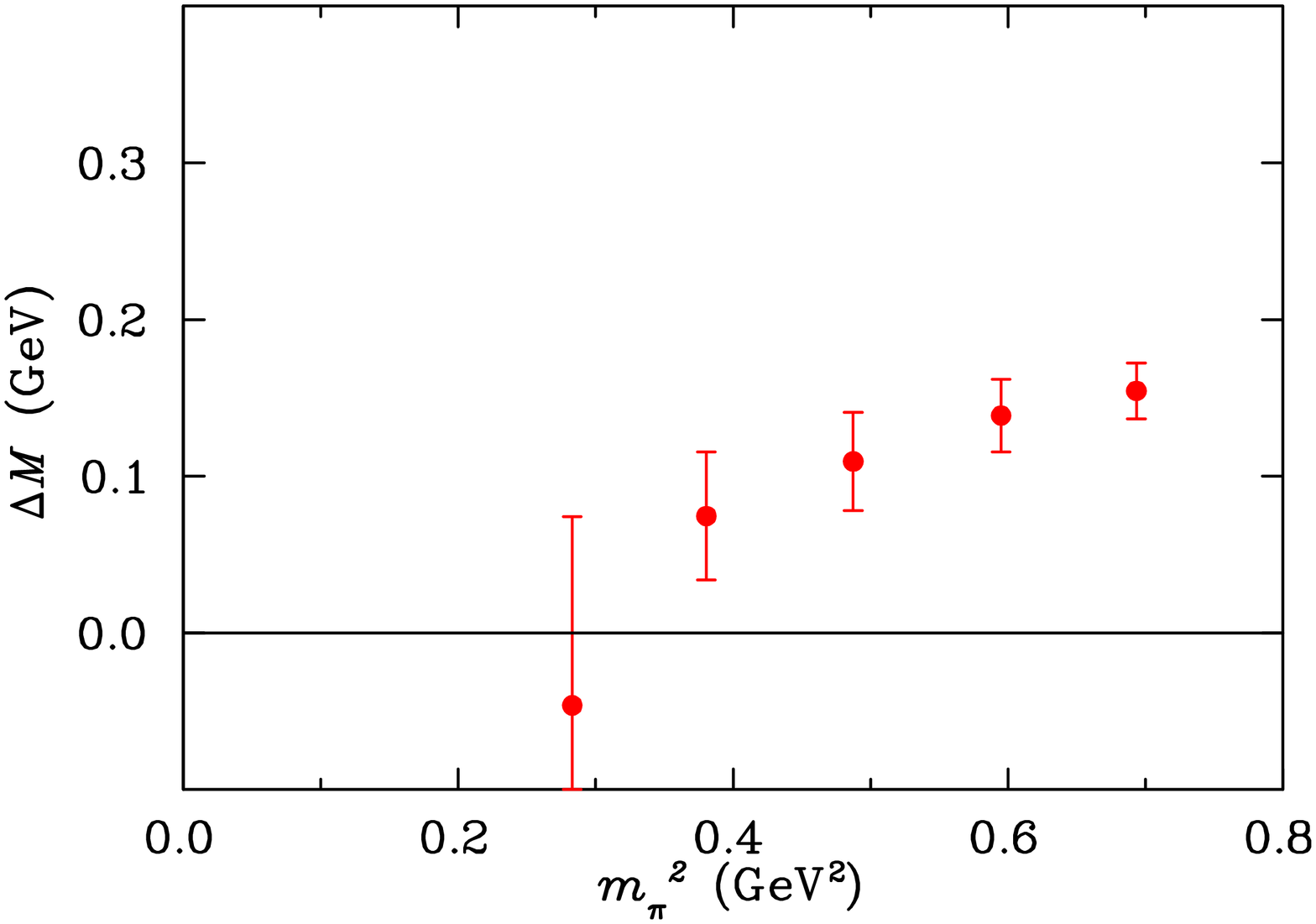} 
\caption{\label{fig:difI1_SS.neg}
	Mass difference between the $I(J^P)=1(\frac{1}{2}^-)$ state extracted with the 
	$SS$-type pentaquark interpolating field and the S-wave $N+K$
	two-particle state.}
\end{figure}

The fitted mass differences between the pentaquark and two-particle state effective masses are
summarised in Table~\ref{tab:difI1.neg}, where we quote the differences
between the $NK$-type ``state 1'' and the S-wave $N+K$, between the
$NK$-type ``state 2'' and the S-wave $N+K^*$, and between the
$SS$-type and the S-wave $N+K$.
These mass differences are illustrated in Figs.~\ref{fig:difI1_NK1.neg}, 
\ref{fig:difI1_NK2.neg} and \ref{fig:difI1_SS.neg}, for the three cases,
respectively.
As for the isoscalar channel, the mass differences for the ground state
% >>>
% are clearly positive, and approximately independent of $m_\pi^2$.
are clearly positive, and weakly dependent on $m_\pi^2$.
% <<<
For both the $NK$-type and $SS$-type ground states, the pentaquark
masses are $\sim 100$~MeV larger than the S-wave $N+K$ two-particle
state.
Similarly, the difference between the excited $NK$-type pentaquark and 
the S-wave $N+K^*$ is $\sim 150$~MeV and approximately constant with
$m_\pi^2$.
% >>>
% The results of Fig.~\ref{fig:difI1_NK1.neg} therefore indicate the presence of repulsion.
% %Moveover, the scalar-scalar diquark interpolator Eq.~(\ref{eq:SS:sing}) fails to isolate the 
% %lowest lying state. 
% In fact, the $\sim 100$~MeV mass splitting obtained at the largest
% quark mass considered in both the $NK$ and $SS$-type analysis is approached from below in Figs.~\ref{fig:difI1_NK1.neg} and \ref{fig:difI1_SS.neg}.
% No evidence of binding is observed. There is no indication of a resonance in
% the $I(J^P)=1({1\over 2}^-)$ channel which could be interpreted as the
% $\Theta^+$.
%
In fact, Figs.~\ref{fig:difI1_NK1.neg} and \ref{fig:difI1_SS.neg}
suggest that the $\sim 100$~MeV mass splitting obtained at the largest
quark mass considered is approached from below.
There is no evidence of binding and no indication
of a resonance in the $I(J^P)=1({1\over 2}^-)$ channel which could be
interpreted as the $\Theta^+$.
% <<<

%%%%%%%%%%%%%%%%%%%%%%%%%%%%%%%%%%%%%%%%%%%%%%%%%%%%%%%%%%%%%%%%%%%%%%%%
% >>>
% \subsection{Isoscalar positive parity states}
\subsection{Positive parity isoscalar states}
% <<<
\label{ssec:I0+}

% >>>
% While each of the pentaquark operators considered above transform
While each of the pentaquark operators considered above transforms
% <<<
negatively under parity, they nevertheless couple to both negative
and positive parity states, as discussed in Sec.~\ref{ssec:2pt}.
Here we consider whether any of the operators $\chi_{NK}$,
$\chi_{\widetilde{NK}}$ or $\chi_{PS}$ couple to a bound state in
the isospin-0, positive parity channel.
We compare the pentaquark states with the masses of the lowest energy
two-particle states, which correspond to the P-wave $N+K$ and $N+K^*$,
and the S-wave $N^*+K$ states.

A two-particle state in a relative P-wave can be constructed on the
lattice by adding one unit of lattice momentum ($p = 2\pi/L$)
to the effective mass,
$E_{\rm eff} = \sqrt{ M_{\rm eff}^2 + p^2 }$, of each particle.
This effectively raises the mass of the two-particle state relative
to the positive parity pentaquark.
If a pentaquark state exists, it should therefore clearly lie
below the lowest P-wave scattering state.

\begin{figure}[tp]
\includegraphics[height=12.0cm,angle=90]{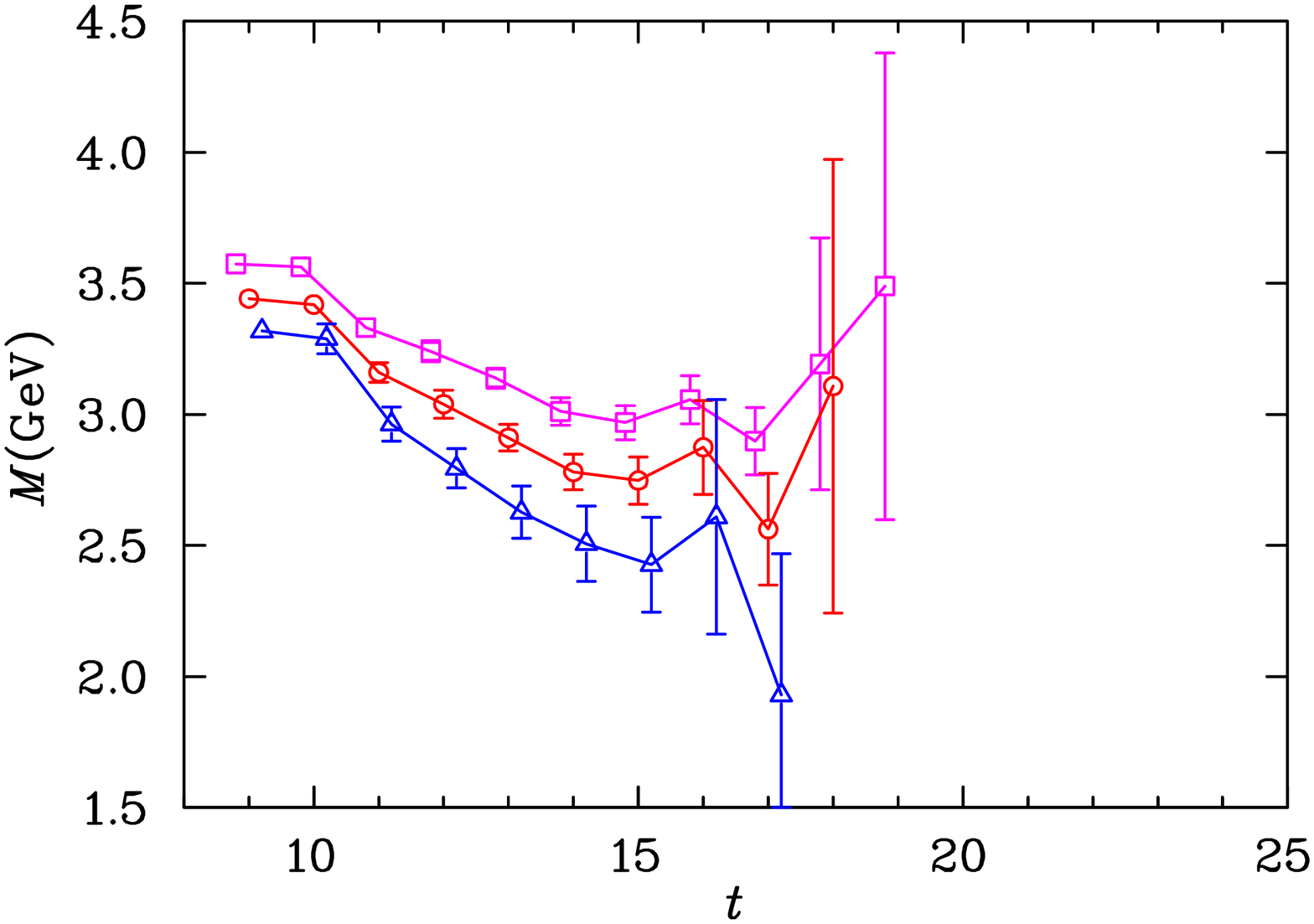}
\caption{\label{fig:I0_NK.pos}
	Effective mass of the $I(J^P)=0(\frac{1}{2}^+)$
% >>>
%	colour singlet $NK$-type pentaquark interpolator, $\chi_{NK}$.
%	pentaquark for $\kappa=1.2780$ (highest), 1.2885 (middle)
%	and 1.2990 (lowest).}
	colour singlet $NK$-type pentaquark interpolator, $\chi_{NK}$,
	for $\kappa=1.2780$ (squares), 1.2885 (circles)
	and 1.2990 (triangles).}
% <<<
\end{figure}

\begin{figure}[tp]
\includegraphics[height=12.0cm,angle=90]{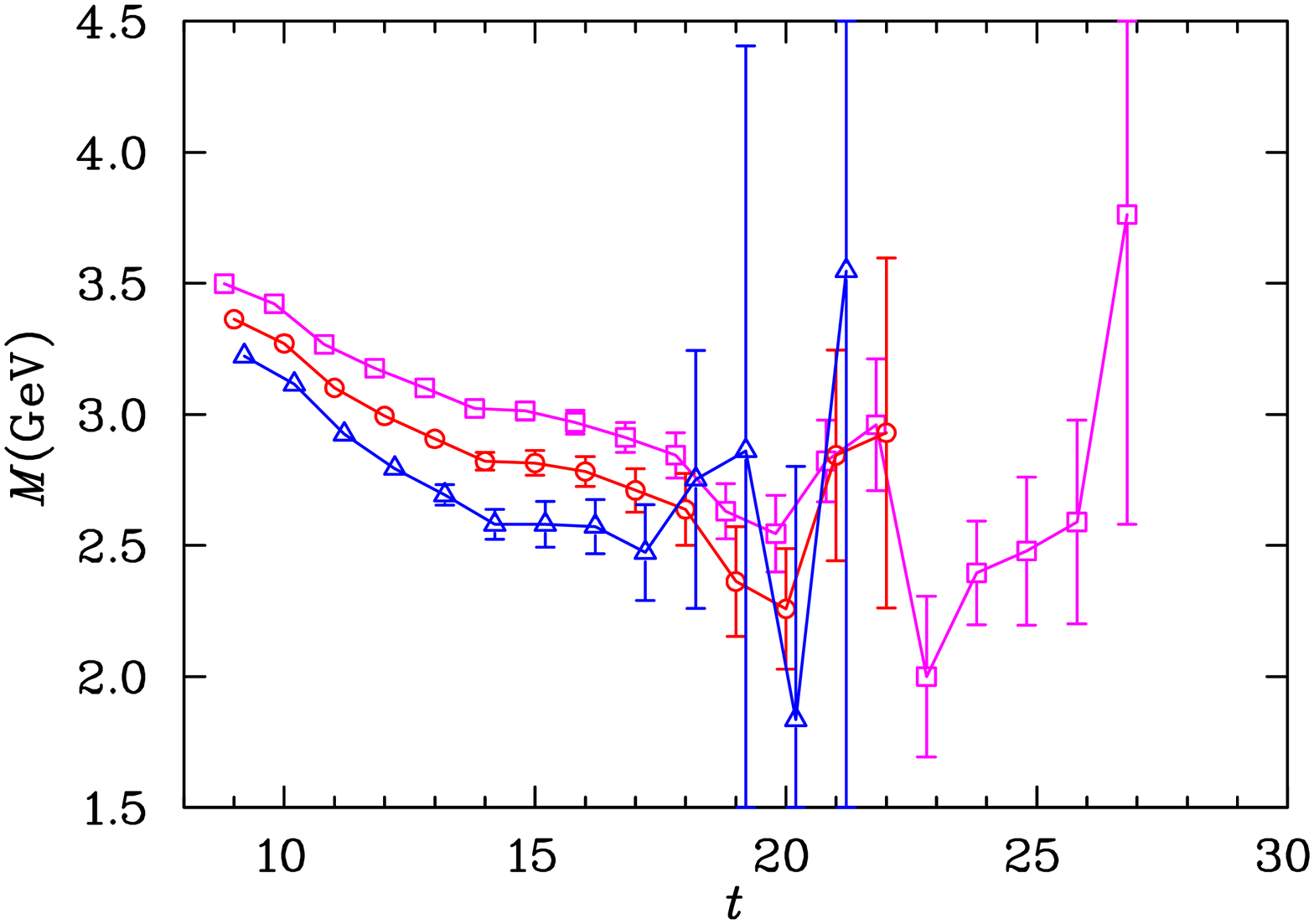}
\caption{\label{fig:I0_PS.pos}
	As in Fig.~\ref{fig:I0_NK.pos}, but for the
	$I(J^P)=0(\frac{1}{2}^+)$ $PS$-type
	pentaquark interpolator, $\chi_{PS}$.}
\end{figure}

As in the negative parity channel, we perform a correlation matrix
analysis using the two $NK$-type fields in order to isolate possible
excited states.
% >>>
% While the analysis suggested the presence of an excited state, the
% signal in the positive parity channel is considerably more noisy
% than in the negative parity.
% Consequently, in practice for this channel we revert back to the
While the analysis suggests the presence of an excited state, the
signal in the positive parity channel is considerably more noisy
than for negative parity.
Consequently, in practice for this channel we revert to the
% <<<
standard analysis method and extract only the ground state.
Since the colour singlet and colour fused operators return the same
ground state mass, we present the results for the colour singlet
operator since the signal here is less noisy.

\begin{table}[tp]     
\caption{\label{tab:I0.pos}
	Masses of the $I(J^P)=0(\frac{1}{2}^+)$ states extracted with the  
	colour singlet $NK$,
	and $PS$-type pentaquark interpolating fields 
	for various values of $\kappa$. }
\begin{ruledtabular} 
\begin{tabular}{c|cc}
$\kappa$ & $aM_{NK}$ & $aM_{PS}$ \cr
\hline
1.2780 & 1.935(40) & 1.721(57) \cr
1.2830 & 1.867(50) & 1.642(76) \cr
1.2885 & 1.782(66) & 1.547(119) \cr
1.2940 & 1.681(90) & 1.458(207) \cr
1.2990 & 1.561(126) &  \cr
1.3025 & 1.421(170) &  \cr
\end{tabular}
\end{ruledtabular}
\end{table}

\begin{table}[tp]     
\caption{\label{tab:2P.pos}
        The masses of the
% >>>
%	P-wave $N+K$, $N+K^*$ and the S-wave $N^{*}K$ two-particle states. }
	P-wave $N+K$, $N+K^*$ and the S-wave $N^* + K$ two-particle states. }
% <<<
\begin{ruledtabular} 
\begin{tabular}{c|ccc}
$\kappa$ & $aM^{\rm P-wave}_{N+K}$          & $aM^{\rm S-wave}_{N+K^{*}}$
         & $aM^{\rm S-wave}_{N^{*}+K}$ 
\cr
\hline
1.2780 & 1.692(7) & 1.891(27) & 1.873(9) \cr
1.2830 & 1.629(8) & 1.830(32) & 1.818(9) \cr
1.2885 & 1.558(8) & 1.760(39) & 1.755(10) \cr
1.2940 & 1.483(10) & 1.684(53) & 1.690(11) \cr
1.2990 & 1.414(13) & 1.594(85) & 1.631(14) \cr
1.3025 & 1.363(17) & 1.433(134) & 1.588(17) \cr
\end{tabular}
\end{ruledtabular}
\end{table}

\begin{figure}[tp]
\includegraphics[height=12.0cm,angle=90]{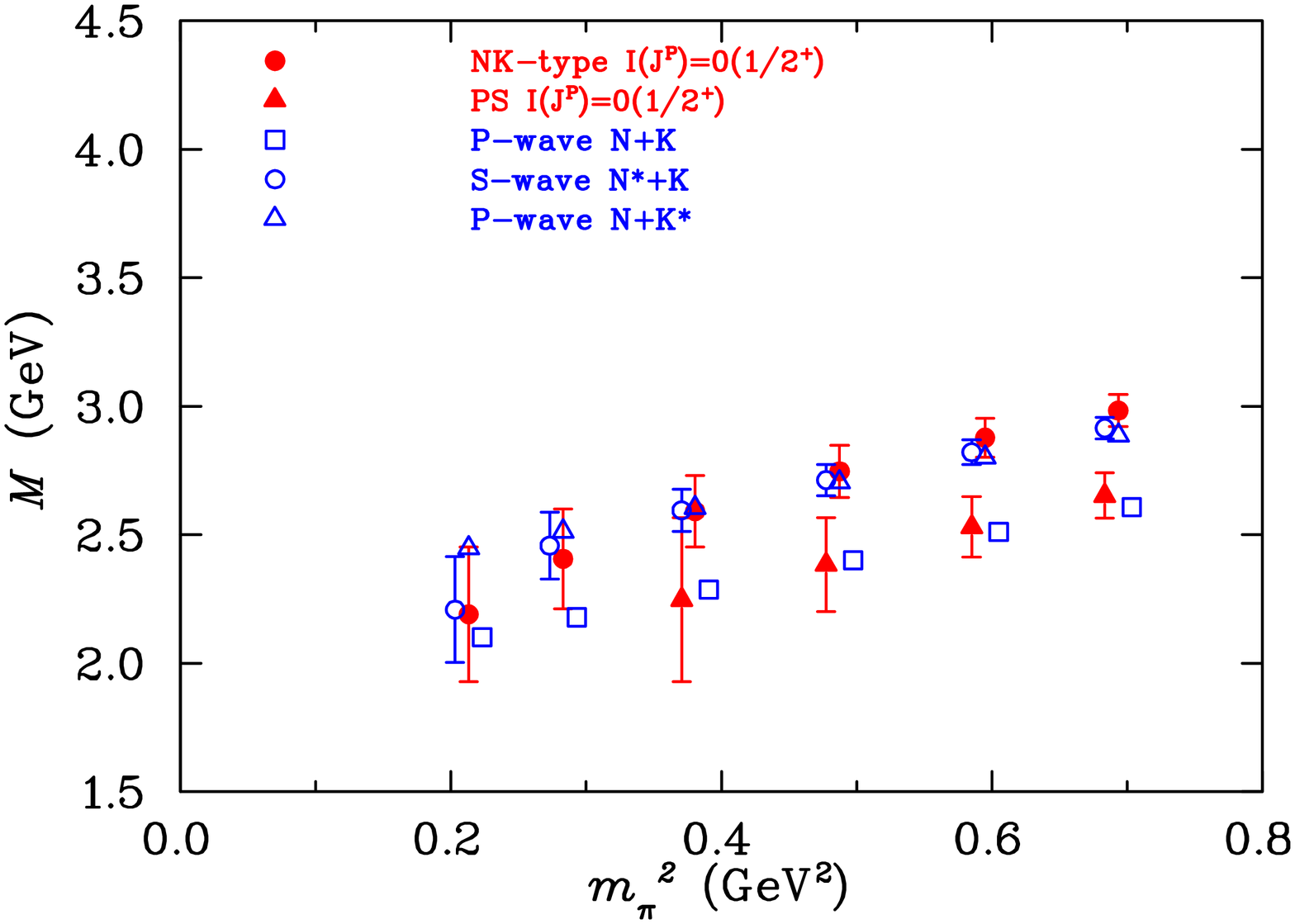}
\caption{\label{fig:I0.pos}
	Masses of the $I(J^P)=0(\frac{1}{2}^+)$ states extracted with the 
	colour singlet $NK$
	and $PS$-type pentaquark interpolating fields
	as a function of $m_\pi^2$.
	For comparison, the masses of the P-wave $N+K$ and $N+K^*$
	and S-wave $N^*+K$ two-particle states are also shown.
        Some of the points have been offset for clarity. }
\end{figure}

The effective masses for the $NK$ and $PS$-type
interpolators are shown in Figs.~\ref{fig:I0_NK.pos} and 
\ref{fig:I0_PS.pos}, respectively.
The signal clearly becomes noisier at earlier times, and we fit the
effective masses for the $NK$-type field at $t=15-17$, while those
for the $PS$-type interpolator  are fit at $t=19-21$.

\begin{table}[tp]     
\caption{\label{tab:difI0_NK.pos}
	Mass differences between the $I(J^P)=0(\frac{1}{2}^+)$ states
	extracted with the colour singlet $NK$ and $PS$-type  pentaquark
	interpolating fields 
	and the P-wave $N+K$ two-particle state.}
\begin{ruledtabular} 
\begin{tabular}{c|cc}
$\kappa$ & $aM_{NK} - aM^{\rm P-wave}_{N+K}$
	 & $aM_{PS} - aM^{\rm P-wave}_{N+K}$ \cr	
\hline
1.2780 & 0.228(38) & 0.035(57) \cr
1.2830 & 0.223(48) & 0.021(78) \cr
1.2885 & 0.209(65) & 0.003(122) \cr
1.2940 & 0.183(90) & -0.003(210) \cr
1.2990 & 0.132(128) &  \cr
1.3025 & 0.040(174) &  \cr
\end{tabular}
\end{ruledtabular}
\end{table}

\begin{figure}[tp]
\includegraphics[height=12.0cm,angle=90]{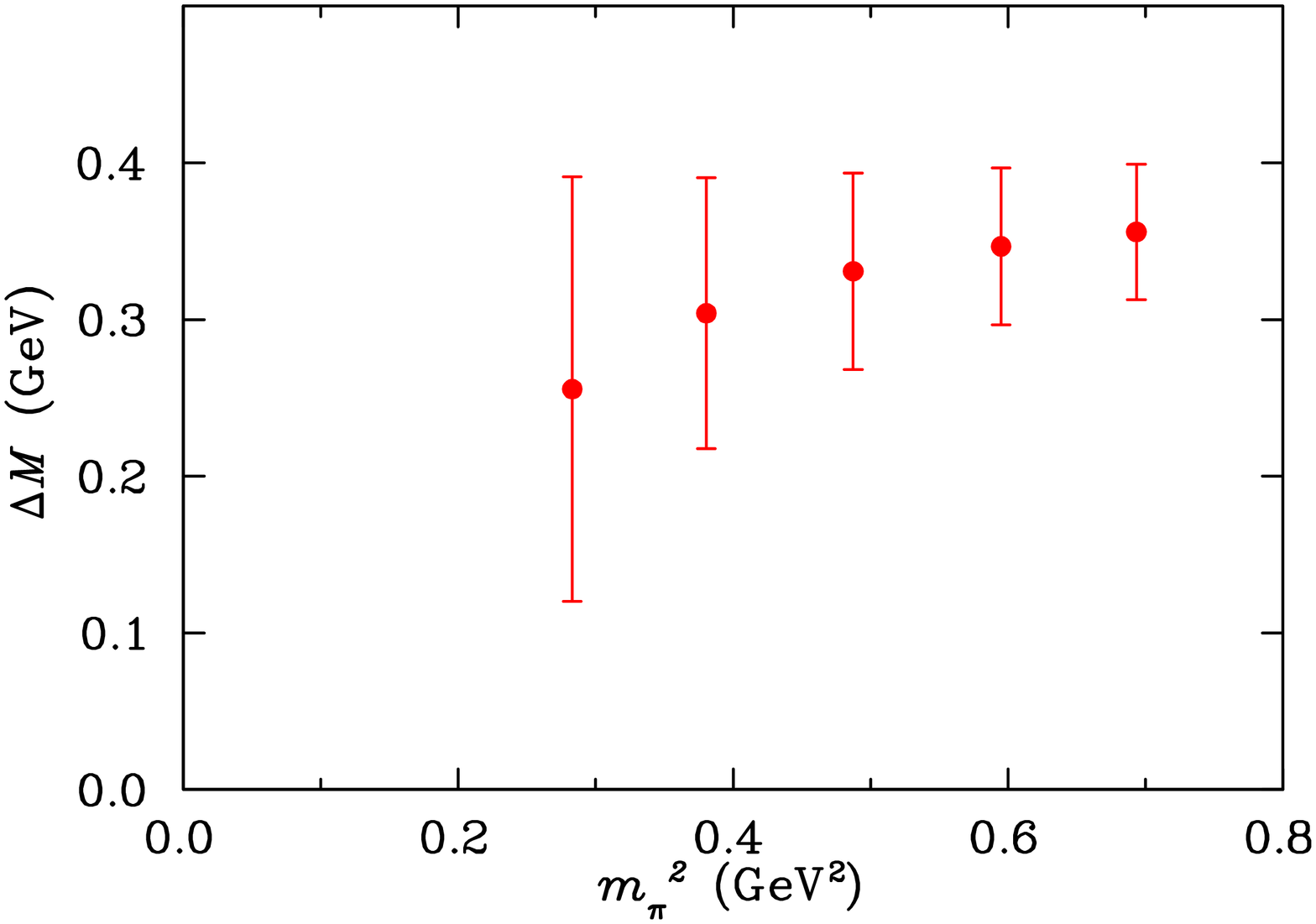} 
\caption{\label{fig:difI0_NK.pos}
	Mass difference between the $I(J^P)=0(\frac{1}{2}^+)$ state extracted with the 
	$NK$-type pentaquark interpolating field and the P-wave $N+K$ two-particle state.}
\end{figure}

\begin{figure}[tp]
\includegraphics[height=12.0cm,angle=90]{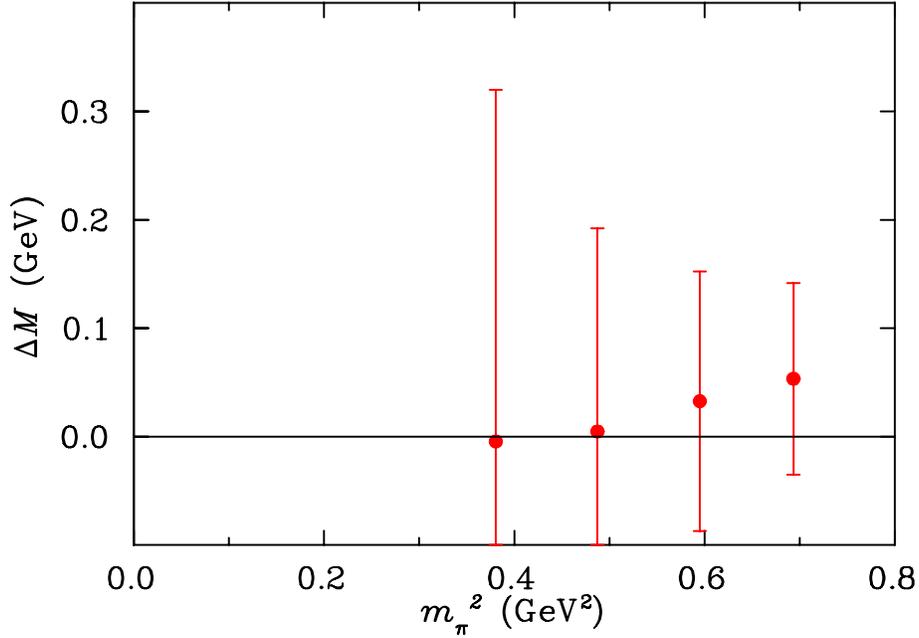} 
\caption{\label{fig:difI0_PS.pos}
	Mass difference between the $I(J^P)=0(\frac{1}{2}^+)$ state extracted with the 
	$PS$-type pentaquark interpolating field and the P-wave
	$N+K$ two-particle state.}
\end{figure}

The results are tabulated in Table~\ref{tab:I0.pos} and shown in
Fig.~\ref{fig:I0.pos}.
%The masses of the positive parity states extracted with the $NK$ and $PS$-type interpolating fields are
%very similar, agreeing within errors for most values of $\kappa$.
The masses of the positive parity states extracted with the $NK$ and $PS$-type interpolating fields are
very different.
The mass extracted with the $NK$-type interpolator is similar to both the S-wave $N^{*}+K$ mass and  P-wave $N+K^*$ energy,
% >>>
% whereas the mass extracted with the  $PS$-type interpolator is consistent with the P-wave $N+K$ energy.
whereas the mass extracted with the  $PS$-type interpolator is consistent
with the P-wave $N+K$ energy, which are given in
Table~\ref{tab:difI0_NK.pos}.
% <<<
%the P-wave $N+K^*$ state,
%whereas the  $PS$-type interpolator has much better access to the P-wave $NK$ state.
%
%Compared with the P-wave $N+K$ two-particle state, however, the
%masses of the pentaquarks are considerably higher.
%In fact, the positive parity ground state masses appear to be
%consistent with the S-wave $N^*+K$ two-particle state, or possibly
%the P-wave $N+K^*$.
The signal obtained with the $PS$-type interpolator is rather noisy where we fit the effective masses,
% >>>
% therefore we only present results for the four largest quark masses for this operator. 
and we therefore only present results for the four largest quark masses
for this operator. 
% <<<
% >>>
% The reason the signal is so poor was mentioned in Sec.~\ref{ssec:I0-}, since we do not expect that our
% operators couple strongly to the P-wave states due to the addtional
As mentioned in Sec.~\ref{ssec:I0-}, the reason the signal is so poor is 
that our operators do not couple strongly to the P-wave states due to the
additional
% <<<
{\it small } component of the interpolating field spinors contributing to this state.

For the differences between the pentaquark and two-particle state masses,
we also fit the effective masses at $t=15-17$ for the $NK$-type field,
and $t=19-21$ for the $PS$-type field.
The results are shown in Table~\ref{tab:difI0_NK.pos}, and in
Figs.~\ref{fig:difI0_NK.pos} and \ref{fig:difI0_PS.pos} for the
differences between the masses extracted with the ($NK$ and $PS$-type) pentaquark interpolating fields and the
P-wave $N+K$ two-particle state.
The mass obtained with the $NK$-type field is $\sim 300$~MeV heavier than
the lowest energy two-particle state (P-wave $N+K$) for all quark masses
considered.
The mass obtained with the $PS$-type field is consistent with the the lowest energy two-particle state (P-wave $N+K$) 
for all quark masses considered.
Once again this suggests that there is no binding in the
$I(J^P)=0(\frac{1}{2}^+)$ channel, and hence no indication of
a $\Theta^+$ resonance.

%%%%%%%%%%%%%%%%%%%%%%%%%%%%%%%%%%%%%%%%%%%%%%%%%%%%%%%%%%%%%%%%%%%%%%%%
% >>>
% \subsection{Isovector positive parity states}
\subsection{Positive parity isovector states}
% <<<

% >>>
% For the isospin-1, positive parity channel analysis, we found that the
% correlation matrix did not produce improved results for the ground
For the isospin-1, positive parity channel analysis, we find that the
correlation matrix does not produce improved results for the ground
% <<<
state masses compared with the standard analysis.
In the case of the largest three $\kappa$ values the algorithm requires
that we step back three or more time slices before the correlation
matrix analysis works.
The use of a correlation matrix analysis on these
data is inappropriate due to large errors in the data.

\begin{figure}[tp]
\includegraphics[height=12.0cm,angle=90]{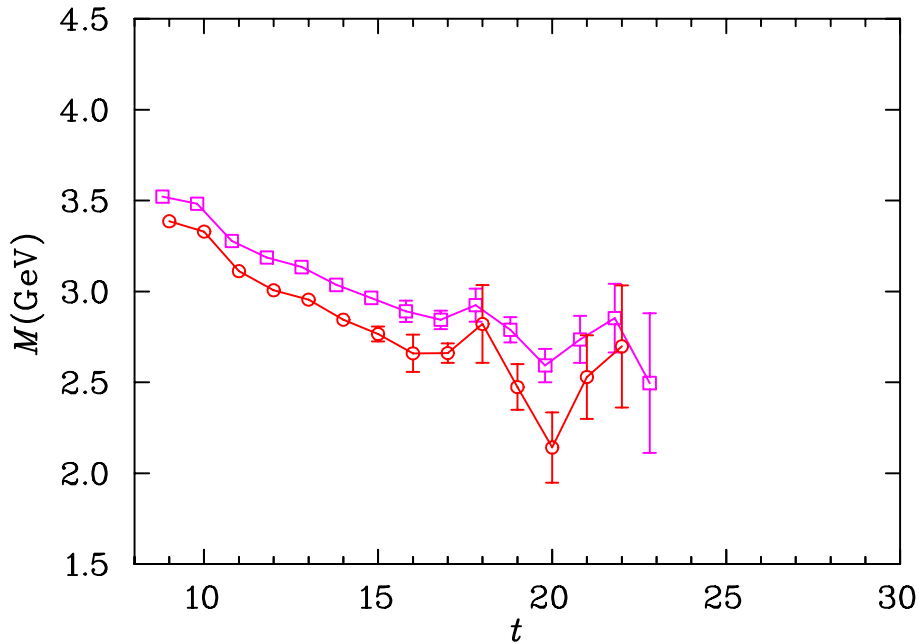}
\caption{\label{fig:I1_NK.pos}
	Effective mass of the $I(J^P)=1(\frac{1}{2}^+)$
	colour singlet $NK$-type pentaquark interpolator, $\chi_{NK}$.
% >>>
%	The data correspond to $\kappa=1.2780$ (squares),
%	1.2885 (circles), and 1.2990 (triangles).}
	The data correspond to $\kappa=1.2780$ (squares)
	and 1.2885 (circles).}
% <<<
\end{figure}

\begin{figure}[tp]
\includegraphics[height=12.0cm,angle=90]{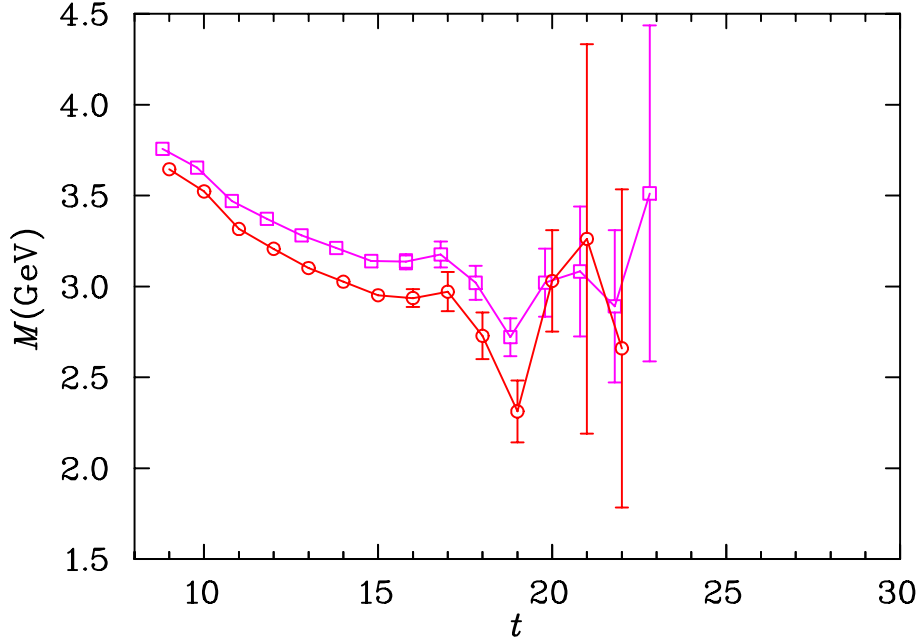}
\caption{\label{fig:I1_SS.pos}
	As in Fig.~\ref{fig:I1_NK.pos}, but for the
	$I(J^P)=1(\frac{1}{2}^+)$ $SS$-type
	pentaquark interpolator, $\chi_{SS}$.}
\end{figure}

The effective masses for the $NK$-type and $SS$-type
interpolating fields are illustrated in Figs.~\ref{fig:I1_NK.pos} and 
\ref{fig:I1_SS.pos}, respectively.
Because the signal for the positive parity is rather more noisy than in 
the corresponding negative parity channel, we only show the effective
mass for the smallest and third-smallest values of $\kappa$.
For the $NK$-type pentaquarks, the colour-singlet $\chi_{NK}$ and
colour-fused $\chi_{\widetilde{NK}}$ fields are found to access the
% >>>
% same ground state, and in Fig.~\ref{fig:I1_NK.pos} we show the
same ground state, and in Fig.~\ref{fig:I1_NK.pos} we only show the
% <<<
results of the former.

\begin{table}[tp]     
\caption{\label{tab:I1.pos}
	Masses of the $I(J^P)=1(\frac{1}{2}^+)$ states extracted with the  
	colour singlet $NK$ and $SS$-type 
	pentaquark interpolating fields for various values of $\kappa$. }
\begin{ruledtabular} 
\begin{tabular}{c|cc}
$\kappa$ & $aM_{NK}$ & $aM_{SS}$ \cr
\hline
1.2780 & 1.732(48) & 1.956(133) \cr
1.2830 & 1.651(57) & 1.939(158) \cr
1.2885 & 1.536(71) & 1.954(214) \cr
\end{tabular}
\end{ruledtabular}
\end{table}

\begin{figure}[tp]
\includegraphics[height=12.0cm,angle=90]{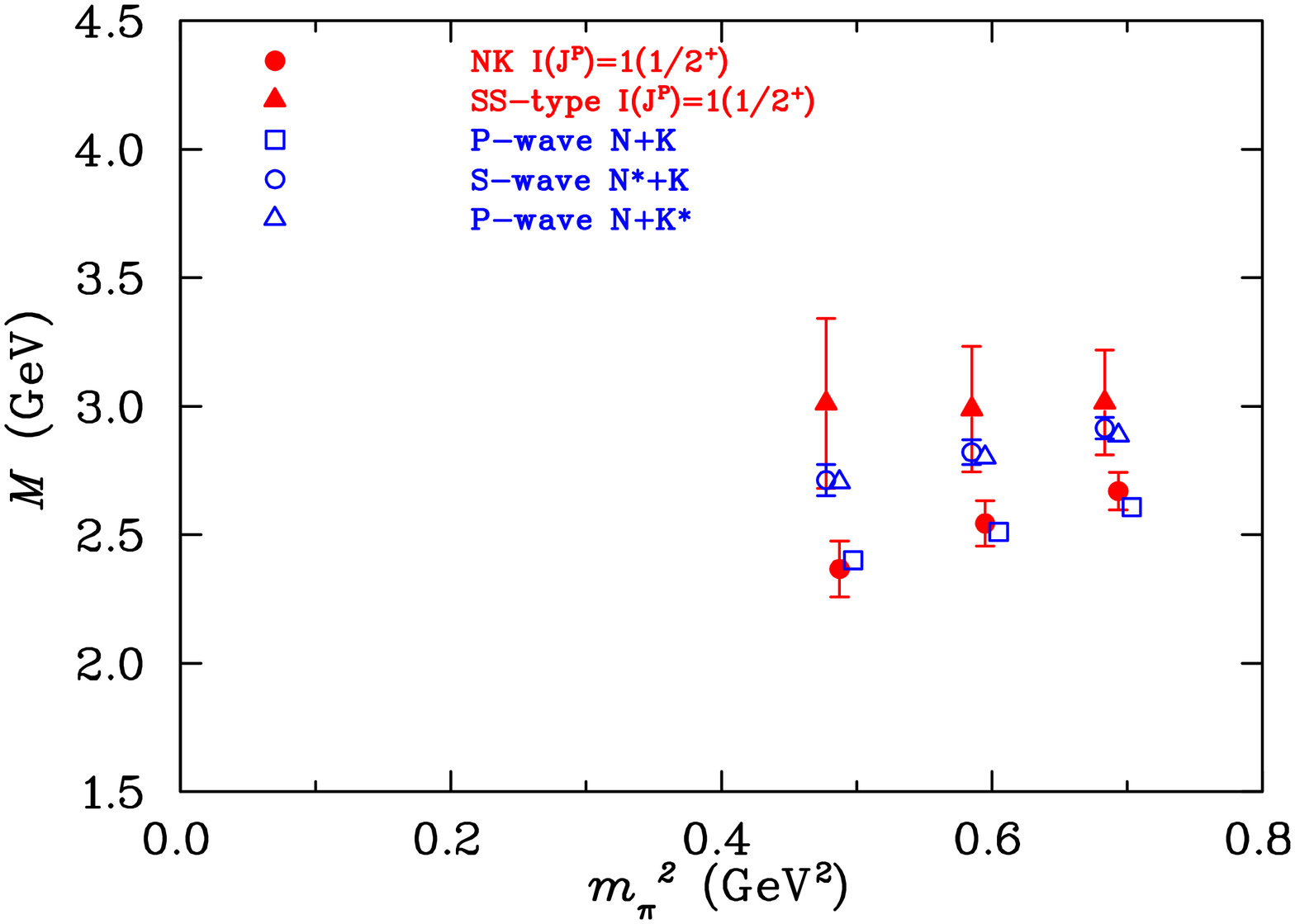} 
\caption{\label{fig:I1.pos}
	Masses of the $I(J^P)=1(\frac{1}{2}^+)$ states extracted with the 
	colour singlet $NK$
	and $SS$-type pentaquark interpolating fields
	as a function of $m_\pi^2$.
	For comparison, the masses of the P-wave $N+K$, S-wave $N^*+K$
	and P-wave $N+K^*$ two-particle states are also shown.
	Some of the points have been offset for clarity.}
\end{figure}

% >>>
% The effective masses for the $NK$ and $SS$-type interpolators were fitted at time slices
The effective masses for the $NK$ and $SS$-type interpolators are fitted
at time slices
% <<<
$t=20-22$ for the three largest quark masses.
The results for the extracted masses and the corresponding two-particle
states are shown in Table~\ref{tab:I1.pos} and in Fig.~\ref{fig:I1.pos}.
%The ground state masses for the $\chi_{NK}$ and $\chi_{SS}$ fields
%agree to within errors, and are consistent with the mass of the lowest-lying
%two-particle state, the P-wave $N+K$.
% >>>
% The ground state masses for the $NK$ and $SS$-type fields are very different.
The ground state masses for the $NK$ and $SS$-type fields are again very
different.
% <<<
The mass extracted with the $NK$-type interpolator is consistent with the P-wave $N+K$ energy,
whereas the mass extracted with the $SS$-type interpolator is consistent with both the S-wave $N^{*}+K$ mass and  P-wave $N+K^*$ energy.

% Stronger coupling to the s-wave states may explain why the NK-type does 
% not access the lower energy state.

%\input TF_I1.pos/MassS/tab.tex
\begin{table}[tp]     
\caption{\label{tab:difI1.pos}
	Mass differences between the $I(J^P)=1(\frac{1}{2}^+)$ states
	extracted with the colour singlet $NK$ and $SS$-type pentaquark
	interpolating fields and P-wave $N+K$ two-particle state. }
\begin{ruledtabular} 
\begin{tabular}{c|cc}
$\kappa$ & $aM_{NK} - aM^{\rm P-wave}_{N+K}$
	 & $aM_{SS} - aM^{\rm P-wave}_{N+K}$ \cr
\hline
1.2780 & 0.093(43) & 0.135(58) \cr
1.2830 & 0.079(53) & 0.117(61) \cr
1.2885 & 0.034(70) & 0.084(69) \cr
\end{tabular}
\end{ruledtabular}
\end{table}

\begin{figure}[tp]
\includegraphics[height=12.0cm,angle=90]{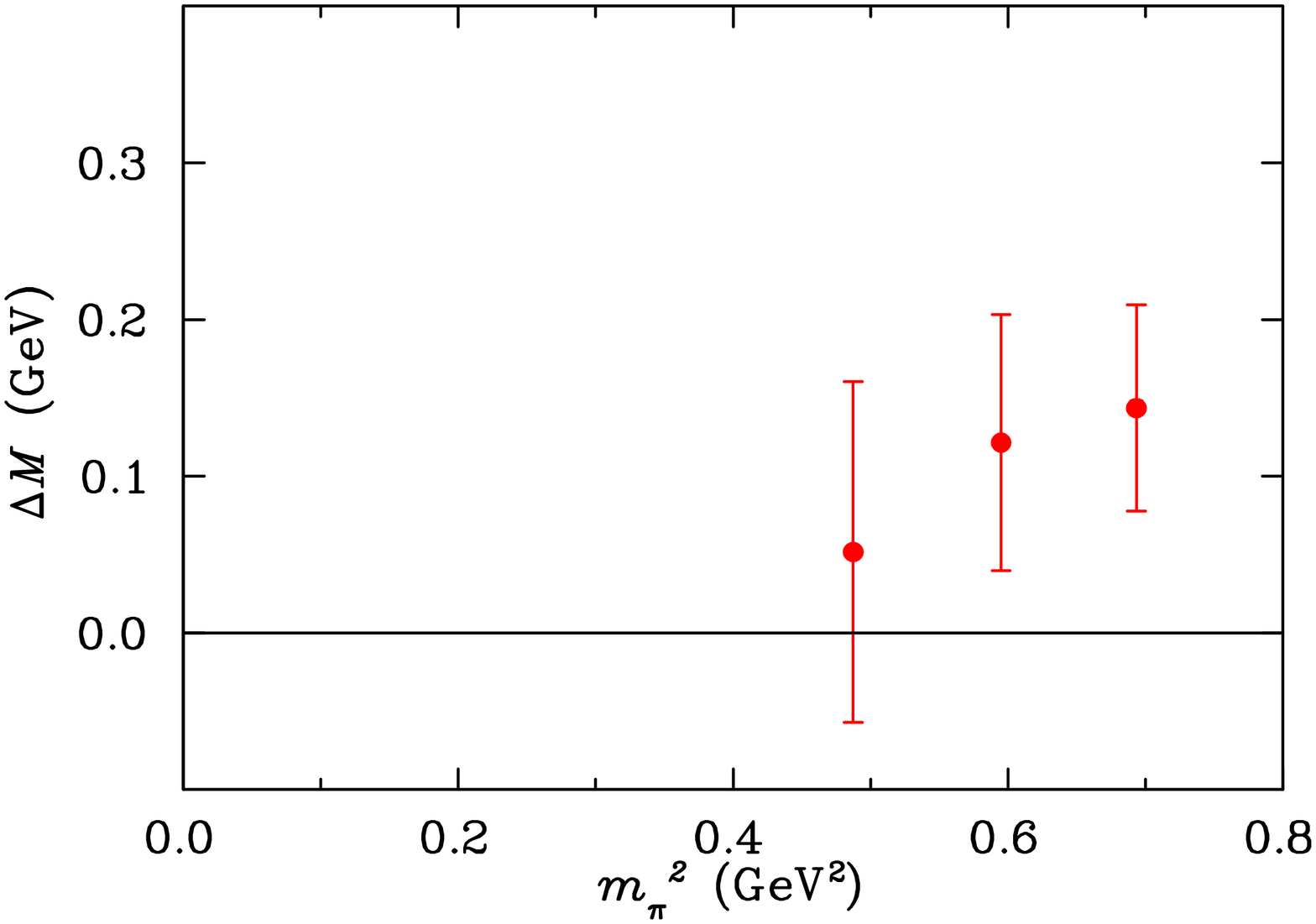} 
\caption{\label{fig:difI1_NK.pos}
	Mass difference between the $I(J^P)=1(\frac{1}{2}^+)$ state extracted with the 
	$NK$-type pentaquark interpolating field and the P-wave $N+K$ two-particle state.}
\end{figure}

\begin{figure}[tp]
\includegraphics[height=12.0cm,angle=90]{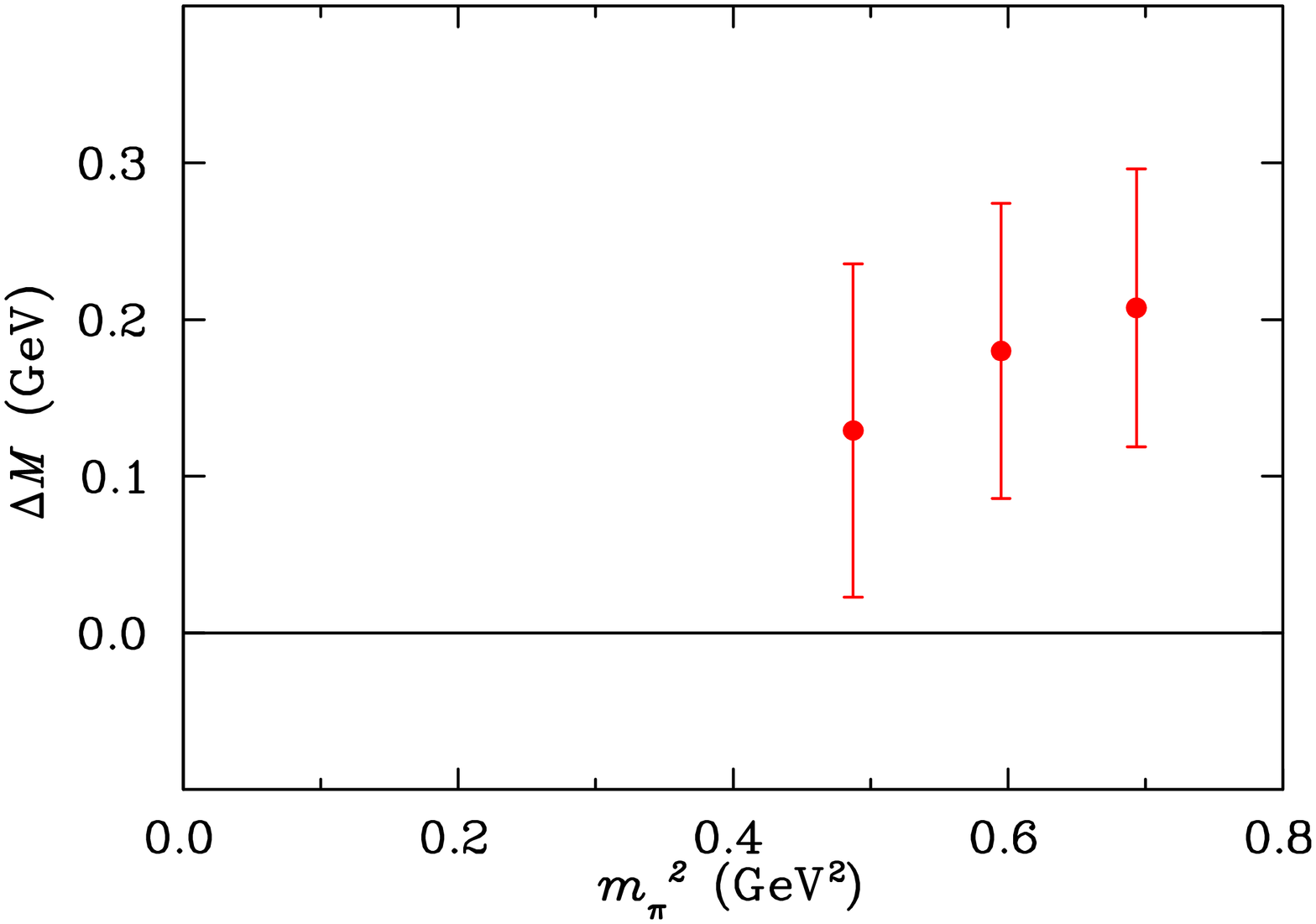} 
\caption{\label{fig:difI1_SS.pos}
	Mass difference between the $I(J^P)=1(\frac{1}{2}^+)$ state extracted with the 
	$SS$-type pentaquark interpolating field and the P-wave
	$N+K$ two-particle state.}
\end{figure}

The results of the mass splitting analysis are shown in
Table~\ref{tab:difI1.pos}, and illustrated in
Figs.~\ref{fig:difI1_NK.pos} and \ref{fig:difI1_SS.pos}.
%The mass difference between $NK$-type pentaquark and the P-wave
%$N+K$ two-particle state is positive for values of $m_\pi^2$
%where a signal is obtained. The splitting increases with $m_\pi^2$ giving an indication of
%repulsion, not attraction.
%The difference between the mass extracted from the $SS$-type pentaquark interpolating field and the P-wave
%$N+K$ is consistent with zero within relatively large errors.
The mass difference between the $NK$ and $SS$-type pentaquarks and the P-wave $N+K$ two-particle state is positive
for the largest quark masses. 
% >>>
% The splitting increases with $m_\pi^2$ giving an indication of repulsion, not attraction.
The splitting increases with $m_\pi^2$, giving an indication of repulsion,
rather than attraction.
% <<<
In all cases, the masses exhibit the opposite behaviour to that
which would be expected in the presence of binding.
We therefore do not see any indication of a resonance that could
% >>>
% be interpreted as the $\Theta^+$.
be interpreted as the $\Theta^+$ in this channel.
% <<<

% We see that the SS-type operator couples more strongly to the lower 
% energy state than the NK-type and that we have access to the lowest 
% energy two particle state.
% Therefore in the positive parity channel the SS-type configuration of 
% quarks is the most tightly bound.

\subsection{ Comparison with previous results }
\label{ssec:prev}
%
%  Summary of the fields used and what people thought
%
To place our results in context, we summarise here the results of
earlier lattice calculations of pentaquark masses, and compare those
with the findings of this analysis.
Table~\ref{litrev:tab} presents a concise summary of published 
lattice simulations, together with the results of this analysis,
including the actions and interpolating fields used, analysis methods
employed, and some remarks on the results. In every case, the general 
% >>>
% features of the simulation results support our findings.
features of the simulation results are consistent with our findings.
% <<<

\begin{figure}[tp]
\includegraphics[height=15.0cm,angle=90]{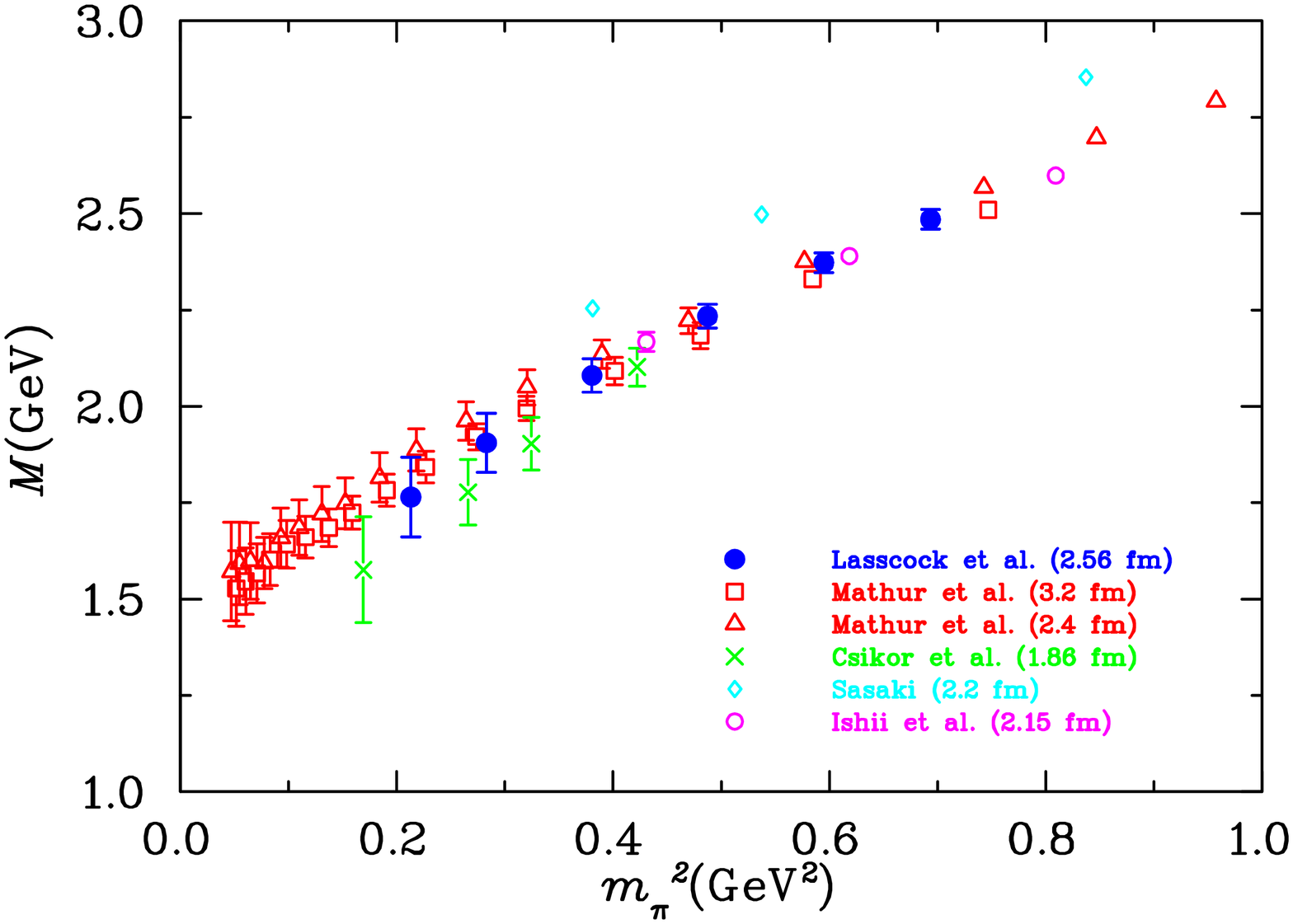} 
\caption{\label{fig:sum.neg}
	Compilation of results for the lowest-lying
	$I(J^P)=0(\frac{1}{2}^-)$ state from lattice QCD pentaquark
	studies.}
\end{figure}

The isoscalar negative parity channel was originally presented by
Csikor {\it et al.} \cite{Csikor:2003ng} and Sasaki \cite{Sasaki:2003gi} as a candidate
for the $\Theta^+$.
We therefore summarise in Fig.~\ref{fig:sum.neg} the results in
this channel from the previous lattice simulations.
At the larger quark masses the results of our analysis are in
% >>>
% excellent agreement with those of the Kentucky group \cite{Mathur:2004jr},
excellent agreement with those of Mathur {\it et al.} \cite{Mathur:2004jr},
% <<<
Csikor {\it et al.} \cite{Csikor:2003ng} and Ishii {\it et al.} \cite{Ishii:2004qe,Ishii:2004ib}.
% >>>
% In our analysis, and also in that of the Kentucky group \cite{Mathur:2004jr},
In our analysis, and also in that of Mathur {\it et al.} \cite{Mathur:2004jr},
% <<<
improved fermion actions were used, and the results are in agreement
at the smaller quark masses.
The results from Csikor {\it et al.} \cite{Csikor:2003ng} lie slightly lower
than the others at small quark masses, which may be due to scaling
violations of the Wilson fermion action.

The central issue in all of these analyses is the interpretation
of the data.
The earlier work of Csikor {\it et al.} \cite{Csikor:2003ng} and Sasaki
\cite{Sasaki:2003gi} identified the $0({1\over 2}^-)$ channel as a possible
candidate for the $\Theta^+$ based on naive linear extrapolations
and comparison of quenched QCD with experiment.
% >>>
% Later work by the Kentucky group \cite{Mathur:2004jr} analysed the volume
Later work by Mathur {\it et al.} \cite{Mathur:2004jr} analysed the volume
% <<<
dependence of the couplings of the operators to this state and
determined that the lowest energy state in this channel was an
$N+K$ scattering state.
Using hybrid boundary conditions Ishii {\it et al.} \cite{Ishii:2004qe,Ishii:2004ib}
% >>>
% also found that this was a $N+K$ scattering state.
% Our work is consitent with the findings of both of these studies.
also found that this was an $N+K$ scattering state.
Our work is consistent with the findings of both of these studies.
% <<<
%
%
% All calculations of the ground state mass in the negative parity $I=0$
% channel are, so the main issue is in the interpretation of the data. 
% \newpage

\begin{sidewaystable}[tp]     
\caption{\label{litrev:tab}
	Summary of published lattice QCD pentaquark studies,
	including the fields used, a brief description of the
	analysis techniques, and some observations from the work.}
\begin{ruledtabular}
\begin{tabular}{l|l|l|l|l}
Group & Action & Operators & Analysis methods & Observations	\cr
\hline
{\small Lasscock {\it et al.} } & {\small FLIC }
  & $\chi_{NK}$, $\chi_{\widetilde{NK}}$,
  & {\small $2 \times 2$ correlation matrix;                }    
  & {\small $I(J^P)=1(\frac{1}{2}^{\pm})$ $NK$ scattering states;}	\cr
 &
  & \ \ \ $\chi_{SS}$, $\chi_{PS}$
  & {\small mass splittings analysis with     }
  & {\small $I(J^P)=0(\frac{1}{2}^{-})$ $NK$ scattering state; } 
										\cr
&
  &
  & {\small \ \ \ $NK$, $NK^*$ }
  & {\small $0(\frac{1}{2}^{+})$ $N^{*}K$ scattering state}			\cr
\hline
{\small Csikor {\it et al.} \cite{Csikor:2003ng} } & {\small Wilson}
  & $\chi_{NK}$, $\chi_{\widetilde{NK}}$
  & {\small $2 \times 2$ correlation matrix;}
  & {\small $0(\frac{1}{2}^-)$ $NK$ degenerate state;}			\cr
&
  &
  & mass ratio with $NK$
  & $0(\frac{1}{2}^-)$ excited state, $0(\frac{1}{2}^+)$ deemed too massive		\cr
\hline
{\small Sasaki \cite{Sasaki:2003gi} } &  {\small Wilson}
& $\chi_{PS}$
  & {\small standard analysis}
  & {\small $0(\frac{1}{2}^-)$ above S-wave $NK$;}				\cr 
&
  &
  &
  & {\small $0(\frac{1}{2}^+)$ above P-wave $NK$}				\cr
\hline
{\small Mathur {\it et al.} \cite{Mathur:2004jr} } & {\small overlap}
  & $\chi_{NK}$, $\chi_{\widetilde{NK}}$
  & {\small volume dependence}
  & {\small $0,1(\frac{1}{2}^-)$ $NK$ scattering state;}			\cr
&
  &
  &
  & {\small $0,1(\frac{1}{2}^+)$ P-wave $NK$ degenerate}			\cr 
\hline
{\small Ishii {\it et al.} \cite{Ishii:2004qe,Ishii:2004ib}} & {\small Wilson}
& $\chi_{PS}$
  & hybrid boundary conditions;
  & {\small $0(\frac{1}{2}^-)$ $NK$ scattering state;}				\cr
&
  &
  & Bayesian analysis
  & $0(\frac{1}{2}^+)$ deemed too massive						\cr
\hline
Alexandrou {\it et al.} \cite{Alexandrou:2004ws} & {\small Wilson} & $\chi_{PS}$ & 
volume dependence & $0({1\over 2}^-)$ more consistent with single particle state; \cr
& & & &  $NK$ scattering state not seen \cr                
\hline
{\small Chiu {\it et al.} \cite{Chiu:2004gg}} & {\small domain wall  }
& $\chi_{NK}$, $\chi_{\widetilde{NK}}$,
\footnote{The $NK$-type fields used by Chiu {\it et al.} \cite{Chiu:2004gg} differ by a
	$\gamma_5$ in the nucleon part of the operator from the other groups listed,
	which effectively reverses the intrinsic parity of the operator.}
  & $3 \times 3$ correlation matrix
  & {\small $0(\frac{1}{2}^-)$ $NK$ scattering state;}				\cr
&
  & \ \ \ $\chi_{PS}$
  &
  & {\small ground state $0(\frac{1}{2}^-)$ less massive than $0(\frac{1}{2}^+)$} \cr
\hline
{\small Takahashi {\it et al.} \cite{Takahashi:2004sc}} & {\small Wilson}
& $\chi_{NK}$, $\chi_{\widetilde{NK}}$
  & $2 \times 2$ correlation matrix;
  & {\small $\frac{1}{2}^-$ $NK$ scattering state;}	\cr
&
  &
  & volume dependence
  & {\small $\frac{1}{2}^-$ excited state; }\cr
&
  &
  &
  & {\small $\frac{1}{2}^+$ $N^*K$ scattering state}
\cr
\hline

\end{tabular}
\end{ruledtabular}
\end{sidewaystable}
\newpage

\section{Conclusion}
% >>>
% We have completed a comprehensive analysis of interpolating fields
We have performed a comprehensive analysis of interpolating fields
% <<<
holding the promise to provide good overlap between the the QCD vacuum
and low-lying pentaquark states.  Central to our analysis is the
search for evidence of attraction between the constituents of
pentaquark states as the input quark masses increase.  Every other
baryon resonance ever studied on the lattice becomes stable on the
lattice at sufficiently large quark masses.  This is the standard
resonance signature in lattice QCD.  The mass of the resonance
becomes less than the sum of the masses of its decay products and is
prevented from decaying by energy conservation.  Attraction is
essential to the formation of a resonance in the light quark mass
regime of QCD.

Our results reveal no evidence of attraction that leads to a bound
pentaquark state at large quark masses.  Rather, evidence of repulsion
is evident in the correlation functions giving rise to the
lowest-lying pentaquark masses.  This is particularly evident in the
$I(J^P) = 1(\frac{1}{2}^-)$ state and 
%to a lesser extent in the
in the more accurate results for the 
% >>>
% $0(\frac{1}{2}^-)$ state.   Similarly both positive parity states show
% an increasing mass splitting between pentaquark and two-particle
% states again suggesting repulsion as opposed to attraction.
$0(\frac{1}{2}^-)$ state.   Similarly, both positive parity states show
an increasing mass splitting between pentaquark and two-particle
states, again suggesting repulsion as opposed to attraction.
% <<<

Moreover, in every case where an interpolating field was constructed
% >>>
% to favor $J^{P}=\frac{1}{2}^{2}$ states more exotic than the colour-singlet paring of a $K$
to favor $J^P=\frac{1}{2}^+$ states which are more exotic than the
colour-singlet paring of a $K$
% <<<
and $N$, the approach to the lowest-lying state was compromised.  In
most cases, the same ground state mass was recovered in the
correlation function analysis, but with increased error bars.  This
% >>>
% provides further evidence that the lowest lying state is simply a $NK$
provides further evidence that the lowest lying state is simply an $NK$
% <<<
scattering state.  

In the case of the $I(J^P) = 1(\frac{1}{2}^-)$ state, the colour-fused
$NK$ interpolator of Eq.~(\ref{eq:NK:fused}) had sufficient overlap
with an exited state to allow a successful correlation matrix
analysis.  Again, the exotic colour-fused $NK$ interpolating failed to
produce evidence of a bound pentaquark state, the signature of a
resonance on the lattice.

Similarly, the scalar diquark-type interpolating field of
Eq.~(\ref{eq:SS:sing}) produced effective masses that lie higher than
those recovered from the colour-singlet $NK$-type interpolating field
of Eq.~(\ref{eq:NK:sing}).  Again, a low-lying pentaquark state was
not accessed, indicating the absence of the standard lattice resonance
signature.
In short, evidence supporting the existence of a spin-${1\over 2}$
pentaquark resonance does not exist in quenched QCD.

This result makes it clear that a similar analysis in full
dynamical-fermion QCD is essential to resolving the fate of the
% >>>
% putative pentaquark resonance.  We have resolved mass splittings the
% order of 100~MeV, and one might wonder what effect the dynamics of
% full QCD could have on this state.  Self-energies the order of 100 MeV
% are common place in chiral effective field theory.
putative pentaquark resonance.  We have resolved mass splittings of the
order of 100~MeV, and one might wonder what effect the dynamics of
full QCD could have on this state.
As differences in self-energies between full and quenched QCD of order
100~MeV or more have been observed \cite{Young:2002cj}, one cannot
yet rule out the possible existence of a pentaquark in full QCD.

% Consider for example the self-energies of the nucleon and Delta masses
% in quenched and full QCD reproduced here from Ref.~\cite{Young:2002cj}
% examining masses from the MILC collaboration \cite{Bernard:2001av}.
% The differences between dashed and solid curves illustrate the
% differences between quenched and full $2+1$ flavour dynamical-fermion
% results.  Even at fairly large quark masses (squared pion masses) the
% differences between quenched and full QCD self energies are large, at
% the order of 100 MeV relevant to the existence of the pentaquark
% resonance.  Moreover the Delta resonance mass drops significantly in
% going from quenched QCD to full QCD whereas the nucleon mass is
% relatively stable.  This feature leaves open the possibility of the
% existence of a pentaquark resonance in full QCD.
%
% \begin{figure}[!t]
% \begin{center}
% \epsfig{file=fqFit.ps, width=11cm, angle=90}
% \caption{Fit of finite-range regularized chiral effective field theory
%   \protect\cite{Young:2002cj} (open squares) to lattice data
%   \protect\cite{Bernard:2001av} (Quenched $\vartriangle$, Dynamical
%   $\blacktriangle$) with adjusted self-energy expressions accounting
%   for finite volume and lattice spacing artifacts.  The
%   infinite-volume, continuum limit of quenched (dashed lines) and
%   dynamical (solid lines) are shown. The lower curves and data points
%   are for the nucleon and the upper ones for the $\Delta$.}
% \label{fig:fqFit}
% \end{center}
% \end{figure}
% <<<

%%%%%%%%%%%%%%%%%%%%%%%%%%%%%%%%%%%%%%%%%%%%%%%%%%%%%%%%%%%%%%%%%%%%%%%%%%
\begin{acknowledgements}
DBL thanks K.~Maltman for interesting discussions on pentaquark models.
This work was supported by the Australian Research Council,
and the U.S. Department of Energy contract \mbox{DE-AC05-84ER40150},
under which the Southeastern Universities Research Association (SURA)
operates the Thomas Jefferson National Accelerator Facility
(Jefferson Lab).
\end{acknowledgements}

%%%%%%%%%%%%%%%%%%%%%%%%%%%%%%%%%%%%%%%%%%%%%%%%%%%%%%%%%%%%%%%%%%%%%%%%%%
%\input bib.tex
\bibliographystyle{apsrev} 
\bibliography{bib}

\end{document}